\documentclass[aps,prx,nofootinbib,twocolumn,latexsym,superscriptaddress,longbibliography]{revtex4-1}
\usepackage{amssymb}

\usepackage{latexsym}
\usepackage{amsmath}
\usepackage{bm}
\usepackage[mathcal]{eucal}
\renewcommand{\Im}{\mathrm{Im}}
\renewcommand{\Re}{\mathrm{Re}}
\newcommand{\be}{\begin{equation}}
\newcommand{\ee}{\end{equation}}
\newcommand{\Tr}{\mathrm{Tr}}

\usepackage{graphicx}
\usepackage{dcolumn}
\usepackage{amsfonts}
\usepackage{bm}
\usepackage{array}
\usepackage{natbib}

\begin{document}

\title{Thermalization of dipole oscillations in confined systems by rare collisions}
\author{Maxim Khodas}
\affiliation{Racah Institute of Physics, Hebrew University of Jerusalem, Jerusalem 91904, Israel}
\author{Alex Levchenko}
\affiliation{Department of Physics, University of Wisconsin--Madison, Madison, Wisconsin 53706, USA}
\begin{abstract}
We study the relaxation of the center-of-mass, or dipole oscillations in the system of interacting fermions confined spatially. 
With the confinement frequency $\omega_{\perp}$ fixed the particles were considered to freely move along one (quasi-1D) or two (quasi-2D) spatial dimensions.
We have focused on the regime of rare collisions, such that the inelastic collision rate, $1/\tau_{in} \ll \omega_{\perp}$.
The dipole oscillations relaxation rate, $1/\tau_{\perp}$ is obtained at three different levels: by direct perturbation theory, solving the integral Bethe-Salpeter equation and applying the memory function formalism.
As long as anharmonicity is weak, $1/\tau_{\perp} \ll 1/ \tau_{in}$ the three methods are shown to give identical results. In quasi-2D case $1/\tau_{\perp} \neq 0$ at zero temperature.
In quasi-1D system $1/\tau_{\perp} \propto T^3$ if the Fermi energy, $E_F$ lies below the critical value, $E_F < 3 \omega_{\perp}/4$. Otherwise, unless the system is close to integrability, the rate  $1/\tau_{\perp}$ has the temperature dependence similar to that in quasi-2D. In all cases the relaxation results from the excitation of particle-hole pairs propagating along unconfined directions resulting in the relationship $1/\tau_{\perp} \propto 1/\tau_{in}$, with the inelastic rate $1/\tau_{in} \neq 0$ as the phase-space opens up at finite energy of excitation, $\hbar \omega_{\perp}$. While $1/\tau_{\perp} \propto \tau_{in}$ in the hydrodynamic regime, $\omega_{\perp} \ll 1/\tau_{in}$, in the regime of rare collisions, $\omega_{\perp} \gg 1/\tau_{in}$, we obtain the opposite trend $1/\tau_{\perp} \propto 1/\tau_{in}$.
\end{abstract}

\date{\today }

\maketitle

\section{Introduction}

Interacting systems confined  to one dimension are often characterized by very slow or approach to equilibrium or even by no relaxation \cite{Dalfovo1999,Giorgini2008,Cazalilla2011,Imambekov2012}.
This stimulates the active research on the mechanisms of thermalization in such systems \cite{Moeckel2008,Rigol2009,Eckstein2009,Kollar2011,Bertini2015,Brandino2015}.

In an extended, 3D systems the relaxation normally is reached on the time scale of an inter-particle collisions.
In contrast, in some confined systems  the relaxation does not take place on the observation scale. 
The collective oscillations of the center of mass, or dipole oscillations (DOs) of cold gases of fermions in harmonic trapping potential show very slow decay \cite{Pezze2004,Altmeyer2007}.
In the ``quantum Newton's cradle'' experiment \cite{Kinoshita2006} the breathing mode excited in the effectively 1D Bose liquid does not thermalize even though the inter-particle interactions are not weak. 
In the very recent experiment \cite{Tang2017} the thermalization of momentum distribution has been studied in the array of dipolar quantum Newton's cradles. The two-time scale decay processes contains fast and short decay followed by slower near-exponential thermalization. In the work \cite{Tang2017} the second, thermalization regime has been linked to the integrability-breaking interactions mediated by the long range magnetic dipole-dipole  interactions. The reported  thermalization rates scale with the square of  integrability breaking interaction strength.

Recently, the numerical solutions of the generalized hydrodynamic equations \cite{Castro-Alvaredo2016,Bertini2016} has been obtained for harmonic and weakly anharmonic trap potentials  \cite{Caux2017}. In the harmonic case the reported change of the phase space distribution function is small. It is not the case for weak anharmonicity for which the relaxation to the almost time-independent state has been observed. The anharmonicity therefore noticeably affects the dynamics. In fact, in purely harmonic trap the DOs do not relax \cite{Brey1989,Yip1991,Dobson1994} according to the generalization of the Kohn theorem (KTh) \cite{Kohn1961}. This property holds for interacting electrons \cite{Wixforth1994,Drexler1992}, as well as systems of cold atoms \cite{Pezze2004,Altmeyer2007,Pantel2012}. The KTh restrictions are also lifted in mixtures of isotopes as different species have different oscillation frequency along the confinement direction.
This scenario has been considered theoretically in Refs.~\cite{Chiacchiera2010,Bamler2015}, and realized as mixture of Bose and Fermi superfluids in experiment \cite{Ferrier-Barbut2014}. 

In this paper we study the basic model designed to help to clarify the role of the weak anharmonicity,
of different kinds of integrability breaking perturbations, and of dimensionality  in approach to equilibrium.
Our goal is to evaluate  the relaxation rate, $1/\tau_{\perp}$ of the DOs. Specifically, we consider interacting fermions in the two settings referred to as quasi-2D (quasi-two-dimensional) and quasi-1D (quasi-one-dimensional). In the first, quasi-2D setting the particles are confined only in one, $\hat{z}$ direction and free to move in the other two. The degree of the lateral confinement is characterized by the finite frequency of oscillations $\omega_{\perp}$ in the confining potential. In the second, quasi-1D setting, the particles move freely in only one direction, confined along $\hat{z}$ direction as before, and have their motion in the third direction is suppressed by strong confinement. Here we do not address the important problem of crossover between integrable and thermalizing behavior, and rather focus on the long time near-exponential in time relaxation as reported in \cite{Tang2017}. 

Once the system is not integrable the inelastic processes with the rate $1/\tau_{in}$ take place causing the local equilibration on a microscopic scale. The hydrodynamic regime sets in when the inelastic rate, $1/\tau_{in}$ exceeds the typical frequency of macroscopic motion $\sim \omega_{\perp}$.
The relaxation on the macroscopic scale of DOs \cite{Iqbal2015} or of collective modes in higher dimensions \cite{Dalfovo1999} is controlled by the viscosity and thermal conductivity coefficients that are non-zero in the non-integrable systems \cite{Gangardt2010,Matveev2017,DeGottardi2018}.

Here we focus on the opposite regime of weak scattering, $1/\tau_{in} \ll \omega_{\perp}$.
Then the constrains imposed by energy-momentum conservations (kinematic constrains), integrability and by the KTh can be understood by applying the Fermi golden rule. As we show in Sec.~\ref{sec:PT} the kinematic constrains in quasi-1D are the same as in 1D only below some critical occupation.
The integrability is broken by virtual inter-band transitions in quasi-1D systems  \cite{Mazets2008}.
Above the critical  filling this effect is stronger as the real inter-band transition become available. 
In the hydrodynamic regime the inter-band scattering drastically modifies the bulk viscosity \cite{DeGottardi2018}.
 
The Paper is organized as follows. In Sec.~\ref{sec:Model} we formulate the model, and introduce the dimensional anharmonicity parameter. The golden rule is employed in Sec.~\ref{sec:PT} in order to estimate the relaxation rate of the DOs, $1/\tau_{\perp}$ based on the perturbation theory and a simple ansatz of the dipole correlation function. In this section we explore the distinct role of the inter-band excitations in quasi-1D and quasi-2D cases. In Sec.~\ref{sec:BS} we go beyond the perturbation theory by summing up all the contributions to the dipole correlation function that are most divergent in powers of $\tau^{-1}_{\perp}/(\omega - \omega_{\perp})$. This program is achieved by solving the integral Bethe-Salpeter equation.
In Sec.~\ref{sec:MF} we compute the relaxation rate, $1/\tau_{\perp}$ using a complementary approach based on memory function formalism \cite{Forster1975}.
The results are discussed in Sec.~\ref{sec:Discussion}.
 
\section{Formulation of the model}
\label{sec:Model}

We consider the system of interacting fermions of a mass $m$ with their motion along $z$-direction being confined and free otherwise.
The Hamiltonian of the system contains four parts,
\be\label{H1}
H = H_0 + H_{0,\perp}+  H_{\epsilon,\perp} + H_{int}\, ,
\ee
where $H_0$ is a kinetic energy of a motion in the $xy$-plane, $H_0 = \sum_j (p_{j,x}^2 + p_{j,y}^2)/ 2 m$, with index $j$ enumerating fermions and $\bm{p}_j$ is the momentum operator of $j$-th fermion. 
Here for definiteness we assume confinement in one spatial dimension. 

The confining potential is weakly anharmonic. In the absence of anharmonicity the motion along the $z$-direction is determined by a single confining frequency, $\omega_{\perp}$,
\be\label{H-harm}
H_{0,\perp} = \sum_j \left[ \frac{p_{j,z}^2}{ 2 m} + \frac{ m \omega_{\perp}^2 z_j^2}{ 2} \right]\, .
\ee
Here the position of the $j$th fermion is given by the vector, $\bm{r}_j = (x_j,y_j,z_j)$.

\begin{figure}[h]
\begin{center}
\includegraphics[width=0.7\columnwidth]{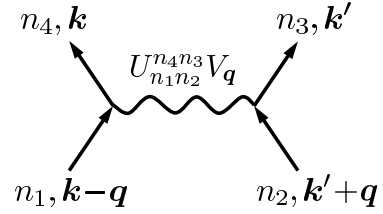}
\caption{The interaction amplitude defined by Eq.~\eqref{H_full}.
}
\label{fig:Def1}
\end{center}
\end{figure}

The energies and eigenstates of $H_0 + H_{0,\perp}$ are respectively,
\begin{align}\label{spectr}
E_{n,\bm{k}} & = (n + 1/2) \omega_{\perp} + \epsilon_k\, , 
\notag \\
\psi_{n,\bm{k}}(\bm{r}) & = \phi_n^{(0)}(z)  e ^{ i (k_x x + k_y y)}\, ,
\end{align} 
where the in-plane momentum, $\bm{k} = (k_x,k_y)$ defines the kinetic energy, $\epsilon_k = k^2/2 m$, $\phi_n(z)$, $n = 0,1,\ldots $ is the $n$-th normalized eigenstate of the harmonic oscillator \eqref{H-harm}, 
\be\label{phi_n}
\phi_n^{(0)}(z) = (2^n n! \ell_{\perp})^{-1/2} \pi^{-1/4} \exp(- z^2/2 \ell_{\perp}^2) h_n(z/\ell_{\perp})
\ee
characterized by the oscillator length $\ell_{\perp} = \sqrt{\hbar/ m \omega_{\perp}}$. Here $h_n$ are the Hermite polynomials.  We suppress the spin index in this work as the spin degrees of freedom are irrelevant to the present discussion. The second quantized form of $H_0 + H_{0,\perp}$ is
\begin{align}\label{def13}
H_0 + H_{0,\perp} = \sum_{n=0}^{\infty} \sum_{\bm{k}} E_{n,\bm{k}}  \psi_{n,\bm{k}}^{\dag} \psi_{n,\bm{k}} \, ,
\end{align}
where the operator $\psi_{n,\bm{k}}^{\dag}$ creates the fermion in a state $\psi_{n,\bm{k}}(\bm{r})$ introduced in Eq.~\eqref{spectr}. In general we characterize the weak anharmonicity by a small  parameter, $\epsilon \ll 1$, such that,
\be\label{def12}
H_{\epsilon,\perp}  = \epsilon \omega_{\perp} \sum_j f(z_j/\ell_{\perp})\, ,
\ee
where $f(x)$ is a dimensionless function of order one for $x \sim 1$.
The typical $z$-value can be taken as the spatial extent of the system.
The parameter $\epsilon$ then is the ratio of the classical frequency shift for amplitude of oscillations of the order of the system size to the base frequency, $\omega_{\perp}$. 
The two-body interaction Hamiltonian,
\be\label{def14}
H_{int}(\bm{r}_j,\bm{r}_{j'}) = U\left(\left|z_j - z_{j'}\right|/\ell_{\perp}\right) V (x_{j} - x_{j'},y_{j} - y_{j'})
\ee
is assumed to be separable as none of our conclusions depends on the exact form of $H_{int}(\bm{r}_j,\bm{r}_{j'}) $ in any essential way. In second quantization notations, Eq.~\eqref{def13} the interaction, Eq.~\eqref{def14} takes the form (see Fig.~\ref{fig:Def1}),
\begin{align}\label{H_full}
H_{int} =&  \frac{1}{2 }\sum_{n_1,n_2,n_3,n_4}  \sum_{\bm{k},\bm{k}',\bm{q}}  U_{n_1n_2}^{n_4n_3}
\notag \\
& \times V_{\bm{q}} \psi^{\dag}_{n_1,\bm{k}-\bm{q}} \psi^{\dag}_{n_2,\bm{k}'+\bm{q}}  \psi_{n_3,\bm{k}'} \psi_{n_4,\bm{k}}\, ,
\end{align}
where the dimensionless matrix elements of the transversal motion interaction
\begin{align}\label{tr_mat}
 U_{n_1n_2}^{n_4n_3} & = \int d z_1 \int d z_2   \phi^*_{n_1}(z_1)   \phi^*_{n_2}(z_2)  U(|z_1 - z_{2}|/\ell_{\perp}) 
 \notag \\
& \times  \phi_{n_3}(z_2)  \phi_{n_4}(z_1)
\end{align}
and the matrix element for an in-plane interaction
\begin{align}\label{in_mat}
V_q = \int dx dy e^{- i (q_x x + q_y y)} V(x,y)
\end{align}
has a dimensionality of the inverse density of states. We emphasize that in the anharmonic confining potential, $\epsilon \neq 0$, and the wave-functions $\phi_n(z)$ in Eq.~\eqref{tr_mat} differ from the wave-functions, $\phi^{(0)}_n(z)$ in Eq.~\eqref{phi_n}. 

Note the following properties of the transversal matrix elements, Eq.~\eqref{tr_mat}
\begin{align}\label{symm_mat}
U_{n_1n_2}^{n_4n_3} = U_{n_2n_1}^{n_3n_4}\, , \quad U_{n_1n_2}^{n_4n_3} = U_{n_4n_2}^{n_1n_3}=U_{n_1n_3}^{n_4n_2}\, .
\end{align}
The first property in Eq.~\eqref{symm_mat} follows from the symmetry of the interaction under the exchange of the coordinates $z_1$ and $z_2$. The second property follows from the time reversal symmetry as the wave-functions and the interaction potential are real.
The properties Eq.~\eqref{symm_mat} allow us to switch rows and columns in the band arguments of the matrix elements of interaction.
We use this property extensively throughout the paper.

\section{Perturbation theory results for a single occupied band}
\label{sec:PT}
In this section we perform the perturbative evaluation of the decay rate of the DOs. 
Strictly speaking the perturbation theory is ill defined at $\omega \approx \omega_{\perp}$ because the excitation energy $\omega_{\perp}$ is infinitely degenerate. Nevertheless, it provides us with the useful insight into mechanisms of DOs relaxation and gives a well defined prediction for the actual rates if interpreted in a physically sensible way. We start with the case of singly occupied band.

Let us introduce the correlation function of the observables $A$ and $B$,  
\begin{align}\label{K}
K_{A,B} (\omega) =  - i \int_0^{\infty} d t e^{ i \omega t } \langle [A^{\dag}(t), B(0)] \rangle\, ,
\end{align}
where $[A,B]=AB - BA$, and the time dependence is specified by the Heisenberg equations of motion, $\dot{A} = i [H,A(t)]$ where
$\dot{A} = \partial A/\partial t$. 
The natural choice of collective variables representing the transitions between the adjacent bands is
\begin{align}\label{PT1}
A_n = \sum_{\bm{k}} \psi_{n,\bm{k}}^{\dag}  \psi_{n-1,\bm{k}} \, ,
\end{align}
where the index $n=1,2,\ldots$. 
In a non-interacting system 
\begin{align}\label{PT27}
K^{(0)}_{A_n,A_{n'}}(\omega) = \delta_{n,n'} \frac{N_{n-1}-N_{n}}{ \omega - \omega_{\perp} + i 0^+}\, ,
\end{align}
where $N_n$ is the total occupation of the $n$-th band and $0^+$ is positive infinitesimal.
The Eq.~\eqref{PT27} shows that only the occupied bands are of importance. 
We, therefore focus on $K_{A_1,A_1}$.
In generic situation, the DOs are induced by the dipole operator,
\be\label{SM}
\mathcal{F}_1 = \sum_{n=1}^{\infty} \sqrt{n} A_n \, .
\ee

Let us compute the correlation function $K_{A_1,A_1}$ up to the second order in the interaction, $K_{A_1,A_1} \approx  K^{(0)}_{A_1,A_1} + K^{(1)}_{A_1,A_1} + K^{(2)}_{A_1,A_1}$.
The first order correction, $K^{(1)}_{A_1,A_1}\propto v_{int}/(\omega - \omega_{\perp})^2$ accounts for the renormalization of $\omega_{\perp}$ 
see Ref.~\cite{Iqbal2014} and we proceed to the analysis of the second order corrections.

\begin{figure}[h]
\begin{center}
\includegraphics[width=0.8\columnwidth]{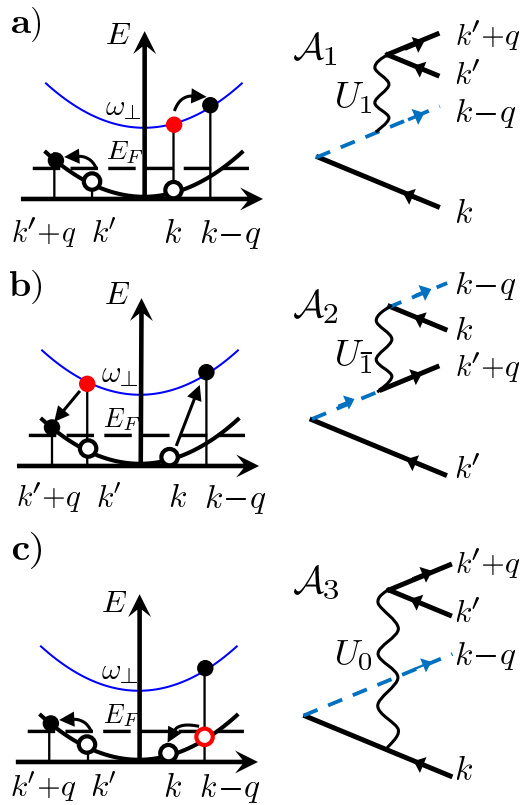}
\caption{ (color online) 
The three scattering processes with the amplitudes $\mathcal{A}_1$, $\mathcal{A}_2$ and $\mathcal{A}_3$ given by Eq.~\eqref{ampl-13} are shown as panels a), b) and c) respectively.
On the left the lowest and the next bands separated by $\omega_{\perp}$ in energy, $E$ are shown as thick(black) and thin(blue) parabolic lines.
The Fermi level at $E=E_F$ crosses only the lowest energy band.
All panels show the same set of particles with momenta $k'+q$ and $k-q$ and holes with momenta $k$ and $k'$ in the final state (black).
The virtual state includes, a) a particle at the momentum $k$ (red), b) a particle at momentum $k'$ (red) and c) a hole at the momentum $k-q$ (red).
On the right the amplitudes of the scattering process is represented diagrammatically.
The dashed(blue) and thick(black) lines denote propagators of the excitations in the upper and lower bands respectively.}
\label{fig:ampl1}
\end{center}
\end{figure}

\subsection{Second order perturbation theory} 
\label{sec:PT_2nd}
\begin{figure}[h]
\begin{center}
\includegraphics[width=0.9\columnwidth]{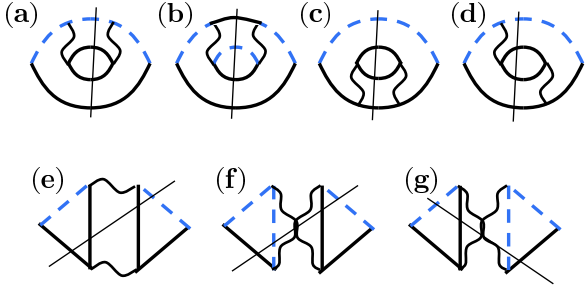}
\caption{ (color online) 
The diagrammatic representation of the second order correction to the correlation function, $K_{A_1,A_1}$, resulting from the scattering processes indicated in Fig.~\ref{fig:ampl1}.
The solid thick (black) lines represents the propagator of particles in the first unoccupied band.
The dashed (blue) lines represents the propagator of holes in the lowest band.
The expression Eq.~\eqref{GR-13} represents the total contribution from the processes with the four quasi-particles in the final state.
The propagators of these quasiparticles are crossed by the thin (black) line.
Panels: (a) and (b) represent two kinds of the self energy of the particles,  (c) represents the self energy of the holes,
(d),(e),(f) and (g) represent all vertex corrections diagrams to the second order in interaction.}
\label{fig:Diagrams1}
\end{center}
\end{figure}

By utilizing the golden rule, we write 
\begin{align}\label{GR-13}
\Im K^{(2)}_{A_1,A_1}&(\omega)  = -\pi \left( 1 - e^{ -\omega/T} \right) \sum_{\bm{k},\bm{k}',\bm{q}} \left| \sum_{i=1}^{3} \mathcal{A}_i \right|^2 
\notag \\
\times & (1-f_{0,\bm{k}' + \bm{q}}) f_{0,\bm{k}' } (1-f_{1,\bm{k} - \bm{q}}) f_{0,\bm{k} } 
\notag \\
\times &  \delta \left( \omega - \xi_{0,\bm{k}'+ \bm{q}} +  \xi_{0,\bm{k}'}  - \xi_{1,\bm{k} - \bm{q} } + \xi_{0,\bm{k}} \right)\, ,
\end{align}
where $\xi_{n,\bm{k}} = E_{n,\bm{k}} - \mu$ is the energy in Eq.~\eqref{spectr} measured relative to the chemical potential $\mu$, and $f_{n,\bm{k}} = (e^{\xi_{n,\bm{k}}/T } -1)^{-1}$ is the Fermi-Dirac function.
Furthermore, ignoring the exchange processes for brevity, the three amplitudes, $\mathcal{A}_i$, $i=1,2,3$ of the scattering processes shown in Fig.~\ref{fig:ampl1} read
\be\label{ampl-13}
\mathcal{A}_1 =  \frac{V_{\bm{q}} U_1} { \omega - \omega_{\perp}}, \, 
\mathcal{A}_2 = \frac{V_{\bm{q}} U_{\bar{1}}}{ \omega - \omega_{\perp}} ,  \,
\mathcal{A}_3 = -\frac{V_{\bm{q}} U_0 }{ \omega - \omega_{\perp}} ,
\ee
where we  have introduced shortened notations, $U_{00}^{00} = U_0$, $U_{01}^{01} = U_1$ and $U_{01}^{10} = U_{\bar{1}}$. Equations \eqref{GR-13} and \eqref{ampl-13} express in a compact form the result of the summation of eight diagrams shown in Fig.~\ref{fig:Diagrams1}. The exchange terms in the scattering amplitude are further discussed in Sec. V. 

To obtain an insight into Eq.~\eqref{GR-13} it is useful to write it in alternative form more closely related to the life time of a single particle excitations.
The inelastic scattering rate of a particle in the band $m=1$ with momentum $\bm{k}$, $1/\tau_{1\bm{k}}$ contains two contributions,
$1/\tau_{1\bm{k}} =1/\tau_{1a\bm{k}} + 1/\tau_{1b\bm{k}}$, represented by the self energy insertions in Fig.~\ref{fig:Diagrams1}a,b respectively.
The quantities $1/\bar{\tau}_{1a\bm{k}} = U_{1}^{-2} /\tau_{1a\bm{k}}$ and $1/\bar{\tau}_{1b\bm{k}} = U_{\bar{1}}^{-2} /\tau_{1b\bm{k}}$ do not depend on the confining potential and are described solely in terms of the motion in the unconfined direction(s). 
Similarly, for the hole relaxation rate  $1/\tau_{0\bm{k}}$, see Fig.~\ref{fig:Diagrams1}c we define the rate $1/\bar{\tau}_{0\bm{k}} = U_{0}^{-2}/\tau_{0\bm{k}} $ which is independent of the form of the confining potential.
These rates are given in terms of the golden rule as well,
\begin{subequations}\label{tau_bar}
\begin{align}\label{tau_bar_a}
\frac{1}{\bar{\tau}_{1a\bm{k}}} & =\frac{2 \pi}{ 1 - f_{1\bm{k}} } \sum_{\bm{k}',\bm{q}} V_{\bm{q}}^2 (1-f_{1,\bm{k}-\bm{q}}) (1 - f_{0,\bm{k}'+\bm{q}}) f_{0,\bm{k}'} 
\notag \\
& \times \delta(\xi_{1,\bm{k}} - \xi_{0,\bm{k}'+\bm{q}} + \xi_{0,\bm{k}'}  - \xi_{1,\bm{k}-\bm{q}})  ,
\end{align}
\begin{align}\label{tau_bar_b}
\frac{1}{\bar{\tau}_{1b\bm{k}}} & =  \frac{2 \pi}{1 - f_{1\bm{k}}} \sum_{\bm{k}',\bm{q}} V_{\bm{q}}^2 (1-f_{0,\bm{k}-\bm{q}})(1 - f_{1,\bm{k}'+\bm{q}}) f_{0,\bm{k}'} 
\notag \\
& \times \delta(\xi_{1,\bm{k}} - \xi_{1,\bm{k}'+\bm{q}} + \xi_{0,\bm{k}'}  - \xi_{0,\bm{k}-\bm{q}})  ,
\end{align}
\begin{align}\label{tau_bar_c}
\frac{1}{\bar{\tau}_{0\bm{k}}} & =  \frac{2 \pi}{f_{0\bm{k}}} \sum_{\bm{k}',\bm{q}} V_{\bm{q}}^2 f_{0,\bm{k}+\bm{q}}(1 - f_{0,\bm{k}'+\bm{q}}) f_{0,\bm{k}'} 
\notag \\
& \times \delta(\xi_{0,\bm{k}}  - \xi_{0,\bm{k}+\bm{q}}+ \xi_{0,\bm{k}'+\bm{q}} - \xi_{0,\bm{k}'} )  .
\end{align} 
Equations \eqref{tau_bar} are further elaborated in App.~\ref{app:qp_rates_A}.
\end{subequations}
With each of the definitions Eq.~\eqref{tau_bar} one can rewrite Eq.~\eqref{GR-13} in one of the three equivalent ways,
\begin{subequations}\label{GR-14}
\begin{align}
\Im K^{(2)}_{A_1,A_1}& = \frac{ (U_0 - U_1 -  U_{\bar{1}})^2}{(\omega - \omega_{\perp})^2}  
 \sum_{\bm{k}} \frac{f_{1,\bm{k}} - f_{0,\bm{k}}}{2 \bar{\tau}_{1a\bm{k}} } 
\\
& =\frac{ (U_0 - U_1 -  U_{\bar{1}})^2}{(\omega - \omega_{\perp})^2  }
 \sum_{\bm{k}}  \frac{ f_{1,\bm{k}} - f_{0,\bm{k}} }{ 2 {\bar\tau}_{1b\bm{k}} } 
 \\
& =\frac{ (U_0 - U_1 -  U_{\bar{1}})^2}{(\omega - \omega_{\perp})^2  }
 \sum_{\bm{k}}  \frac{ f_{1,\bm{k}} - f_{0,\bm{k}} }{ 2 \bar{\tau}_{0\bm{k}} }  \label{K_tau_c},
 \end{align}
\end{subequations}
where the identity $(1 - e^{ - \omega_{\perp}/T} ) f_{0\bm{k}} ( 1 - f_{1\bm{k}})= f_{0\bm{k}} - f_{1\bm{k}}$ has been used.

In order to relate $K^{(2)}_{A_1,A_1}$ to the actual decay rate of the DOs, $1/\tau_{\perp}$ in this section we adopt a simple guess
\begin{align}\label{PT27-A}
K_{A_{1},A_{1}}(\omega) \approx \frac{N_0-N_{1}}{ \omega - \omega_{\perp} + i/\tau^{(2)}_{\perp}}\, .
\end{align}
which as we show in Sec.~\ref{sec:BS} holds under conditions of weak anharmonicity, $\epsilon \ll 1$. 
By matching the result, Eq.~\eqref{GR-14} with the expansion of Eq.~\eqref{PT27-A} in $1/\tau_{\perp}$ we obtain to the second order,
\begin{align}\label{PT-res1}
\frac{1}{\tau_{\perp}^{(2)}} = \frac{ (U_0 - U_1 -  U_{\bar{1}})^2}{N_0 - N_1} \sum_{\bm{k}} \frac{ f_{0,\bm{k}} - f_{1,\bm{k}} }{ 2 \bar{\tau}_{0,a,b,\bm{k}} }\, .
\end{align} 
Alternative form of this equation relates $1/\tau_{\perp}$ to the average quasiparticle relaxation rate,
\begin{align}\label{PT-res1_A}
\frac{1}{\tau_{\perp}^{(2)}} = (U_0 - U_1 -  U_{\bar{1}})^2
\frac{       \sum_{\bm{k}} ( f_{0,\bm{k}} - f_{1,\bm{k}} ) (2 \bar{\tau}_{0,a,b,\bm{k}} )^{-1}       }{  \sum_{\bm{k}} (f_{0,\bm{k}} - f_{1,\bm{k}})  }\, .
\end{align} 
Equation \eqref{PT-res1_A} up to a prefactor $(U_0 - U_1 -  U_{\bar{1}})^2$ and replacement $f_{1,\bm{k}} \rightarrow f_{2,\bm{k}}$ is identical to the expression for the decay rate of quadruple excitations in a quasi-two-dimensional system of fermionic polar molecules created by the lattice modulation pulse used in Ref.~\cite{Babadi2011}. We stress that the prefactor in Eq.~\eqref{PT-res1_A} is absolutely crucial as it results from finite state interactions, and in our case guarantees the KTh.

Below we point out two immediate consequences of Eq.~\eqref{PT-res1}.
First, KTh is satisfied in view of the identity, $U_0 - U_1 -  U_{\bar{1}}=0$ valid for harmonic confinement for any interaction. For anharmonic potential $U_0 - U_1 -  U_{\bar{1}} \neq 0$ and in general, $1/\tau_{\perp} \neq 0$. So that $1/\tau_{\perp}^{(2)} $ can be non-zero only in an anharmonic trap.
Secondly, according to Eq.~\eqref{tau_bar_c} the relaxation rate of the hole, $1/\bar{\tau}_{0\bm{k}}$ is determined by the intra-band processes, and hence identical to the inelastic rate of the hole in the pure 1D or 2D case.
In the 2D systems this rate is non-zero and as a result $1/\tau_{\perp}^{(2)}\neq 0$.
In contrast, in 1D systems with quadratic dispersions the inelastic rate vanishes, in fact to all orders in interactions, see Ref.~\cite{Khodas2007a}. Therefore, according to Eq.~\eqref{PT-res1} in quasi-1D to the second order in interaction the DOs have vanishing relaxation rate even in anharmonic trap at least at $T=0$.

By representing Eq.~\eqref{GR-13} as a convolution of the polarization operators, 
\begin{align}\label{Pi}
\Pi_{nm}(\bm{q},\omega) = \sum_{\bm{k}} \frac{ f_{m,\bm{k}} -  f_{n,\bm{k}+\bm{q} }} {  \omega+i0 - \xi_{n,\bm{k}+\bm{q} } + \xi_{m,\bm{k} }}\, ,
\end{align}
we obtain an alternative form of the result, Eq.~\eqref{PT-res1},
\begin{align}\label{rate}
\frac{ 1}{\tau_\perp^{(2)}} & =
 \frac{(U_{1} + U_{\bar{1}} - U_{0})^2 }{N_0 - N_1} 
 \notag \\
& \times \!\!  \int_0^{\omega_{\perp}} \frac{d \Omega}{\pi} \sum_{\bm{q}} V_{\bm{q}}^2 \Im \Pi_{10}( \bm{q},\omega_{\perp} - \Omega)  \Im \Pi_{00}( \bm{q},\Omega)\, .  
\end{align}

\begin{figure}[h]
\begin{center}
\includegraphics[width=1.0\columnwidth]{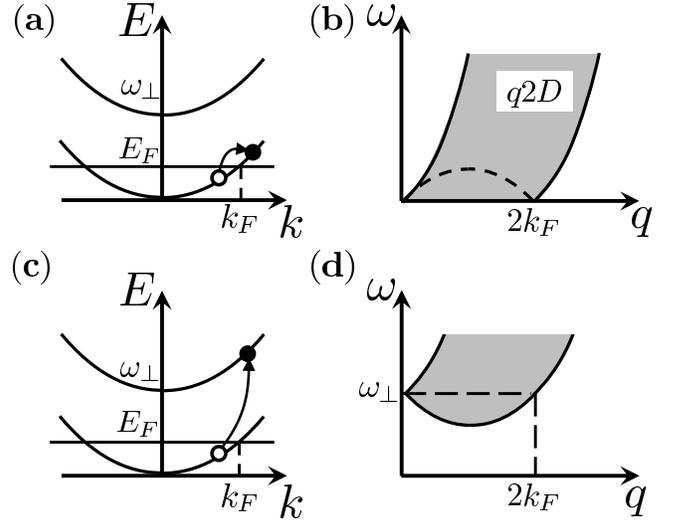}
\caption{ (color online) 
(a) The two parabolas show energy as a function of an in-plane momentum of the two lowest bands of transversal quantization. 
The energy splitting between bands is $\omega_{\perp}$.
Only the lowest band is occupied.
All the states up to the Fermi momentum, $k_F$ and up to Fermi energy, $E_F$ are populated at zero temperature.
The continuum of intra-band particle-hole excitations is formed by promoting the particle above $E_F$.
The upper edge of the continuum is the line $\omega = q ^2/2 m+ k_Fq$, the lower edge is $\omega=\max \{ 0, q ^2/2 m - k_Fq\}$. 
(b) The intra-band particle-hole continuum in quasi-2D. 
The shaded area in the $(q,\omega)$-plane includes all the possible pairs of momentum and energy $(q,\omega)$ of intra-band excitations such as the one shown in (a).
(c) The same as (a) with inter-band particle-hole excitation shown instead of the intra-band one.
(d) The continuum of inter-band particle-hole excitations is represented by the shaded area in the $(q,\omega)$-plane. 
}
\label{fig:cont1}
\end{center}
\end{figure}

\begin{figure}[h]
\begin{center}
\includegraphics[width=1.0\columnwidth]{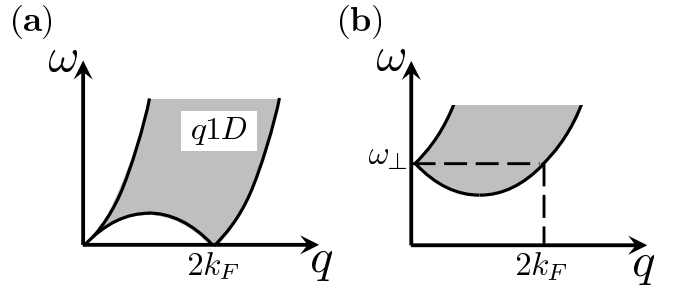}
\caption{ (color online) 
(a) The particle-hole continuum  of intra-band excitations in quasi-1D
(b) The particle-hole continuum  of inter-band excitations in quasi-1D
}
\label{fig:cont2}
\end{center}
\end{figure}

\begin{figure}[h]
\begin{center}
\includegraphics[width=0.8\columnwidth]{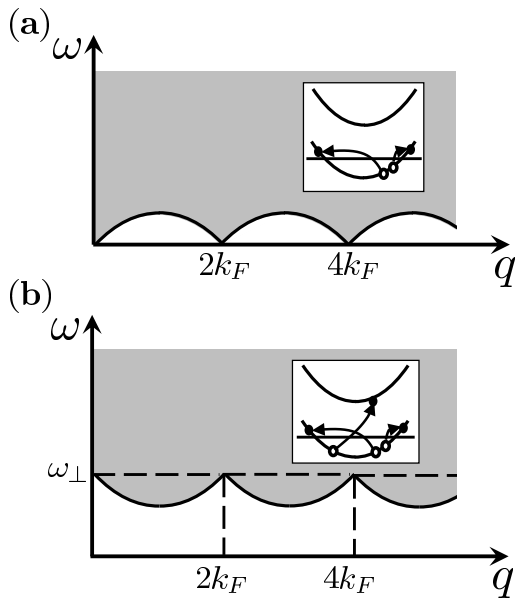}
\caption{ (color online) 
(a) The intra-band continuum of excitations in quasi-1D.
The shaded area bounded from below by $\omega^{0\rightarrow0}(q)$ represents all possible pairs of momentum and energy, $q$ and $\omega$ of all intra-band excitations. 
(b) 
The continuum of excitations in quasi-1D with exactly one inter-band particle-hole excitation and arbitrary number of intra-band pairs.
The shaded area represents bounded from below by $\omega^{0\rightarrow1}(q)$, Eq.~\eqref{min01} includes all points $(q,\omega)$ of kinematically allowed excitations of this kind.
}
\label{fig:cont3}
\end{center}
\end{figure}

\begin{figure}[h]
\begin{center}
\includegraphics[width=0.7\columnwidth]{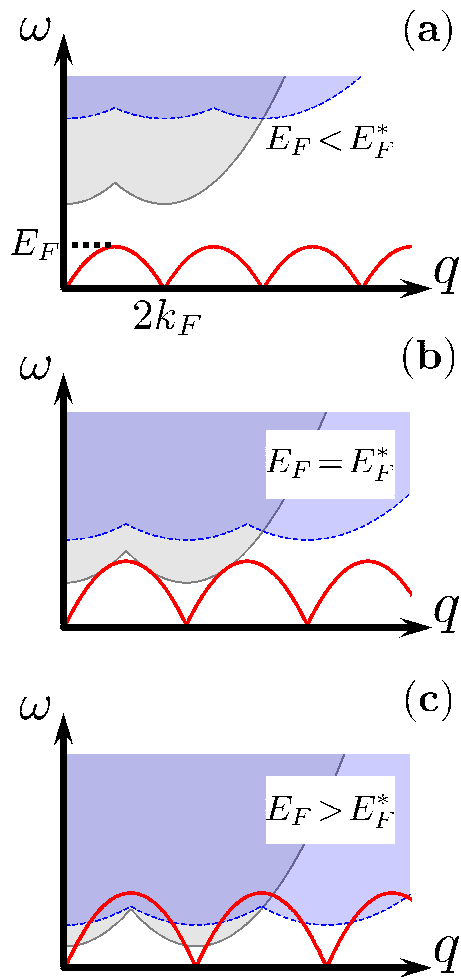}
\caption{The thick solid (red) line shows the lower edge of the intra-band continuum, $\omega^{0\rightarrow0}(q)$.
The thin solid (grey) line indicates the lower edge  $\omega^{0\rightarrow1}_{\{2\}}(q)$ ($\omega^{0\rightarrow1}_{\{4\}}(q)$) of the continuum of two (four) inter-band pairs excitations, see Eq.~\eqref{boundary}. 
Panel (a): $E_F = \omega_{\perp}/2$ and $1/\tau_{\perp}=0$; panel (b): $E_F = 3\omega_{\perp}/4$ and $1/\tau_{\perp}=0$; panel (c): $E_F = 7\omega_{\perp}/8$ and $1/\tau_{\perp} \neq 0$.
At critical filling $E_F = E_F^*=3 \omega_{\perp}/4$ the edges $\omega^{0\rightarrow0}(q)$ and $\omega^{0\rightarrow1}_{\{2\}}(q)$ touch, panel (b).
}
\label{fig:cont4}
\end{center}
\end{figure}

Equation \eqref{rate} has a transparent meaning. The finite relaxation rate of DOs is obtained when the two particle-hole excitations can be found: one inter- and the other intra-band such that their total momentum is zero and their total energy is $\omega_{\perp}$. Each of the two kinds of particle-hole excitations forms a continuum shown in Figs.~\ref{fig:cont1} and \ref{fig:cont2} in quasi-1D and quasi-2D systems respectively.
In these graphs each excitation is shown on a momentum-energy plane forming
the shaded regions on these graphs. Below we address the quasi-1D and quasi-2D cases separately based on the above interpretation.

\subsection{$1/\tau^{(2)}_{\perp}$ in quasi-2D systems}

We start with verifying the statement $1/\tau^{(2)}_{\perp} \neq 0$ based on Eq.~\eqref{rate}.
Fig.~\ref{fig:cont1}b(d) represents the familiar intra(inter)-band continuum in 2D. 
The intra-band excitations with momentum $q < 2 k_F$ and energy $\omega  = v_F q - q^2/2m$ are represented by the dashed line in Fig.~\ref{fig:cont1}b. The  inter-band excitations with momentum $-q$ and energy $\omega = \omega_{\perp} - v_F q + q^2/ 2 m$ are shown as the lowest boundary of the inter-band continuum in Fig.~\ref{fig:cont1}d. Clearly, when the total momentum of two such pairs is zero, the total energy is $\omega_{\perp}$. Since there are intra-band excitations both below and above the dashed line on Fig.~\ref{fig:cont1}b, the frequency $\omega_{\perp}$ falls within the continuum of two particle-hole excitations.
This implies the finite life-time of DOs already in the second order of perturbation theory.

For completeness we have computed this rate starting from Eq.~\eqref{rate}, 
\begin{align}\label{rate_eval}
\frac{ 1}{\tau_\perp} & = \bar{C} 
\epsilon^2  (V m)^2  E_F\, ,
\end{align}
where $\bar{C}$ is non-universal constant, which for a quartic anharmonicity and for short-range interaction evaluates to $\bar{C} \approx 0.94  \left( 3/4 \right)^2/(2 \pi)^3$, (see App.~\ref{app:rates} for details).
The result, Eq.~\eqref{rate_eval} corresponds to the Fermi liquid relaxation rate if the energy window available for the decay is identified as $E_F$.
It is however smaller by a factor of $\epsilon^2$ because of the matrix elements suppression imposed by the KTh.

\subsection{Relaxation of DOs in quasi-1D systems at $T=0$}

We now discuss the relaxation of the DOs in quasi-1D. The $T=0$ result, $1/\tau_{\perp}^{(2)}=0$ can be understood in two ways. First, based on the representation \eqref{PT-res1} we have traced it back to the vanishing of the hole inelastic relaxation rate in pure 1D \cite{Khodas2007a}.
Alternative interpretation we adopt in this section is based on the representation \eqref{rate}.
The latter is formulated in terms of the intra- and inter-band continua shown in Fig.~\ref{fig:cont2}a and Fig.~\ref{fig:cont2}b, respectively.
The excitations available at the second order in interaction contain one intra-band pair with momentum $q$ and one intra-band pair with the opposite momentum, $-q$.
For $|q|> 2k_F$ the inter-band pairs have energy exceeding $\omega_{\perp}$.
Therefore the total energy of the two pairs is above $\omega_{\perp}$.
For $|q| < 2k_F$ the lowest possible energy of the inter-band pairs is $\omega = \omega_{\perp} - v_F q + q^2/ 2 m$ while the lowest energy of the intra-band pair is $\omega =  v_F q - q^2/ 2 m$.
As a result, in contrast to the quasi-2D case, $\omega_{\perp}$ is a lower edge of the continuum of the two particle-hole excitations, and $1/\tau_{\perp}^{(2)}=0$.  

It is natural to ask whether $1/\tau_{\perp}=0$ to all orders in interaction.
Below we argue that it is the case provided that $E_F< E_F^* = 3 \omega_{\perp}/4$, and in general
the rate $1/\tau_{\perp}$ is finite otherwise.

As the confinement potential is assumed to be symmetric the parity is conserved.
The perturbation $A_1$, Eq.~\eqref{PT1} connects neighboring states of transversal quantization of opposite parity and therefore is parity odd. As a result, the sum of the band indices of all the excitations in the finite state has to be likewise odd. Furthermore, the excitations to the second or higher unoccupied bands are not of interest as they carry energy which exceeds $\omega_{\perp}$.
We therefore limit the consideration to the excitation to the first unoccupied band.
The allowed finite states include any odd number of particles in the first band. 
Below we separately consider the finite states with one and three particles in the first band.

\subsubsection{Contribution of finite states with a single particle in the first unoccupied band to $1/\tau_{\perp}$}
\label{sec:PT_tau1}
It is illuminating to think of these finite states as composites of the two types of excitations.
The first type includes all the possible intra-band particle-hole excitations.
These excitations form a continuum which is identical to the particle-hole continuum in the pure 1D case, see Fig.~\ref{fig:cont3}a.
It is unbounded from above and bounded from below by 
$\omega^{0\rightarrow0}(q) = \max_{n=-\infty}^{\infty}[v_F(q - 2 k_F n) + (q - 2 k_F n)^2/2 m ]$, \cite{Khodas2006,Khodas2007a}.
The second category includes a single inter-band particle-hole pair in addition to an arbitrary number of intra-band ones.
These excitations likewise form a continuum unbounded from above and shown in Fig.~\ref{fig:cont3}b.
It's lower boundary is 
\be\label{min01}
\omega^{0\rightarrow1}(q) =\omega_{\perp} - \omega^{0\rightarrow0}(q)\, .
\ee

An excitation with one particle in the unoccupied band has energy $\omega > \min_q [ \omega^{0\rightarrow0}(q) + \omega^{0\rightarrow1}(q) ]= \omega_{\perp}$.
Therefore, $\omega_{\perp}$ forms a boundary of a continuum considered, and DOs remain protected against the decay. The same conclusion would be reached if one would consider just a single inter-band particle-hole pair without any intra-band excitations with the lower boundary, $\omega_{\{1\}}^{0\rightarrow1}(q)= \omega_{\perp} - v_F q + q^2 / 2 m$.
Here the subscript in curly brackets denote the total number of  particle-hole excitations. 

\subsubsection{Contribution of finite states with more than one  particle in the first unoccupied band to $1/\tau_{\perp}$}
\label{sec:PT_tau2}
Now we explore the possibility of the relaxation into excitations which contain 3, 5 or more particles in the first unoccupied band. As the Fermi energy, $E_F$ approaches the inter-band separation, $\omega_{\perp}$ the energy cost of an inter-band excitation $\omega_{\perp} -E_F$ decreases.
This makes it necessary to consider the excitations with more than one particle in the unoccupied band as a possible decay channel.

Consider the lower boundary of these continua, $\omega^{0\rightarrow1}_{\{3\}}(q)$, $\omega^{0\rightarrow1}_{\{5\}}(q)$ or in general $\omega^{0\rightarrow1}_{\{2 l +1\}}(q)$ where as before, $l$ denotes the total number of particle-hole excitations, al of which are inter-band.
The rate $1/\tau_{\perp}$ remains zero provided $\omega_{\perp} \leq \omega^{0\rightarrow1}_{\{2 l +1\}}(q\!=\!0)$ for some $l$, and $1/\tau_{\perp} \neq 0$ otherwise.

It is again expedient to view an arbitrary excitation with $2l+1$ as composite of two,
\begin{align}\label{min03}
\omega^{0\rightarrow1}_{\{2 l +1\}}(q\!=\!0) = \min_q\left[ \omega^{0\rightarrow1}_{\{1\}}(-q) + \omega^{0\rightarrow1}_{\{2 l \}}(q)\right]\, .
\end{align}
By combining Eqs.~\eqref{min01} and \eqref{min03} we conclude that $1/\tau_{\perp}$ is finite if and only if there exists a wave-number $q^*$ such that $\omega^{0\rightarrow1}_{\{2 l \}}(q^*) <  \omega^{0\rightarrow0}(q^*)$. In fact since for $n>0$, $\omega^{0\rightarrow1}_{\{2 (l+n) \}}(q) > \omega^{0\rightarrow1}_{\{2 l \}}(q)$ our condition is $\omega^{0\rightarrow1}_{\{2 \}}(q^*) <  \omega^{0\rightarrow0}(q^*)$.

The boundaries $\omega^{0\rightarrow1}_{\{2  \}}(q)$ and $\omega^{0\rightarrow1}_{\{4  \}}(q)$ are presented in Fig.~\ref{fig:cont4} for different filling of the lowest band.
Explicitly, (see App.~\ref{app:edges})
\begin{align}\label{boundary}
\omega^{0\rightarrow1}_{\{2l \}}(q) = 2 l (\omega_{\perp} - E_F) + \min_{u=0,\pm1,\ldots,\pm l} \frac{(q - 2 u k_F)^2}{ 4 l m}\, .
\end{align}
The important observation is that
for low enough fillings, $E_F < E_F^* = 3 \omega_{\perp}/4$, Fig.~\ref{fig:cont4}a,  $\omega^{0\rightarrow1}_{\{2 \}}(q) >  \omega^{0\rightarrow0}(q)$ for all  $q$, and $1/\tau_{\perp} = 0$.
The two continua touch at $E_F = E_F^*$, Fig.~\ref{fig:cont4}b
At $E_F>E_F^*$ the inequality $\omega^{0\rightarrow1}_{\{2  \}}(q) <  \omega^{0\rightarrow0}(q)$ is satisfied for a finite range of momenta, and $1/\tau_{\perp} \neq 0$ as a result.

So far we have obtained the following results. According to KTh if the confining potential is strictly harmonic, $1/\tau_{\perp} = 0$ regardless of other details. We have observed that it results from the cancellation of the matrix elements of arbitrary translationally invariant interaction between states of the harmonic oscillator. Once the confinement is anharmonic the DOs in general have a finite lifetime. In quasi-2D it decays even at $T=0$ regardless of the band occupations, and is weakly temperature dependent. In quasi-1D at $T=0$ the decay is possible only if the occupation of the lowest band is large enough to open the phase space  for the inter-band processes. In this case the relaxation of the DOs in quasi-1D and quasi-2D is qualitatively similar.
Below the critical filling only intra-band processes are available. These processes are inefficient in relaxing the DOs. At finite $T$ the relaxation rate due to the intra-band processes is finite.

Let us remark that in quasi-1D case $1/\tau_{\perp} =0$ if the transverse matrix elements are equal to the same constant and fermion interact via the singular interaction, $V_{q} \propto q^2$.
The latter is the case of Cheon-Shigehara model \cite{Cheon1999}, dual to the Lieb-Liniger model.
As a result in such a model the relaxation does not occur \cite{Lunde2007,AL-PRB11}.

\subsection{Relaxation of DOs in quasi-1D systems at $T \neq 0$}

Let us now briefly consider the relaxation of the DOs in quasi-1D system at finite yet low temperatures, $T \ll E_F$. The scaling of $1/\tau_{\perp}$ with temperature readily follows from the representation, Eq.~\eqref{PT-res1} based on the results of Refs.~\cite{Karzig2010,Micklitz2011} on the energy-relaxation rate of the hot particles in quantum wires. The finite hole energy-relaxation rate, $1/\bar{\tau}_{0k}$ is obtained at fourth order perturbation theory interaction. Ignoring any dependence of the scattering amplitudes $\mathcal{A}$ of these processes on energy we have the estimates, 
$1/\bar{\tau}_{0,k} \propto |\mathcal{A}|^2 m T [\xi_{0,k}/v_F]^2$ for $\xi_{0,k} \ll \sqrt{E_F T}$ and  $1/\bar{\tau}_{0,k} \propto |\mathcal{A}|^2 m^2 E_F T^3/\xi_{0,k}^2$ for $|\xi_{0,k}| \gg \sqrt{E_F T}$, \cite{Karzig2010,Micklitz2011,Imambekov2012}.
These results follow from the scaling of the phase space available for relaxation of holes.
In our situation all the holes contribute to $1/\tau_{\perp}$ additively.
Clearly, the deeper the hole the larger relaxation rate it has.
For typical hole at energy $ |\xi_{0,k}| \lesssim E_F$ the phase space scales as $T^3$.
Indeed, there are five momenta parametrizing the final states: three for holes and two for particles.  
With two integrations removed by the conservation laws the remaining three integrations produce the $T^3$ scaling.
This arguments in fact requires that the hole gives up energy less than $\xi_{0,k}$, i.e. in small portions.

As $\sqrt{E_F T } \ll E_F$, only small part of all holes have the relaxation rate scaling linearly with $T$, and we obtain the estimate,
\begin{align}\label{hole_1D}
\frac{1}{\tau_{\perp}(T)} \propto \epsilon^2 |\mathcal{A}|^2 m^2 \frac{T^3}{E_F}\, ,
\end{align}
where we have reintroduced the prefactor $\epsilon^2$ in compliance with the KTh.

\section{Relaxation rate from the solution of the integral Bethe-Salpeter equation}
\label{sec:BS}
In this section we compute the life-time of the DOs in the case of a single occupied band by solving the integral Bethe-Salpeter equation. In this approach we analyze the integral vertex functions closely linked to the correlation functions of $A_{n}$ that satisfies the integral Bethe-Salpeter equation to find the location of the poles of the vertex function in the complex frequency plain. The integral equation of the kind we discuss here has been introduced in the works \cite{Eliashberg1961,Eliashberg1962} studying the universal long-wavelength properties of Fermi liquid. The integral equation on vertices and density correlation function has been obtained and solved for the marginal Fermi liquid \cite{Shekhter2009}.
This approach has been implemented in computing the $T^2$ resistivity correction in Fermi liquid in the presence of periodic lattice potential in Ref.~\cite{Yamada1986}.
The processes violating the KTh are treated as a perturbation in a vector space of the vertex functions in a manner similar to the analysis of the the diffusion constant in the anisotropic conductors \cite{Wolfle1984}. Similar approach has been applied earlier to the study of the transversal spin oscillations across a quasi-1d channel, \cite{Berman2014}. In some sense, our calculations include the ideas used in Ref.~\cite{Ipatova1963} for the calculation of the transversal spin relaxation time, $T_2$ due to the spin non-conserving magnetic dipole interaction in Fermi liquid.

It is convenient to employ the finite-temperature, Matsubara formalism for the present discussion.
The Matsubara Green functions defined in the standard way,
\begin{align}\label{K_Mats}
\mathcal{K}_{A,B} (\omega_n) =  - \int_0^{\beta} d \tau e^{ i \omega_n \tau } \langle A^{\dag}(\tau)B(0) \rangle\, ,
\end{align}
$\omega_n = 2 \pi T n$, with integer $n$ that gives the retarded correlation function, Eq.~\eqref{K} upon the analytic continuation, $i \omega_n \rightarrow \omega + i 0$.
As we consider the case of a single occupied band, the correlation function we specifically study is $\mathcal{K}_{A_1,A_1}(i \omega_n)$.
The amputated Matsubara vertex function, $\Gamma_{\bm{k}}(\epsilon_m,\omega_n)$ is defined by the relation,
\begin{align}\label{def_G}
\mathcal{K}_{A_1,A_1} (i\omega_n) = T \sum_{\epsilon_m,\bm{k}} \mathcal{G}_{1,\bm{k}}(\epsilon_m + \omega_n) \mathcal{G}_{0,\bm{k}}(\epsilon_m )    \Gamma_{\bm{k}}(\epsilon_m,\omega_n)\, ,
\end{align}
where the single-particle Green function read,
\begin{align}\label{GrF}
\mathcal{G}_{0(1),\bm{k}}(\epsilon_m) = - \int_0^{\beta} d \tau' e^{ i \epsilon_m \tau'} \langle \psi_{0(1),\bm{k}}(\tau') \psi_{0(1),\bm{k}}^{\dag} \rangle\, .
\end{align}

\begin{figure}
\begin{center}
\includegraphics[width=0.8\columnwidth]{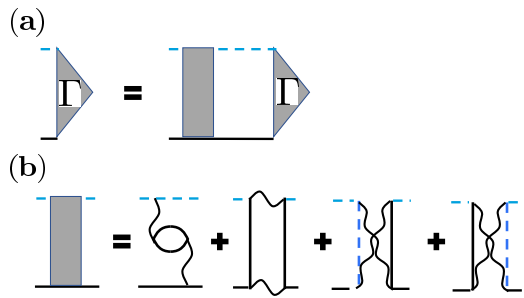}
\caption{(color on line)
The solid thick (black) lines represents the propagator of particles in the first unoccupied band.
The dashed (blue) lines represents the propagator of holes in the lowest band.
(a) The graphical representation of Bethe-Salpeter equation, \eqref{BS13} satisfied by the vertex function, $\Gamma_{k}(\omega)$ introduced in Eq.~\eqref{def_G}.
(b) the scattering amplitude contains four contributions that are in one to one correspondence to the vertex correction shown in Fig.~\ref{fig:Diagrams1}d,e,f,g respectively.}
\label{fig:BSE}
\end{center}
\end{figure}

The Bethe-Salpeter equation is presented graphically in Fig.~\ref{fig:BSE} and can be written in the form,
\begin{align}\label{BS13}
T&\sum_{\epsilon} \mathcal{G}_{1,\bm{k}}(\epsilon + \omega)  \mathcal{G}_{0,\bm{k}}(\epsilon) \Gamma_{\bm{k},\epsilon}(\omega)
=
\notag \\
  T^2& \sum_{\epsilon,\epsilon'} \sum_{\bm{k}'} \mathcal{G}_{1,\bm{k}}(\epsilon + \omega)  \mathcal{G}_{0,\bm{k}}(\epsilon) 
 \notag \\ 
& \times \big[ U_{0} U_{1} T_{1\bm{k}\bm{k}'}(\epsilon-\epsilon') + U_{0} U_{\bar{1}} T_{2\bm{k}\bm{k}'}(\epsilon-\epsilon') 
\notag \\
&\, \,
+ U_{1} U_{\bar{1}} T_{3\bm{k}\bm{k}'}(\epsilon'+\epsilon+\omega)
+ U_{1} U_{\bar{1}} T_{4\bm{k}\bm{k}'}(\epsilon'+\epsilon+\omega)\big]
\notag \\
& \times \mathcal{G}_{1,\bm{k}'}(\epsilon' + \omega)  \mathcal{G}_{0,\bm{k}'}(\epsilon')  \Gamma_{\bm{k}',\epsilon'}(\omega),
\end{align}
where the four scattering amplitudes, 
\begin{align}\label{BS15}
T_{1\bm{k}\bm{k}'}(\Omega) &=  T \sum_{\bm{p},\bar{\epsilon}}|V_{\bm{k}-\bm{k}'}|^2 \mathcal{G}_{0,\bm{p}-\bm{k} +\bm{k}'}(\bar{\epsilon} - \Omega) \mathcal{G}_{0,\bm{p}}(\bar{\epsilon}),
\notag \\
T_{2\bm{k}\bm{k}'}(\Omega) &=  T \sum_{\bm{p},\bar{\epsilon}}|V_{\bm{k}-\bm{p}}|^2 \mathcal{G}_{0,\bm{p}-\bm{k} +\bm{k}'}(\bar{\epsilon} - \Omega) \mathcal{G}_{0,\bm{p}}(\bar{\epsilon}),
\notag \\
T_{3\bm{k}\bm{k}'}(\Omega) &=  T \sum_{\bm{p},\bar{\epsilon}}|V_{\bm{k}-\bm{p}}|^2 \mathcal{G}_{0,\bm{p}}(\bar{\epsilon}) \mathcal{G}_{1,-\bm{p}+\bm{k}+\bm{k}'}(-\bar{\epsilon}+\Omega) ,
\notag \\
T_{4\bm{k}\bm{k}'}(\Omega) &=  T \sum_{\bm{p},\bar{\epsilon}}|V_{\bm{k}-\bm{p}}|^2 \mathcal{G}_{1,\bm{p}}(\bar{\epsilon}) \mathcal{G}_{0,-\bm{p}+\bm{k}+\bm{k}'}(-\bar{\epsilon}+\Omega)\, . 
\end{align}

The central object is the vertex function $\Gamma_{\bm{k}}(\omega)$ constructed from the Matsubara vertex in the following way.
The vertex, $\Gamma_{\bm{k},\epsilon_n}(\omega_n)$ appearing in Eq.~\eqref{BS13} is first analytically continued in the frequency plane,
$i \epsilon_n \rightarrow \epsilon - i 0$, $i \epsilon_n + i\omega_n\rightarrow \epsilon + \omega + i 0$ which signifies the pair of Green functions adjacent to the vertex as so called RA (retarded-advanced) pair playing the central role in the theory of Fermi liquid.
The next step is the reduction to the on-mass-shell vertex function achieved by setting $\epsilon = \xi_{0,\bm{k}}$.
The second possible choice,  $\epsilon + \omega = \xi_{1,\bm{k}}$ is equivalent to the first one.
Indeed, since we consider $\omega \approx \omega_{\perp}$, $\xi_{1,\bm{k}} = \xi_{0,\bm{k}} + \omega_{\perp}$, and the analytically continued vertex is assumed to be smooth on the scale, $|\omega - \omega_{\perp}| \approx 1/\tau_{\perp}$ the two on-mass-shell conditions introduced above are equivalent.
Obtained in this way the RA, on-mass-shell vertex is a function of momentum $\bm{k}$ and the frequency $\omega$, which we denote $\Gamma_{\bm{k}}(\omega)$.

The self-energies are also expressible through the functions, Eq.~\eqref{BS15}.
The holes at the $n=0$ band has a single type of self-energy, that can be written in the two alternative forms,
\begin{align}\label{Sigma_0}
\Sigma_{0,\bm{k}}(\epsilon) & = U_0^2 T \sum_{\bm{k}',\epsilon_{n'}}  T_{1\bm{k}\bm{k}'}(\epsilon_{n} - \epsilon_{n'}) \mathcal{G}_{0,\bm{k}'}(\epsilon_{n'})
\notag \\
& =
U_0^2 T \sum_{\bm{k}',\epsilon_{n'}}  T_{2\bm{k}\bm{k}'}(\epsilon_{n} - \epsilon_{n'}) \mathcal{G}_{0,\bm{k}'}(\epsilon_{n'}) \, .
\end{align}
The self energy of the particles in the $n=1$ band contains the two distinct contributions,
\begin{align}\label{Sigma_1ab}
\Sigma_{1,\bm{k}}(\epsilon) & = \Sigma_{1,\bm{k}}^{(a)}(\epsilon) +\Sigma_{1,\bm{k}}^{(b)}(\epsilon)\, ,
\end{align}
where
\begin{align}\label{Sigma_1a}
\Sigma_{1,\bm{k}}^{(a)}(\epsilon_n + \omega_n) & =
U_1^2 T \sum_{\bm{k}',\epsilon_{n'}}  T_{1\bm{k}\bm{k}'}(\epsilon_{n} - \epsilon_{n'}) \mathcal{G}_{1,\bm{k}'}(\epsilon_{n'}+ \omega_n)
\notag \\
 =
U_1^2 T  \sum_{\bm{k}',\epsilon_{n'}} & T_{4\bm{k}\bm{k}'}(\epsilon_{n} + \epsilon_{n'}+\omega_n) \mathcal{G}_{0,\bm{k}'}(\epsilon_{n'}) \, ,
\end{align}
and
\begin{align}\label{Sigma_1b}
\Sigma_{1,\bm{k}}^{(b)}(\epsilon+ \omega) & =
U_{\bar{1}}^2 T \sum_{\bm{k}',\epsilon_{n'}}  T_{2\bm{k}\bm{k}'}(\epsilon_{n} - \epsilon_{n'}) \mathcal{G}_{1,\bm{k}'}(\epsilon_{n'}+ \omega_n)
\notag \\
 =
U_{\bar{1}}^2 T &\sum_{\bm{k}',\epsilon_{n'}}  T_{3\bm{k}\bm{k}'}(\epsilon_{n} + \epsilon_{n'}+\omega_n) \mathcal{G}_{0,\bm{k}'}(\epsilon_{n'}) \, .
\end{align}
The quasi-particle life time,
\begin{align}\label{life_time}
\frac{1}{\tau_{n,\bm{k}}} = - 2 \Im \Sigma^R_{n,\bm{k}}(\xi_{n,\bm{k}})
\end{align}
is a meaningful concept as long as the self-energy on the right hand side of Eq.~\eqref{life_time} is smooth function of the energy variable, $\epsilon$ at the mass shell, $\epsilon = \xi_{n,\bm{k}}$ on a scale defined by Eq.~\eqref{life_time}.
This condition is satisfied as the scale associated with the self-energy frequency dependence is typically of the order of $\omega_{\perp}$.
Corresponding to the separation of the particle self-energy, Eq.~\eqref{Sigma_1ab} we introduce the two contributions to the hole life time,
\begin{align}\label{life_t13}
\frac{1}{\tau_{1,\bm{k}}} &=  \frac{1}{\tau_{1a,\bm{k}}} + \frac{1}{\tau_{1b,\bm{k}}}\, ,
\notag \\  
\frac{1}{\tau_{1a,b,\bm{k}}} & = - 2 \Im [\Sigma^{(a,b)}_{1,\bm{k}}]^R(\xi_{1,\bm{k}})\, .
\end{align}
The frequency summation in the Eq.~\eqref{Sigma_0} followed by the analytic continuation to the real frequency yields (see App.~\ref{app:qp_rates}),
\begin{subequations}\label{rates}
\begin{align}\label{rate_0}
\frac{1}{\tau_{0,\bm{k}}}  = & U_0^2  \sum_{\bm{k}'} \left[ \tanh\frac{\xi_{\bm{k}'}}{2 T} - \coth\frac{\xi_{\bm{k}'} - \xi_{\bm{k}}}{ 2 T } \right]  
\notag \\
& \times \Im T^R_{1(2)\bm{k}\bm{k}'}(\xi_{\bm{k}} - \xi_{\bm{k}'})\, .
\end{align}
Similarly we have for each of the two inelastic relaxation rates introduced in Eq.~\eqref{life_t13} two equivalent expressions,
\begin{align}\label{qp_rate1a}
\frac{1}{\tau_{1a,\bm{k}}} = & U_1^2 \sum_{\bm{k}'} \left[ \tanh\frac{\xi_{\bm{k}'} + \omega_{\perp}}{2 T} - \coth\frac{\xi_{\bm{k}'} - \xi_{\bm{k}}}{ 2 T } \right] 
\notag \\
& \times  \Im T^R_{1\bm{k}\bm{k}'}(\xi_{\bm{k}} - \xi_{\bm{k}'})
\notag \\
= & U_1^2 \sum_{\bm{k}'} \left[ \tanh\frac{\xi_{\bm{k}'}}{2 T} - \coth\frac{\xi_{\bm{k}} + \xi_{\bm{k}'}+ \omega_{\perp}}{ 2 T} \right] 
\notag \\
& \times \Im T^R_{4\bm{k}\bm{k}'}(\xi_{\bm{k}}+\xi_{\bm{k}'} + \omega_{\perp}) \, ,
\end{align}
and
\begin{align}\label{qp_rate1b}
\frac{1}{\tau_{1b,\bm{k}}} = & U_{\bar{1}}^2 \sum_{\bm{k}'} \left[ \tanh\frac{\xi_{\bm{k}'} + \omega_{\perp}}{2 T} - \coth\frac{\xi_{\bm{k}'} - \xi_{\bm{k}}}{ 2 T } \right]  
\notag \\
& \times \Im T^R_{2\bm{k}\bm{k}'}(\xi_{\bm{k}} - \xi_{\bm{k}'})
\notag \\
= & U_{\bar{1}}^2 \sum_{\bm{k}'} \left[ \tanh\frac{\xi_{\bm{k}'}}{2 T} - \coth\frac{\xi_{\bm{k}} + \xi_{\bm{k}'}+ \omega_{\perp}}{ 2 T} \right] 
\notag \\
& \times \Im T^R_{3\bm{k}\bm{k}'}(\xi_{\bm{k}}+\xi_{\bm{k}'} + \omega_{\perp})\, .
\end{align}
\end{subequations}

The Bethe-Salpeter equation satisfied by $\Gamma_{\bm{k}}(\omega)$ is obtained from Eq.~\eqref{BS13} by performing the summation over the frequencies 
$\epsilon$ and $\epsilon'$ within the interval, $-  \omega < \epsilon, \epsilon' < 0$ by the standard method of analytical continuation \cite{Abrikosov1963}.
The other frequencies make a contribution that is smaller than the one we retain in the parameter, $(\omega_{\perp} \tau_{in})^{-1}  \ll 1$.
We write the Bethe-Salpeter equation in the matrix form,
\begin{align}\label{BS17}
\sum_{\bm{k}'} & D_{\bm{k}\bm{k}'}(\omega)   \Gamma_{\bm{k}'} = \sum_{\bm{k}'} \big[ U_0 U_1 \bar{T}_{1\bm{k}\bm{k}'}(\omega) + U_0 U_{\bar{1}} \bar{T}_{2\bm{k}\bm{k}'}(\omega)  
\notag \\
& + U_1 U_{\bar{1}} \bar{T}_{3\bm{k}\bm{k}'}(\omega) + U_1 U_{\bar{1}} \bar{T}_{4\bm{k}\bm{k}'}(\omega)\big]
 \Gamma_{\bm{k}'}\,. 
\end{align}
The functions  $D_{\bm{k}\bm{k}'}(\omega)$ and $ \bar{T}_{l\bm{k}\bm{k}'}(\omega)$ in  Eq.~\eqref{BS17} are obtained form the left and right side of Eq.~\eqref{BS13} respectively.
Our strategy will be to consider Eq.~\eqref{BS17} at a given frequency, $\omega$ as a homogeneous linear, matrix equation on the vector  $\Gamma$ with components labeled by the subscript $\bm{k}$.
In this way the functions $D_{\bm{k}\bm{k}'}(\omega)$ and $ \bar{T}_{l\bm{k}\bm{k}'}(\omega)$ are to be understood as matrices acting on a vector $\Gamma$.
The explicit evaluation of these matrices is relegated to the App.~\ref{app:M2}.
The matrix $D$ is diagonal,
\begin{align}\label{M_1}
D_{\bm{k}\bm{k}'}(\omega)=  \frac{\delta_{\bm{k},\bm{k}'}}{2}  \frac{  \tanh \frac{\xi_{\bm{k}} + \omega }{ 2 T } - \tanh \frac{\xi_{\bm{k}}} { 2 T }  }{ \omega - \omega_{\perp} + i / 2 \tau_{1,\bm{k}} + i / 2 \tau_{0,\bm{k}} }\, .
\end{align}
For the matrices $\bar{T}_{1\bm{k}\bm{k}'}(\omega)$ and $\bar{T}_{2\bm{k}\bm{k}'}(\omega)$ we have,
\begin{subequations}
\begin{align}\label{Gamma_1_13}
\bar{T}_{1(2)\bm{k}\bm{k}'}&(\omega) = 
\frac{i}{4} 
\frac{  \tanh \frac{\xi_{\bm{k}} + \omega}{ 2 T} -  \tanh \frac{\xi_{\bm{k}}}{ 2 T} }{ \omega - \omega_{\perp} + i / 2 \tau_{1,\bm{k}} + i / 2 \tau_{0,\bm{k}} }
\notag \\
& \times 
 \mathrm{Im}T^R_{1(2)\bm{k}\bm{k}'}(\xi_{\bm{k}}  - \xi_{\bm{k}'} ) 
\notag \\
& \times  
\frac{ \tanh \frac{\xi_{\bm{k}'} + \omega}{ 2 T} +  \tanh \frac{\xi_{\bm{k}'}} { 2 T} - 2 \coth \frac{\xi_{\bm{k}'} - \xi_{\bm{k}}}{ 2 T}  }{ \omega - \omega_{\perp} + i / 2 \tau_{1,\bm{k}'} + i / 2 \tau_{0,\bm{k}'} } \, .
\end{align}
Similarly, for the matrices $\bar{T}_{3\bm{k}\bm{k}'}(\omega)$ and $\bar{T}_{4\bm{k}\bm{k}'}(\omega)$,
\begin{align}\label{T3}
\bar{T}_{3(4)\bm{k}\bm{k}'} 
&(\omega) = 
-\frac{i}{4} 
\frac{  \tanh \frac{\xi_{\bm{k}} + \omega}{ 2 T} -  \tanh \frac{\xi_{\bm{k}}}{ 2 T} }{ \omega - \omega_{\perp} + i / 2 \tau_{1,\bm{k}} + i / 2 \tau_{0,\bm{k}} }
\notag \\
 \times &
 \mathrm{Im}T^R_{3(4)\bm{k}\bm{k}'}(\xi_{\bm{k}}  + \xi_{\bm{k}'} + \omega ) 
\notag \\
 \times & 
\frac{ \tanh \frac{\xi_{\bm{k}'} + \omega}{ 2 T} +  \tanh \frac{\xi_{\bm{k}'}} { 2 T} - 2 \coth \frac{\xi_{\bm{k}'} + \xi_{\bm{k}}+  \omega}{ 2 T}  }{ \omega - \omega_{\perp} + i / 2 \tau_{1,\bm{k}'} + i / 2 \tau_{0,\bm{k}'} } \, .
\end{align}
\end{subequations}
It is crucial for the forthcoming discussion that the matrices, $\bar{T}_{n\bm{k}\bm{k}'}$ in Eqs.~\eqref{Gamma_1_13} and \eqref{T3} are symmetric,
$\bar{T}_{n\bm{k}\bm{k}'} = \bar{T}_{n\bm{k}'\bm{k}}$.
Even though this property is not transparent from the above equations, it follows from the definitions of these matrices and can be easily checked numerically. 
Before we address the DOs attenuation we have to establish the consistency of the developed formalism with the KTh to the first place.
This task is implemented in the following Sec.~\ref{sec:KT_BS}.
\subsection{KTh from the Bethe-Salpeter equation}
\label{sec:KT_BS}
In order to verify the consistency of the Bethe-Salpeter equation, \eqref{BS17} one has to show that the latter possesses a non-trivial solution for $\omega=\omega_{\perp}$ provided the confining potential is harmonic. 
As we demonstrate below such a solution reads,
\begin{align}\label{G_K}
\Gamma^{\perp}_{\bm{k}} = \frac{i}{  2 \tau_{1,\bm{k}} }+ \frac{ i }{ 2 \tau_{0,\bm{k}}} \, .
\end{align}
The solution, Eq.~\eqref{G_K} has a transparent meaning.
As we discussed previously the DOs are a coherent superposition of the inter-band particle-hole excitation consistent with the Fermi statistics.
To reach this excitation the external driving has to be stronger for those particle-hole pairs that decay faster.
The form, Eq.~\eqref{G_K} expresses is the direct expression of this observation.  

To show that Eq.~\eqref{G_K} solves Eq.~\eqref{BS17} at $\omega = \omega_{\perp}$ we first note that according to Eq.~\eqref{M_1},
\begin{align}\label{BS_KT13}
\sum_{\bm{k}'} D_{\bm{k}\bm{k}'}(\omega_{\perp}) \Gamma^{\perp}_{\bm{k}'}  =   \frac{1}{2} \left[ \tanh \frac{\xi_{\bm{k}} + \omega_{\perp} }{ 2 T } - \tanh \frac{\xi_{\bm{k}}} { 2 T } \right].
\end{align}
Next we consider the first term on the right hand side of Eq.~\eqref{BS17} for the choice Eq.~\eqref{G_K},
\begin{align}\label{BS_KT15}
U_0 &U_1 \sum_{\bm{k}'}  \bar{T}_{1\bm{k}\bm{k}'}(\omega_{\perp}) \Gamma^{\perp}_{\bm{k}}
= U_0 U_1 \sum_{\bm{k}'}  \frac{i}{4} 
\frac{  \tanh \frac{\xi_{\bm{k}} + \omega_{\perp}}{ 2 T} -  \tanh \frac{\xi_{\bm{k}}}{ 2 T} }{ i / 2 \tau_{1,\bm{k}} + i / 2 \tau_{0,\bm{k}} }
\notag \\
& \times 
 \mathrm{Im}T^R_{1\bm{k}\bm{k}'}(\xi_{\bm{k}}  - \xi_{\bm{k}'} ) 
\notag \\
& \times  
\left[\tanh \frac{\xi_{\bm{k}'} + \omega_{\perp}}{ 2 T} +  \tanh \frac{\xi_{\bm{k}'}} { 2 T} - 2 \coth \frac{\xi_{\bm{k}'} - \xi_{\bm{k}}}{ 2 T}  \right].
\end{align}
We then make use of the relations, Eqs.~\eqref{rate_0} and \eqref{qp_rate1a} to bring the last equation to the form,
\begin{subequations}\label{vert}
\begin{align}\label{BS_KT17}
U_0 U_1 \sum_{\bm{k}'} & \bar{T}_{1\bm{k}\bm{k}'}(\omega_{\perp}) \Gamma^{\perp}_{\bm{k}}
=
\frac{1}{2}\left[ \tanh \frac{\xi_{\bm{k}} + \omega_{\perp}}{ 2 T} -  \tanh \frac{\xi_{\bm{k}}}{ 2 T} \right]
\notag \\
\times & \frac{ U_0^{-1} U_1 / \tau_{0,\bm{k}} + U_0 U_1^{-1}  / \tau_{1a,\bm{k}}  }{1 / \tau_{1,\bm{k}} + 1 /  \tau_{0,\bm{k}}}\, .
\end{align}
The other terms in Eq.~\eqref{BS17} are obtained in a very similar way,
\begin{align}\label{BS_KT19}
U_0 U_{\bar{1}} \sum_{\bm{k}'} & \bar{T}_{2\bm{k}\bm{k}'}(\omega_{\perp}) \Gamma^{\perp}_{\bm{k}}
=
\frac{1}{2}\left[ \tanh \frac{\xi_{\bm{k}} + \omega_{\perp}}{ 2 T} -  \tanh \frac{\xi_{\bm{k}}}{ 2 T} \right]
\notag \\
\times & \frac{ U_0^{-1} U_{\bar{1}} / \tau_{0,\bm{k}} + U_0 U_{\bar{1}}^{-1}  / \tau_{1b,\bm{k}}  }{1 / \tau_{1,\bm{k}} + 1 /  \tau_{0,\bm{k}}}\, ,
\end{align}
\begin{align}\label{BS_KT21}
U_1 U_{\bar{1}} \sum_{\bm{k}'} & \bar{T}_{3\bm{k}\bm{k}'}(\omega_{\perp}) \Gamma^{\perp}_{\bm{k}}
=
-\frac{1}{2}\left[ \tanh \frac{\xi_{\bm{k}} + \omega_{\perp}}{ 2 T} -  \tanh \frac{\xi_{\bm{k}}}{ 2 T} \right]
\notag \\
\times & \frac{ U_1 U^{-1}_{\bar{1}} / \tau_{1b,\bm{k}}  }{1 / \tau_{1,\bm{k}} + 1 /  \tau_{0,\bm{k}}}\, ,
\end{align}
\begin{align}\label{BS_KT23}
U_1 U_{\bar{1}} \sum_{\bm{k}'} & \bar{T}_{4\bm{k}\bm{k}'}(\omega_{\perp}) \Gamma^{\perp}_{\bm{k}}
=
-\frac{1}{2}\left[ \tanh \frac{\xi_{\bm{k}} + \omega_{\perp}}{ 2 T} -  \tanh \frac{\xi_{\bm{k}}}{ 2 T} \right]
\notag \\
\times & \frac{ U_{\bar{1}} U^{-1}_{1} / \tau_{1a,\bm{k}}  }{1 / \tau_{1,\bm{k}} + 1 /  \tau_{0,\bm{k}}}\, .
\end{align}
\end{subequations}
The Eqs.~\eqref{BS_KT21} and \eqref{BS_KT23} compared to Eqs.~\eqref{BS_KT17} and \eqref{BS_KT19} have opposite sign and contain only one term.
The reason for the presence of only one contribution in Eqs.~\eqref{BS_KT21} and \eqref{BS_KT23} is that the remaining part vanishes because it describes the excitations of holes in the unoccupied band, $n=1$.

Adding up Eqs.~\eqref{vert} reproduces Eq.~\eqref{BS_KT13} thanks to the relationship between the matrix elements $U_0 - U_1 - U_{\bar{1}}=0$ listed in Eq.~\eqref{ident} which holds for strictly parabolic confining potential. 
This shows how the KTh is obtained in the framework of the Bethe-Salpeter equation formalism.
In the next Sec.~\ref{sec:tau_BS} we compute the relaxation rate for the non-parabolic confining potential.

\subsection{Life time, $\tau_{\perp}$ in a weakly anharmonic confinement potential from Bethe-Salpeter equation}
\label{sec:tau_BS}
At finite anharmonicity, $\epsilon \neq 0$ the matrix element do not satisfy the relationships, Eq.~\eqref{ident}.
In result, Eq.~\eqref{G_K} ceases to be a solution of Bethe-Salpeter Eq.~\eqref{BS17}.
Instead, a different solution exists for complex frequency, 
\begin{align}\label{identify}
\omega^* = \omega_{\perp} - i/\tau_{\perp}
\end{align}
on the physical sheet in accord with the causality. 
Hence knowledge of frequency $\omega^*$ provides us with the relaxation rate we are after.
Our strategy is to rewrite the Bethe-Salpeter equation, Eq.~\eqref{BS17} in the form of the eigenvalue problem, 
\begin{align}\label{BS_KT31}
\left[\hat{D}(\omega) - \hat{W}(\omega,\epsilon)\right] |\Gamma \rangle = \lambda(\omega,\epsilon) |\Gamma \rangle\, ,
\end{align}
where the operators on the left hand side act in the space of functions of the quasi-particle momentum, $\langle k | \Gamma \rangle = \Gamma_k$.
The matrix elements of the diagonal operator, $\hat{D}$, are given explicitly by the Eq.~\eqref{M_1}, and those of the operator $\hat{W}$,
by the right hand side of Eq.~\eqref{BS17}.

The equivalent of the KTh is
\begin{align}\label{BS_KT36}
\left[\hat{D}(\omega_{\perp}) - \hat{W}(\omega_{\perp},0)\right] |\Gamma^{\perp} \rangle = \lambda(\omega_{\perp},0) |\Gamma^{\perp} \rangle,
\end{align}
namely at $\omega = \omega_{\perp}$  and $\epsilon = 0$, $|\Gamma^{\perp} \rangle$ is the eigenvector with the eigenvalue $\lambda(\omega_{\perp},0)=0$.
Our goal is to find the complex frequency, $\omega^*$ such that there exists a vector $|\Gamma^* \rangle$ satisfying 
\begin{align}\label{BS_KT39}
\left[\hat{D}(\omega^*) - \hat{W}(\omega^*,\epsilon)\right] |\Gamma^* \rangle = \lambda(\omega^*,\epsilon) |\Gamma^* \rangle\, 
\end{align}
under the condition $\lambda(\omega^*,\epsilon)=0$.
The latter translates to the requirement,
\begin{align}\label{BS_KT41}
& \langle  \Gamma^{\perp}\left| \left[\hat{D}(\omega^*)- \hat{W}(\omega^*,\epsilon)\right] \right| \Gamma^{\perp}  \rangle
\notag \\
& =
\langle \Gamma^{\perp} \left| \left[\hat{D}(\omega_{\perp}) - \hat{W}(\omega_{\perp},0)\right] \right| \Gamma^{\perp}  \rangle 
\end{align}
obtained in the standard perturbation theory by treating the difference between the operators in Eqs.~\eqref{BS_KT36} and \eqref{BS_KT39} as small perturbation.
In view of Eq.~\eqref{BS_KT36} with $\lambda(\omega_{\perp},0)=0$, Eq.~\eqref{BS_KT41} reduces to 
\begin{align}\label{BS_KT43}
\langle & \Gamma^{\perp}\left| \left[\hat{D}(\omega^*)- \hat{W}(\omega^*,\epsilon)\right] \right| \Gamma^{\perp}  \rangle
 = 0.
\end{align}
Making the series expansion of Eq.~\eqref{BS_KT43} for $\omega \approx \omega_{\perp}$ we obtain the condition,
\begin{align}\label{BS_KT46}
\langle & \Gamma^{\perp}\left| \left[\hat{D}(\omega_{\perp})- \hat{W}(\omega_{\perp},\epsilon)\right] \right| \Gamma^{\perp}  \rangle
\notag \\
= - & (\omega^* - \omega_{\perp}) 
\langle\Gamma^{\perp}\left| \left[\frac{ d \hat{D}}{d \omega} - \frac{d\hat{W}}{ d \omega}\right] \right| \Gamma^{\perp}  \rangle ,
\end{align}
where the derivatives can be evaluated at $\omega= \omega_{\perp}$ and $\epsilon=0$.

We start with the left hand side of Eq.~\eqref{BS_KT46}.
From \eqref{M_1} and \eqref{G_K}, 
\begin{align}\label{BS55}
\langle \Gamma^{\perp} | \hat{D}(\omega_{\perp}) | \Gamma^{\perp} \rangle
& =
-\frac{i}{2} \sum_{\bm{k}} \left[ \tanh \frac{\xi_{\bm{k}} + \omega_{\perp} }{ 2 T } - \tanh \frac{\xi_{\bm{k}}} { 2 T } \right] 
\notag \\
& \times \left( \frac{U_0^2}{  2 \bar{\tau}_{0,\bm{k}}} + \frac{U_1^2}{  2 \bar{\tau}_{1a,\bm{k}}}  +   \frac{U^2_{\bar{1}} } {  2 \bar{\tau}_{1b,\bm{k} } } \right)\, ,
\end{align}
where the rates $1/\bar{\tau}_{0,a,b,\bm{k}}$ have been introduces previously in Sec.~\ref{sec:PT_2nd}.
From Eqs.~\eqref{vert} 
\begin{align}\label{BS56}
\langle & \Gamma^{\perp}\left|  \hat{W}(\omega_{\perp},\epsilon) \right| \Gamma^{\perp}  \rangle
=
 - \frac{i}{2} \sum_{\bm{k}} \left[ \tanh \frac{\xi_{\bm{k}} + \omega_{\perp} }{ 2 T } - \tanh \frac{\xi_{\bm{k}}} { 2 T } \right] 
\notag \\ 
& \times  \left( \frac{ U_0 U_1 + U_0 U_{\bar{1}} }{ 2\bar{\tau}_{0,\bm{k}}} +  \frac{ U_0 U_1 - U_1 U_{\bar{1}} }{ 2\bar{\tau}_{1a,\bm{k}}} 
+  \frac{ U_0 U_{\bar{1}} - U_1 U_{\bar{1}}}{ 2\bar{\tau}_{1b,\bm{k}}}
\right)
\end{align}
To make a connection to the perturbation theory analysis given in Sec.~\ref{sec:PT} we 
make use of Eq.~\eqref{PT-res1}
Eqs.~\eqref{BS55} and \eqref{BS56} we obtain for the left hand side of Eq.~\eqref{BS_KT46}
\begin{align}\label{BS57}
\langle & \Gamma^{\perp}\left| \left[\hat{D}(\omega_{\perp})- \hat{W}(\omega_{\perp},\epsilon)\right] \right| \Gamma^{\perp}  \rangle
= - i \frac{N_0 - N_1}{\tau^{(2)}_{\perp}}.
\end{align}

Turning to the right hand side of Eq.~\eqref{BS_KT46} we notice that the strong frequency dependence originates from the energy denominators, $(\omega - \omega_{\perp} + i / 2 \tau_{0,k} + i / 2 \tau_{1,k})^{-1}$.
As a result, the derivatives with respect to $\omega$ operate on these denominators.
With this remark we readily obtain
\begin{align}\label{BS45}
\langle \Gamma^{\perp} | \frac{d \hat{D}}{d \omega}  | \Gamma^{\perp} \rangle & =
\frac{1}{2}\sum_{\bm{k}}  \left[ \tanh \frac{\xi_{\bm{k}} + \omega_{\perp} }{ 2 T } - \tanh \frac{\xi_{\bm{k}}} { 2 T } \right]
\notag \\
& = N_0 - N_1\, .
\end{align}
In computing the contribution of the $d \hat{W}/d \omega$ we notice that because of the symmetry $W_{\bm{k}\bm{k}'} = W_{\bm{k}'\bm{k}}$ one may take the derivative of a one of the two energy denominators and multiplying the result by a factor of two.
The convenient choice is to differentiate the leftmost of the two denominators in Eqs.~\eqref{Gamma_1_13} and \eqref{T3}.
This gives 
\begin{align}\label{BS61}
&\langle \Gamma^{\perp} | \frac{d \hat{W}}{d \omega}  | \Gamma^{\perp} \rangle  = -
\sum_{\bm{k}}  \frac{ \tanh \frac{\xi_{\bm{k}} + \omega_{\perp} }{ 2 T } - \tanh \frac{\xi_{\bm{k}}} { 2 T }}{1/\tau_{0,\bm{k}} + 1/\tau_{1,\bm{k}}}
\notag \\
& \times \left( \frac{ U_0 U_1 + U_0 U_{\bar{1}} }{ 2\bar{\tau}_{0,\bm{k}}} +  \frac{ U_0 U_1 - U_1 U_{\bar{1}} }{ 2\bar{\tau}_{1a,\bm{k}}} 
+  \frac{ U_0 U_{\bar{1}} - U_1 U_{\bar{1}}}{ 2\bar{\tau}_{1b,\bm{k}}}
\right)  \, .
\end{align}
Since the expression Eq.~\eqref{BS61} may be evaluated at $\epsilon = 0$, in view of the relation $U_0 - U_1 - U_{\bar{1}}=0$ we use the relations
$U_0 U_1 + U_0 U_{\bar{1}} = U_0^2$, $U_0 U_1 - U_1 U_{\bar{1}}  = U_1^2$ and $U_0 U_{\bar{1}} - U_1 U_{\bar{1}} = U_{\bar{1}}^2$
to bring it to the simple form,
\begin{align}\label{BS63}
&\langle \Gamma^{\perp} | \frac{d \hat{W}}{d \omega}  | \Gamma^{\perp} \rangle  =
\sum_{\bm{k}}  \left[ \tanh \frac{\xi_{\bm{k}} + \omega_{\perp} }{ 2 T } - \tanh \frac{\xi_{\bm{k}}} { 2 T }\right] 
\notag \\
& = 2 (N_0 - N_1).
\end{align}
Note that contribution Eq.~\eqref{BS63} is twice larger than Eq.~\eqref{BS45}.
The cancellation of the matrix elements does not hold for the matrix elements derivatives, because $\hat{W}$ contain two singular energy denominators while $\hat{D}$ has only one.
Subtracting Eq.~\eqref{BS63} from Eq.~\eqref{BS45} gives
\begin{align}\label{BS66}
\langle\Gamma^{\perp}\left| \left[\frac{ d \hat{D}}{d \omega} - \frac{d\hat{W}}{ d \omega}\right] \right| \Gamma^{\perp}  \rangle = -(N_0 - N_1)\, .
\end{align}
Substituting Eqs.~\eqref{BS57} and \eqref{BS66} in Eq.~\eqref{BS_KT46} we obtain,
\begin{align}\label{BS_KT46_A}
\frac{1}{\tau^{(2)}_{\perp}}
= i (\omega^* - \omega_{\perp}),
\end{align}
which in view of the definition, Eq.~\eqref{identify} gives a simple result,
\begin{align}\label{result_BS}
\frac{1}{\tau_{\perp}} = \frac{1}{\tau^{(2)}_{\perp}}\, ,
\end{align}
namely the relaxation rate obtained by solving the Bethe-Salpeter equation agrees with the naive guess, Eq.~\eqref{PT-res1} based on  the perturbation theory in Sec.~\ref{sec:PT}.
We stress that the condition $1/\tau_{\perp} \ll 1/\tau_{in}$ needed for the result Eq.~\eqref{result_BS} to hold is not obvious from the perturbation theory.
The result, Eq.~\eqref{result_BS} is also confirmed within the memory function formalism in Sec.~\ref{sec:MF} below.

Let us emphasize that the above discussion as is does not apply to quasi-1D case.
The logarithmic singularities crucial in 1D are absent in the present analysis.
Yet, based on the representation Eq.~\eqref{rate} we conjecture  that the inter-band correlation function in 1D acquire the power law frequency dependence,
$K_{A_1,A_1} \propto \int_{0}^{2 k_F} d q (\omega - \omega_{\perp} + i/2 \bar{\tau}_{0,-k_F + q})^{ - 1 - \mu_-(q)}$. Here  the exponent $\mu_-(q)$ introduced in \cite{Khodas2006} controls the power law frequency dependence of the (intra-band) density correlation function at the lower edge of the particle-hole continuum, and $1/\bar{\tau}_{0,k}$ is the hole life time, Eq.~\eqref{hole_1D}.

\section{Memory function approach applied to the calculation of $1/\tau_{\perp}$.}
\label{sec:MF}
The memory function is a method alternative to the solution of the integral Bethe-Salpeter equation.
This formalism outlined in Sec.~\ref{sec:summaryMF} is exact.
To make a further progress one has to relate it to the more familiar correlation functions. 
This is done to the leading order in Sec.~\ref{sec:summaryMF} were we discuss the approximations and their region of validity.

Although these approximations are physically transparent, the whole approach is harder to justify as compared to the Bethe-Salpeter equation. The great advantage of the present formalism is that its physically intuitive. Here we are not trying to emulate all its virtues. Rather we limit the discussion to the life time of the DOs. In a sense, it replaces the integral Bethe-Salpeter equation by the algebraic one.
The latter is similar to the Dyson equation for a single particle excitations, and gives the life time of a particle in terms of the self-energy evaluated on a mass shell. The memory function is an analog of the self energy for the collective excitations. 

\subsection{Summary of the memory function formalism}
\label{sec:summaryMF}

The basic quantity  central to the MF formalism is the Kubo correlation function,
\begin{align}\label{Mem57}
C_{ij}(t) =\langle A_i(t) | A_j \rangle 
\end{align}
which is a matrix on the space of slow variables $A_{i}$ labeled by index $i$.
In our problem this set is given by Eq.~\eqref{PT1}.
The scalar product in Eq.~\eqref{Mem57} is by definition 
\be\label{Mem55}
\langle A | B \rangle = \int_0^{\beta} d \lambda 
\left[ \langle A^{\dag} B( i \hbar \lambda) \rangle - \langle A^{\dag} \rangle \langle B( i \hbar \lambda) \rangle \right]\, ,
\ee
where $\langle \ldots \rangle$ stands for the thermodynamic average.
Strictly speaking, Eq.~\eqref{Mem55} defines a scalar product for normal ordered operators, i.e. with their average subtracted, (see Ref.~\cite{Mori1965} and App.~\ref{app:corr-func}). 
This, however, is inconsequential because all the final results are formulated in terms of commutators.
The time evolution in this space is specified by the Liouvillian super-operator defined by its action on an observable $X$, 
\be\label{Mem67a}
 \hat{L} [X] =   \hbar^{-1} [H,X] \, .
\ee
Equation \eqref{Mem67a} is formally solved by 
\be\label{Mem67_A}
X(t) = \exp( - i \hat{L} t ) X(t=0)\, .
\ee
Consider the Laplace transformation,
\be\label{Mem59}
\hat{C}(z) = \int_0^{\infty} d t e^{ i z t} \hat{C}(t)\, ,
\ee
where $\hat{C}(t)$ is the matrix in the space of slow variables with matrix elements defined by Eq.~\eqref{Mem57}.
With the definition Eq.~\eqref{Mem67_A}, Eq.~\eqref{Mem59} formally evaluates to
\be\label{Mem60}
C_{ij}(z) = \langle A_i |  \frac{ i }{ z - L} | A_j \rangle 
\ee
Though formally exact the expression \eqref{Mem60} is not always practically useful.
The Kubo function is related to the retarded correlation functions introduced in Eq.~\eqref{K}.
\begin{align}\label{Mem47_A}
C_{ij}(z)  = \frac{ 1 }{ i z } \left[ K_{A_i,A_j}(z = i 0^+) - K_{A_i,A_j}(z) \right]\, ,
\end{align}
where $\Im z > 0$, $0^+$ is a positive infinitesimal.

We make use the relationship \eqref{Mem47_A}  on the real axis $z = \omega + i 0^+$. 
The detailed proof of Eq.~\eqref{Mem47_A} is given in App.~\ref{app:corr-func}.

The memory function formalism builds on separation of scales.
Once the set of slow variables are identified the Dyson-equation-like representation of the Kubo matrix exists,
\be\label{Mem63}
\hat{C}(z) = i [z - \hat{\Omega} + i \hat{\Sigma}(z) ]^{-1}  \hat{\chi}_s\, , 
\ee
where $z$ is meant to multiply a unit matrix and the matrices are defined below.
The matrix $\hat{\chi}^s$ is defined by the equal time scalar product and as shown in App.~\ref{app:corr-func} is the negative of the static, $\omega=0$ response function, 
\be\label{Mem61}
[ \hat{\chi}_s ]_{ij} = \langle A_i | A_j \rangle = - K_{A_i,A_j}(\omega = 0) \, .
\ee
The matrices $\hat{\Omega}$ and $\hat{\Sigma}$ referred to as frequency renormalization and memory matrices respectively.
They represent the frequency and relaxation of collective modes. 
These quantities are of the central interest.
We have,
\begin{align}\label{Mem65}
\hat{\Omega} &= \hat{\omega} \hat{\chi}_s^{-1}\, , \quad \omega_{ij} = i \langle \dot{A}_i | A_j \rangle \, .
\end{align}
Similarly to the two equivalent expressions in Eq.~\eqref{Mem61} for $\hat{\chi}$, the $\hat{\omega}$
is equivalently represented as the retarded correlation function,
\begin{align}\label{Mem66}
\hat{\omega}_{ij}  =  i K_{\dot{A}_i,A_j}(\omega =0)
\end{align}
of the forces $\dot{A}$ with the variables $A$ themselves evaluated in the static limit.

To introduce the memory function the projection onto the space of slow variables is defined as 
\be\label{Mem67}
\hat{P} = \sum_{i} \frac{ |A_{i} \rangle \langle A_i |}{ \langle A_i | A_i \rangle }\, .
\ee
Then the operator projecting out the slow dynamics is $\hat{Q} = 1 - \hat{P}$.
The relaxation of the slow modes due to their coupling to the thermal bath of fast degrees of freedom is contained in the memory matrix, 
\begin{align}\label{Mem69}
\hat{\Sigma}(z) & = \hat{\sigma}(z) \hat{\chi}_s^{-1}\, , 
\end{align}
where 
\begin{align}\label{Mem71}
\langle A_i | \hat{\sigma}(z)| A_j \rangle & = \int_0^{\infty} d t e^{ i z t}  \langle \dot{A}_i | \hat{Q} e^{- i \hat{Q} \hat{L} \hat{Q} t} \hat{Q} |\dot{A}_j \rangle 
\notag \\
& =\langle \dot{A}_i| \hat{Q} \frac{i }{z - \hat{Q}\hat{L}\hat{Q}} \hat{Q} |\dot{A}_j \rangle\, .
\end{align}

\subsection{The DOs relaxation rate to the second order in interaction}
\label{sec:SM_approx} 

The above equations are formally exact. However, to turn them into the useful tool usually further approximations are required based on the specifics of the regime studied.
We focus on the regime of week scattering.
So that the inverse inelastic life time of a particle is smaller than the confinement frequency, $1/\tau_{in} \ll \omega_{\perp}$.
As the DOs relaxation is hindered by the KTh we have $1/\tau_{\perp} < 1/\tau_{in}$ such that $1/\tau_{\perp} \ll \omega_{\perp}$.
Now based on this input we first consider the evaluation of the generalized forces $\dot{A}_{i} = i \hbar^{-1} [H,A_i]$ entering the memory matrix, Eq.~\eqref{Mem69}.
Following the decomposition, Eq.~\eqref{H1} we note that $[H_0,A_i]=0$.
Furthermore, $[H_{0,\perp},A_i] \propto A_i$ is filtered out by the projection operator $\hat{Q}$.   
So that the only relevant parts of the Hamiltonian for the calculation of the velocities are $H_{\epsilon,\perp}$ and $H_{int}$.
In the absence of interactions we may in principle include the anharmonicity as a perturbation and work with the operators \eqref{PT1} written in terms of the renormalized operators $\psi_{n,\bm{k}}$.
The latter annihilates the particles in the state that is a superposition of harmonic oscillator states, Eq.~\eqref{phi_n}. As the anharmonicity is assumed to be weak, for a given $n$, the state $\phi_n(z)$ is renormalized by small admixture of states $\phi_{n'}$ with $n' \neq n$. 
Upon the renormalization, the presence of the projection operators, $\hat{Q}$ in Eq.~\eqref{Mem69} excludes the contributions of the anharmonicity to the generalized forces entering the memory function.    
The contribution to the generalized forces $A_i$ describing the dissipation comes therefore from interactions.

Once the interactions are present the effect of anharmonicity cannot be completely eliminated by the above renormalization of the single particle spectrum.
Crucially, matrix elements of interaction defined by \eqref{tr_mat} change upon the inclusion of the anharmonicity effects on the wave-functions of the transversal motion. 
This breaks the KTh and allows for the DOs relaxation.

Unlike the standard situation, where the frequency renormalization matrix, Eq.~\eqref{Mem65} plays no special role, in the present discussion it is essential.
Note first that although all the observables $A_i$ are time reversal even, the matrix  $\hat{\omega}$ is non-zero.
This is because unlike the standard situation, operators $A_i$ are non-Hermitian. 
The importance of careful study of the matrix $\hat{\omega}$ stems from the immense degeneracy of the space of the slow variables.
Specifically, the variables $A_i$ are degenerate with all the observables of the type $\sum_{\bm{k}} g(\bm{k}) \psi_{n+1,\bm{k}}^{\dag} \psi_{n\bm{k}} $ with suitably chosen functions $g(\bm{k})$ to ensure the mutual orthogonality of slow variables.
Let us stress that the space of all such variables is infinite as even for a given $n$  there are infinitely many choices for the functions $g(\bm{k})$. 
For instance, the choice $g(\bm{k}) = \delta(\bm{k}-\bm{k}_0)$ defines a single particle-hole excitation.

The interaction of generic type lifts this degeneracy leaving the DOs unaffected.
In Sec.~\ref{sec:split_mem} we illustrate this point by computing the splitting of the center of mass and breathing oscillations in the system with two occupied bands.
In Sec.~\ref{sec:split_diag} the same results are obtained by performing the direct RPA-like calculation.

To the first order in interaction, the DOs are unaffected by interaction in the harmonic potential and is represented by the special linear superposition of the slow variables. We denote them as $\mathcal{F}_n$, $n =1,2,\ldots$. The dipole operator, Eq.~\eqref{SM} is one such variable.
In practice, the summation in Eq.~\eqref{SM} can be limited to the occupied bands because the unoccupied bands support no particle-hole excitations at frequency $\approx \omega_{\perp}$.
The dipole operator $\mathcal{F}_1$, Eq.~\eqref{SM} is the only slow variable satisfying, $\langle \mathcal{F}_1(t) \rangle \propto e^{ - i \omega_{\perp} t}$.
All the other linear combinations of $\mathcal{F}_n$ that are orthogonal to $\mathcal{F}_1$ are affected by interactions and split off by an amount of the order $\omega_{\perp} v_{int}$.  
Provided that this splitting is big enough we can reduce the space of slow variables to a single variable, namely the dipole operator, Eq.~\eqref{SM}.
The calculation shows that the life time $1/\tau_{\perp} \propto \epsilon^2 (1 / \tau_{in} ) $ and $1/\tau_{in} \propto \omega_{\perp} v_{int}^2$.
We have therefore the following hierarchy of scales, $1/\tau_{\perp} \ll 1/\tau_{in} \ll \omega_{\perp} v_{int} \ll \omega_{\perp}$ that follows from a single assumption of weak interaction, $v_{int} \ll 1$. 
Since the scale of variation of the operator $\hat{\Sigma}$ is $1/\tau_{in} \gg 1/ \tau_{\perp}$ the definition \eqref{Mem63} yields a pole at $z = \omega_{\perp} + i / \tau_{\perp}$.
In other words, the relaxation rate of DOs, $|\mathcal{F}_1\rangle$, Eq.~\eqref{SM} is given by
\be\label{tau_M}
\frac{1}{ \tau_{\perp}} = \Re \langle \mathcal{F}_1 |\hat{\Sigma}(\omega = \omega_{\perp} ) |\mathcal{F}_1 \rangle \, . 
\ee
The relationship \eqref{tau_M} holds provided the rate $1/\tau_{\perp}$ as defined by \eqref{tau_M} is much smaller than the typical frequency range of variation of $\Re \hat{\Sigma} (\omega)$.
In our problem the latter is determined by the single-particle relaxation rate due to the interactions, $1/\tau_{in}$.
As the KTh is violated only weakly the rate $1/ \tau_{\perp}$ indeed turns out to be parametrically smaller than the relaxation rate of individual quasi-particles.

We now evaluate the relaxation rate, Eq.~\eqref{tau_M} to the leading order in interaction.
Based on the representation, Eq.~\eqref{Mem69} we discuss the static susceptibility, $\hat{\chi}_s$  and the matrix $\hat{\sigma}$ separately.
The static susceptibility is real and therefore the imaginary contribution to the memory matrix originates from the matrix $\hat{\sigma}$. 
Hence, to the leading order the susceptibility matrix can be approximated by its value in the non-interacting system, 
\be\label{Mem_chi_s}
\hat{\chi}_s \approx \hat{\chi}^{(0)}_s\, .
\ee
The corrections to $\hat{\chi}_s$, scale as $\propto v_{int}^2$ and therefore are anyway small. 
In Eq.~\eqref{Mem71} defining $\hat{\sigma}$ the projection operators $\hat{Q}$ allow us to compute the force, $\dot{A}_{\perp}$ by commuting ${A}_{\perp}$ only with the interaction part of the Hamiltonian.
In result, to the second order in interaction we have   
\begin{align}\label{Mem100}
\langle \mathcal{F}_1 | \hat{\sigma}| \mathcal{F}_1\rangle\! \approx \!& \int_0^{\infty}\!\!\! d t e^{ i z t}  \langle [\mathcal{F}_1,H_{int}] |  e^{- i \hat{L}^{(0)} t}  |[\mathcal{F}_1,H_{int}] \rangle,
\end{align}
where $\hat{L}^{(0)}$ is the Liouvillian of the non-interacting systems. 
The relationship \eqref{Mem100} has a form of a non-interacting Kubo correlation function, Eq.~\eqref{Mem57}.
Following the relationship Eq.~\eqref{Mem47_A} we write 
\begin{align}\label{Mem72}
\langle \mathcal{F}_1 | \hat{\sigma}| \mathcal{F}_1\rangle \approx    \frac{ 1 }{ i z } &\Big[ K^{(0)}_{[\mathcal{F}_1,H_{int}],[\mathcal{F}_1,H_{int}]}(z = i 0)
\notag \\
&  - K^{(0)}_{[\mathcal{F}_1,H_{int}],[\mathcal{F}_1,H_{int}]}(z) \Big]\, , 
\end{align}
where $K^{(0)}$ denote the correlation function in Eq.~\eqref{K} computed in the non-interacting system.

Summarizing the above arguments, the DOs relaxation rate is based on Eqs.~\eqref{Mem61},  \eqref{Mem69},  \eqref{tau_M}, \eqref{Mem_chi_s} and \eqref{Mem72} leads us to the approximate expression based on the memory function formalizm,
\be\label{tau_lB}
\frac{1}{ \tau_{\perp}} = \frac{ 1 }{ \omega_{\perp} } \frac{ \Im \left[K^{(0)}_{[\mathcal{F}_1,H_{int}],[\mathcal{F}_1,H_{int}]}(\omega_{\perp}) \right] }{ K^{(0)}_{\mathcal{F}_1,\mathcal{F}_1}(\omega=0) }\, .
\ee
The KTh is built into Eq.~\eqref{tau_lB} since in the harmonic confinement, the mode $\mathcal{F}_1$ as defined by Eq.~\eqref{SM} satisfies the Heisenberg equation of motion, $\dot{A}_{\perp} = i \omega_{\perp} \mathcal{F}_1$ to all orders in interaction, which in particular means that $[\mathcal{F}_1,H_{int}]=0$ exactly.
Moreover, the DOs decouple from the rest of the excitations at different frequencies.

\subsection{One occupied band}
\label{sec:MF1}
In this section we implement the program outlined in Sec.~\ref{sec:SM_approx} to compute the life time of DOs in the simplest situation of one occupied band.
In this case the only possible resonant inter-band transitions are from the lowest to the nearest unoccupied band. 
Therefore the slow variable, Eq.~\eqref{SM} excited by the sloshing is $\mathcal{F}_1 = A_1$.
We discuss its frequency and life time in the Secs.~\ref{sec:MF1_frequency} and \ref{sec:MF1_life_time} respectively.
The more detailed discussion of the mode splitting is given in Sec.~\ref{sec:MF2} for the case of two occupied bands.
\subsubsection{Frequency of DOs}
\label{sec:MF1_frequency} 
In this section we demonstrate the consistency of the memory function formalism in the simplest case of single occupied band.
For clarity, here we consider the harmonic confining potential.
In the noninteracting system all the particle-hole excitations carrying zero-momentum and energy $\omega_{\perp}$ are degenerate.
According to the degenerate perturbation theory, the interaction induced energy splitting is of the order $\propto v_{int}$.
This splitting allows us to focus on a one-dimensional space of observables spanned by a single variable, Eq.~\eqref{SM} as discussed above.
Furthermore, as only the lowest band is occupied we can truncate Eq.~\eqref{SM}, $\mathcal{F}_1 = A_1$.
To all orders in interactions, $\dot{A}_1 = i \omega_{\perp} A_1$ and therefore by the definitions Eq.~\eqref{Mem61} and Eq.~\eqref{Mem65},
\be\label{Mem78}
\langle A_1 | \hat{\Omega}| A_1 \rangle = \frac{ \langle A_1 |\hat{\omega} |A_1\rangle}{ \langle A_1 |\hat{\chi}_s |A_1\rangle }
=\omega_{\perp}
\ee

In a one-dimensional space of observables spanned by $A_1$, the Kubo relaxation matrix, Eq.~\eqref{Mem63} reduces to the scalar function of frequency.
The static response for weak interactions,
\begin{align}\label{Mem80}
\chi_s^{(0)} = \frac{N_0-N_{1}}{  \omega_{\perp} }
\end{align}
is found from Eq.~\eqref{Mem61} by setting $\omega =0$ in the expression  \eqref{PT27}.
Substitution of Eqs.~\eqref{Mem78} and \eqref{Mem80} along with $\Sigma(z)=0$ in Eq.~\eqref{Mem63} yields 
\begin{align}\label{Mem85}
C(z)  = 
i \frac{N_0-N_1}{ \omega_{\perp} } \frac{ 1 }{z - \omega_{\perp} }\, .
\end{align}
The Eq.~\eqref{Mem85} can also be obtained by substitution of Eq.~\eqref{PT27} into the relation Eq.~\eqref{Mem47_A}.
This demonstrates the consistency of the present formalism.

\subsubsection{Lifetime of the DOs}
\label{sec:MF1_life_time}
We now turn to the evaluation of the life time of the DOs in the case of a single occupied band.
As in Sec.~\ref{sec:PT},
we project the full interaction Hamiltonian, \eqref{H_full} onto a space of the first two lowest bands. 
The projected Hamiltonian contains three terms,
\begin{align}\label{mem37}
H_{int,1}'  =&   U_{1} \sum V_{\bm{q}} \psi_{1,\bm{k}-\bm{q}}^{\dag} \psi_{0,\bm{k}'+\bm{q}}^{\dag}  \psi_{0,\bm{k}'}  \psi_{1,\bm{k}}
\notag \\
& + U_{\bar{1}}  \sum  V_{\bm{q}} \psi_{1,\bm{k}-\bm{q}}^{\dag}  \psi_{0,\bm{k}'+\bm{q}}^{\dag}  \psi_{1,\bm{k}'} \psi_{0,\bm{k}}
\notag \\
&+ U_{0} \sum \frac{V_{\bm{q}}}{2}  \psi_{0,\bm{k}-\bm{q}}^{\dag} \psi_{0,\bm{k}'+\bm{q}}^{\dag} \psi_{0,\bm{k}'}   \psi_{0,\bm{k}}\, .
\end{align}
To apply the relation, \eqref{tau_lB} we compute the commutator relations of the observable $A_1$ with the projected Hamiltonian, \eqref{mem37}.
These commutation relations are summarized in App.~\ref{app:commute}, (see Eqs.~\eqref{mem45}, \eqref{mem38} and \eqref{V2a}), and give 
\begin{align}\label{mem39}
&[A_1, H_{int,1}']  =  U_{0}  \sum V_{\bm{q}} \psi_{0,\bm{k}-\bm{q}}^{\dag} \psi_{1,\bm{k}'+\bm{q}}^{\dag}    \psi_{0,\bm{k}'} \psi_{0,\bm{k}} 
\notag \\
&+U_{1}  \sum V_{\bm{q}} \left( \psi_{1,k-\bm{q}}^{\dag} \psi_{1,\bm{k}'+\bm{q}}^{\dag}\psi_{0,\bm{k}'} \psi_{1,\bm{k}} \right. \notag \\  
&\left.-\psi_{1,\bm{k}-\bm{q}}^{\dag} \psi_{0,\bm{k}'+q}^{\dag} \psi_{0,\bm{k}'}  \psi_{0,\bm{k}} \!\right)
\notag \\
&+
U_{\bar{1}}  \sum V_{\bm{q}} \left(
\psi_{1,\bm{k}-\bm{q}}^{\dag} \psi_{1,\bm{k}'+\bm{q}}^{\dag}  \psi_{1,\bm{k}'} \psi_{0,\bm{k}} \right. \notag \\ 
&\left. - \psi_{1,k-\bm{q}}^{\dag} \psi_{0,\bm{k}'+\bm{q}}^{\dag}  \psi_{0,\bm{k}'} \psi_{0,\bm{k}}\right)
\end{align}
Note that the terms in Eq.~\eqref{mem39} proportional to $\psi_{1,k-\bm{q}}^{\dag} \psi_{1,\bm{k}'+\bm{q}}^{\dag}\psi_{0,\bm{k}'} \psi_{1,\bm{k}}$ do not contribute to the correlation function in \eqref{tau_lB} as only the lowest, $n=0$ band is occupied, while the next, $n=1$ band is empty, and cannot accommodate holes.
The remaining terms form the combination, $(U_{0}  - U_{\bar{1}}  - U_{1} )\bar{M}$, with the operator 
\begin{align}\label{H_tr}
\bar{M}= 
\sum_{\bm{k},\bm{k}',\bm{q}}  V_{\bm{q}}  \psi_{1,\bm{k}-\bm{q}}^{\dag} \psi_{0,\bm{k}'+\bm{q}}^{\dag} \psi_{0,\bm{k}'}  \psi_{0,\bm{k}}\, ,
\end{align}
It follows that,
\begin{align}\label{FF}
\Im K^{(0)}_{[\mathcal{F}_1,H_{int}],[\mathcal{F}_1,H_{int}]}= &
(U_{0}  - U_{\bar{1}}  - U_{1} )^2
\notag \\
& \times \Im K^{(0)}_{\bar{M},\bar{M}}(\omega = \omega_{\perp}) 
\end{align}
The correlation function on the right hand side of Eq.~\eqref{FF} is (see App.~\ref{der:MM} for details),
\begin{align}\label{MM}
\Im K^{(0)}_{\bar{M},\bar{M}}&(\omega)  = -\pi \left( 1 - e^{ -\omega/T} \right) \sum_{\bm{k},\bm{k}',\bm{q}} \frac{1}{2}(V_q-V_{q-k+k'})^2
\notag \\
\times & (1-f_{0,\bm{k}' + \bm{q}}) f_{0,\bm{k}' } (1-f_{1,\bm{k} - \bm{q}}) f_{0,\bm{k} } 
\notag \\
\times &  \delta \left( \omega - \xi_{0,\bm{k}'+ \bm{q}} +  \xi_{0,\bm{k}'}  - \xi_{1,\bm{k} - \bm{q} } + \xi_{0,\bm{k}} \right)\, .
\end{align}
As before, we omit the contribution of the exchange processes to the correlation functions by replacing in Eq.~\eqref{MM}, $(V_q-V_{q-k+k'})^2/2$ with $V_q^2$.
By comparison of Eq.~\eqref{MM} with Eq.~\eqref{GR-13} we obtain
\begin{align}\label{MA}
\Im K^{(0)}_{[\mathcal{F}_1,H_{int}],[\mathcal{F}_1,H_{int}]} = (\omega - \omega_{\perp})^2 \Im K^{(2)}_{A_1,A_1}(\omega_{\perp}) 
\end{align}
Substituting Eq.~\eqref{MA} along with the expression for the static susceptibility, Eq.~\eqref{Mem80} in the basic Eq.~\eqref{tau_lB} we write the latter in the form,
\be\label{tau_lB1}
\frac{1}{ \tau_{\perp}} =- \frac{ (\omega - \omega_{\perp})^2 }{ N_0 - N_1}   \Im K^{(2)}_{A_1,A_1}(\omega_{\perp}) \, .
\ee
Which exactly corresponds to the naive guess, Eq.~\eqref{PT27-A} and we conclude that the life tine obtained within the memory function formalism coincides with the results obtained by the naive perturbation theory in Sec.~\ref{sec:PT}, namely,
\be
\frac{1}{\tau_{\perp}} = \frac{1}{\tau^{(2)}_{\perp} } \, .
\ee

\subsection{Two occupied bands}
\label{sec:MF2}
Before we discuss the effect of interaction let us summarize the results in the non-interacting case and harmonic confining potential, $\epsilon=0$.
\subsubsection{Non-interacting fermions}
\label{sec:NI}

Compared to the previously considered case of a single occupied band  there are at least two independent slow variables in the problem,
\be
A_1 = \sum_{\bm{p}} \psi_{1,\bm{p}}^{\dag} \psi_{0,\bm{p}}\, ,  A_2 = \sum_{\bm{p}} \psi_{2,\bm{p}}^{\dag} \psi_{1,\bm{p}}\, .
\ee
In the absence of interactions the two-by-two memory matrix, $\hat{\Sigma}$ vanishes and we focus on the frequency renormalization matrix $\hat{\Omega}$, Eq.~\eqref{Mem65}. 

As in the case of a single occupied band considered in Sec.~\ref{sec:MF1_frequency}, $\hat{\Omega}$ trivially follows from the Heisenberg equation of motion, $\dot{A}_{1,2} = i \omega_{\perp} A_{1,2}$.
In this case the comparison of Eq.~\eqref{Mem61} and \eqref{Mem65} gives the relation,
$\hat{\omega} = \omega_{\perp} \hat{\chi}_s$ which from Eq.~\eqref{Mem65} gives $\hat{\Omega} =  \omega_{\perp} $, as expected.
Clearly for any number $N_m$ of occupied bands the matrix $\hat{\Omega}$ in the space of variables $A_n$, $n=1,2,\ldots, N_m$ is the product of $\omega_{\perp}$ and a unit matrix of dimension $N_m$.
In the interacting system the frequency renormalization matrix, $\hat{\Omega}$ in general is non-diagonal.
Even in this case the dipole operator, \eqref{SM} with the summation extending up to $N_m$ remains an eigenstate of $\hat{\Omega}$ if the confining potential is parabolic.

In the subsequent sections we will also need the static response matrix, Eq.~\eqref{Mem61}.
The latter is obtained from Eqs.~\eqref{Mem61} and \eqref{PT27}, 
\be\label{static}
\hat{\chi}_s^{(0)} = 
\frac{1}{\omega_{\perp} }
\begin{pmatrix}
N_0 - N_1 & 0
\\
0 & N_1
\end{pmatrix}\, .
\ee

We have considered the pair of slow variables $A_{1,2}$ which define two degenerate long-lived coherent sloshing oscillations.
Let us stress again that the space of all such variables is infinite.
As discussed in Sec.~\ref{sec:SM_approx}  the splitting of the DOs off the rest of the excitations by the interaction is both generic and essential to the present formulation.  
For this reason we address the mode splitting before discussion of the life time of DOs in Sec.~\ref{sec:LT}.

The mode splitting can be analyzed in the limit of the harmonic confinement as the dimensionless interaction, $v_{int}$ is assumed to be larger than the anharmonicity parameter, $\epsilon$.
In Sec.~\ref{sec:split_mem} we show  that generic interaction lifts the mode degeneracy leaving the DOs, Eq.~\eqref{SM} unaffected.
The same task is achieved in Sec.~\ref{sec:split_diag} using the standard diagrammatic technique.
The two approaches give identical results.

\subsubsection{In- and out-of-phase modes splitting: memory function formalism}
\label{sec:split_mem}

Here we focus on the splitting between the two modes $A_1$ and $A_2$.
These modes represent the two independent sloshing oscillations of particles in $n=0$ and $n=1$ bands respectively.
Interaction mixes the two types of sloshing oscillations.
The mixed modes are in- and out-of-phase oscillations in $n=0$ and $n=1$ bands that are split in frequency.
The former, in-phase combination is the dipole operator of entire system that according to KTh retains the unrenormalized frequency, $\omega_{\perp}$.
The out-of-phase oscillations are expected to have a renormalized frequency and finite life-time.

Our goal is to find the renormalized frequencies of collective modes to the first order in interaction.
In the memory function formalism it amounts to calculating the frequency renormalization matrix, $\hat{\Omega}$ to the first order in interaction.
From Eq.~\eqref{Mem65}, 
\begin{align}\label{Mem_s}
[\hat{\Omega}]_{i,j} &= \sum_{l=1,2}\langle [H_0,A_i]| A_{l}\rangle [\chi_s^{-1}]_{lj} +
\langle [H_{int},A_i]| A_{l}\rangle [\chi_s^{-1}]_{lj} 
\end{align}
Since $[H_0,A_i] = \omega_{\perp} A_i$ and in view of the definition Eq.~\eqref{Mem61} the fist term in Eq.~\eqref{Mem_s}
\begin{align}
\sum_{l=1,2}\langle [H_0,A_i]| A_{l}\rangle [\chi_s^{-1}]_{lj} = \omega_{\perp} \delta_{ij}
\end{align}
even in the presence of interactions.
This observation is akin to the cancellation of disconnected graphs in the diagrammatic technique.
It remains to evaluate the second term in Eq.~\eqref{Mem_s}.
As a result, to the first order in interactions 
\begin{align}\label{Mem_s1}
[\hat{\Omega}]_{i,j} & \approx \sum_{l=1,2} \langle [H_{int},A_i]| A_{l}\rangle [\chi^{(0)}_s]^{-1}_{lj}\,  ,
\end{align}
where both the correlation function $\langle [H_{int},A_i]| A_{l}\rangle$ should be evaluated in the non-interacting system, and the non-interacting static correlation function $\chi^{(0)}_s$ is explicitly given by Eq.~\eqref{static}.
In result we have reduced the calculation of the frequency renormalization matrix, $\hat{\Omega}$ to the evaluation of the non-interacting static correlation functions of the observables $A_i$ and their commutation relations with the interaction Hamiltonian, $[H_{int},A_j]$, 
$i,j=1,2$.
Since the observables $A_{1,2}$ operate within the first three bands, the commutation relations $[H_{int},A_i]$ can be evaluated with the interaction projected onto these three bands,
\begin{align}\label{H_trunc3_A}
H_{int}' = H_{int,a}' + H_{int,b}' + H_{int,c}',
\end{align}
\begin{subequations}\label{H_trunc3}
\begin{align}
H_{int,a}' &= U^{01}_{01} \sum{}^{'} V_{\bm{q}}  \psi_{1,\bm{k}-\bm{q}}^{\dag}  \psi_{0,\bm{k}'+\bm{q}}^{\dag}  \psi_{0,\bm{k}'}  \psi_{1,\bm{k}}
\notag \\
+ &
U^{01}_{10}  \sum{}^{'} V_{\bm{q}} \psi_{1,\bm{k}-{\bm{q}}}^{\dag}  \psi_{0,\bm{k}'+{\bm{q}}}^{\dag}  \psi_{1,\bm{k}'} \psi_{0,\bm{k}} 
\notag \\
+ & \frac{U^{00}_{00}}{2} \sum{}^{'} V_{\bm{q}}  \psi_{0,\bm{k}'+{\bm{q}}}^{\dag}\psi_{0,\bm{k}-{\bm{q}}}^{\dag}  \psi_{0,\bm{k}} \psi_{0,\bm{k}'} 
\notag \\
+ &
\frac{U^{11}_{11}}{2} \sum{}^{'} V_{\bm{q}} \psi_{1,\bm{k}'+{\bm{q}}}^{\dag}  \psi_{1,\bm{k}-{\bm{q}}}^{\dag}  \psi_{1,\bm{k}} \psi_{1,\bm{k}'},
\end{align}
\begin{align}
H_{int,b}' &= 
U^{11}_{02} \sum{}^{'} V_{\bm{q}} \psi_{2,\bm{k}-{\bm{q}}}^{\dag} \psi_{0,\bm{k}'+{\bm{q}}}^{\dag}  \psi_{1,\bm{k}'}  \psi_{1,\bm{k}} 
\notag \\
+ &
U^{02}_{11} \sum{}^{'} V_{\bm{q}} \psi_{1,\bm{k}'}^{\dag}  \psi_{1,\bm{k}}^{\dag}  \psi_{2,\bm{k}-{\bm{q}}} \psi_{0,\bm{k}'+{\bm{q}}} 
\end{align}
\begin{align}
H_{int,c}' &= 
U^{20}_{20} \sum{}^{'} V_{\bm{q}} \psi_{0,\bm{k}-{\bm{q}}}^{\dag} \psi_{2,\bm{k}'+{\bm{q}}}^{\dag}  \psi_{2,\bm{k}'}  \psi_{0,\bm{k}} 
\notag \\
+ &
U^{20}_{02} \sum{}^{'} V_{\bm{q}} \psi_{2,\bm{k}-{\bm{q}}}^{\dag} \psi_{0,\bm{k}'+{\bm{q}}}^{\dag}  \psi_{2,\bm{k}'}  \psi_{0,\bm{k}} 
\notag \\
+ &
U^{21}_{21} \sum{}^{'} V_{\bm{q}}  \psi_{2,\bm{k}'+{\bm{q}}}^{\dag}\psi_{1,\bm{k}-{\bm{q}}}^{\dag} \psi_{1,\bm{k}}   \psi_{2,\bm{k}'} 
\notag \\
+ &
U^{21}_{12} \sum{}^{'} V_{\bm{q}}  \psi_{2,\bm{k}-{\bm{q}}}^{\dag} \psi_{1,\bm{k}'+{\bm{q}}}^{\dag}  \psi_{2,\bm{k}'}  \psi_{1,\bm{k}} \, ,
\end{align}
\end{subequations}
where $\sum{}^{'}$ stands for $\sum_{\bm{k},\bm{k}',\bm{q}}$.
The commutation relations with each of the terms appearing in Eq.~\eqref{H_trunc3} are computed in the App.~\ref{app:commute}.
Summarizing this straightforward calculation we obtain,
\begin{align}\label{A1}
[ A_1,\! & H_{int}'] \!= \! (U_{00}^{00}\!-\!U_{01}^{01}\! -\!U_{01}^{10})\!\!\sum{}^{'} V_{\bm{q}} \psi_{1,{\bm{k}}-{\bm{q}}}^{\dag} \psi_{0,{\bm{k}}'+{\bm{q}}}^{\dag} \psi_{0,{\bm{k}}'}  \psi_{0,{\bm{k}}} 
\notag \\
+ & (U_{01}^{01}\!+\!U_{01}^{10}\! -\!U_{11}^{11})\!\!\sum{}^{'} V_{\bm{q}} \psi_{1,{\bm{k}}-{\bm{q}}}^{\dag} \psi_{1,{\bm{k}}'+{\bm{q}}}^{\dag} \psi_{0,{\bm{k}}'}  \psi_{1,{\bm{k}}}
\notag \\
- &U^{11}_{02}  \sum{}^{'} V_{\bm{q}} \psi_{2,{\bm{k}}-{\bm{q}}}^{\dag}  \psi_{0,{\bm{k}}'+{\bm{q}}}^{\dag}  \psi_{1,{\bm{k}}'}\psi_{0,{\bm{k}}}
\notag \\
+&U^{11}_{02}  \sum{}^{'} V_{\bm{q}} \psi_{2,{\bm{k}}-{\bm{q}}}^{\dag}  \psi_{1,{\bm{k}}'+{\bm{q}}}^{\dag}  \psi_{1,{\bm{k}}'} \psi_{1,{\bm{k}}}
\notag \\
+&(U^{02}_{20}\! -\!U_{12}^{21}) \!\!\sum{}^{'} V_{\bm{q}} \psi_{2,{\bm{k}}-{\bm{q}}}^{\dag} \psi_{1,{\bm{k}}'+{\bm{q}}}^{\dag}  \psi_{2,{\bm{k}}'}  \psi_{0,{\bm{k}}}
\notag \\
-&(U^{02}_{02}\! - \!U^{12}_{12}) \!\!\sum{}^{'} V_{\bm{q}}
\psi_{1,{\bm{k}}-{\bm{q}}}^{\dag} \psi_{2,{\bm{k}}'+{\bm{q}}}^{\dag}  \psi_{2,{\bm{k}}'}  \psi_{0,{\bm{k}}},
\end{align}
\begin{align}\label{A2}
[ A_2, H_{int}'] &=  (U^{01}_{01} - U^{02}_{02})\sum{}^{'}  V_{\bm{q}}  \psi_{2,{\bm{k}}-{\bm{q}}}^{\dag} \psi_{0,{\bm{k}}'+{\bm{q}}}^{\dag} \psi_{0,{\bm{k}}'}  \psi_{1,{\bm{k}}} 
\notag \\
+ &
(U^{01}_{10} - U^{02}_{20})\sum{}^{'}  V_{\bm{q}}  \psi_{2,{\bm{k}}-{\bm{q}}}^{\dag} \psi_{0,{\bm{k}}'+{\bm{q}}}^{\dag} \psi_{1,{\bm{k}}'}  \psi_{0,{\bm{k}}} 
\notag \\
 -&
U^{02}_{11} \sum{}^{'}  V_{\bm{q}}  \psi_{1,{\bm{k}}-{\bm{q}}}^{\dag} \psi_{1,{\bm{k}}'+{\bm{q}}}^{\dag} \psi_{0,{\bm{k}}'}  \psi_{1,{\bm{k}}} 
\notag \\
+ & 
U^{02}_{11} \sum{}^{'} V_{\bm{q}}  \psi_{1,{\bm{k}}-{\bm{q}}}^{\dag} \psi_{2,{\bm{k}}'+{\bm{q}}}^{\dag} \psi_{0,{\bm{k}}{\bm{k}}'}  \psi_{2,{\bm{k}}} 
\notag \\
+ &
U^{02}_{11} \sum{}^{'}  V_{\bm{q}}  \psi_{2,{\bm{k}}-{\bm{q}}}^{\dag} \psi_{1,{\bm{k}}'+{\bm{q}}}^{\dag} \psi_{0,{\bm{k}}'}  \psi_{2,{\bm{k}}} 
\notag \\
+  (U_{11}^{11}\!  - & U^{12}_{12} \! - \! U^{12}_{21})\! \!  \sum{}^{'}  V_{\bm{q}}  \psi_{2,{\bm{k}}-{\bm{q}}}^{\dag} \psi_{1,{\bm{k}}'+{\bm{q}}}^{\dag} \psi_{1,{\bm{k}}'}  \psi_{1,{\bm{k}}} 
\notag \\
- (U_{22}^{22}\! - & U^{12}_{12}\!  -\!  U^{12}_{21})\! \!  \sum{}^{'}  V_{\bm{q}}  \psi_{2,{\bm{k}}-{\bm{q}}}^{\dag} \psi_{2,{\bm{k}}'+{\bm{q}}}^{\dag} \psi_{2,{\bm{k}}'}  \psi_{1,{\bm{k}}} \, .
\end{align}
To illustrate the formalism we compute the $\omega_{11}$ entry of the matrix $\hat{\omega}$.
As the correlation functions are computed in the non-interacting system, all the terms in Eq.~\eqref{A1} which contain the operators acting on the band $n=2$ do not contribute to $\omega_{11}$.
The remaining, first two terms in Eq.~\eqref{A1} give,
\begin{align}\label{contrib1}
\omega_{11}\!& =\! (U_{00}^{00}\!-\!U_{01}^{01}\! -\!U_{01}^{10})\!\!\sum{}^{'}  \!\!
\langle \psi_{1,{\bm{k}}-{\bm{q}}}^{\dag} \psi_{0,{\bm{k}}'+{\bm{q}}}^{\dag} \psi_{0,{\bm{k}}'}  \psi_{0,{\bm{k}}} | \psi_{1\bm{p}}^{\dag} \psi_{0\bm{p}} \rangle
\notag \\
+ 
(&U_{01}^{01}\! +\!U_{01}^{10}\! -\!U_{11}^{11})\!\!\sum{}^{'}\!\!  \langle \psi_{1,{\bm{k}}-{\bm{q}}}^{\dag} \psi_{1,{\bm{k}}'+{\bm{q}}}^{\dag} \psi_{0,{\bm{k}}'}  \psi_{1,{\bm{k}}}| \psi_{1\bm{p}}^{\dag} \psi_{0\bm{p}} \rangle.
\end{align}
In the non-interacting systems, 
\begin{align}\label{W}
\langle  \psi_{0,{\bm{k}}}^{\dag} & \psi_{0,{\bm{k}}'}^{\dag} \psi_{0,{\bm{k}}'+{\bm{q}}}  \psi_{1,{\bm{k}}-{\bm{q}}} | \psi_{1p}^{\dag} \psi_{0p} \rangle =
\notag \\
 = &\langle \psi_{0,{\bm{k}}'+{\bm{q}}}^{\dag} \psi_{0,{\bm{k}}'} \rangle
\langle  \psi_{0,{\bm{k}}}^{\dag}   \psi_{1,{\bm{k}}-{\bm{q}}} | \psi_{1p}^{\dag} \psi_{0p} \rangle
\notag \\
& - \langle  \psi_{0,{\bm{k}}}^{\dag} \psi_{0,{\bm{k}}'+{\bm{q}}} \rangle
\langle \psi_{0,{\bm{k}}'}^{\dag}  \psi_{1,{\bm{k}}-{\bm{q}}}|  \psi_{1p}^{\dag} \psi_{0p} \rangle\, 
\end{align}
as can be deduced from Eq.~\eqref{Mem61}.  
The two terms in Eq.~\eqref{W} represent the direct and exchange contributions to the frequency renormalization respectively.
For the sake of clarity we exclude the direct contributions by setting $V_{{\bm{q}}=0}=0$.
It is straightforward to include it.
In addition, we make another simplifying but not essential assumptions of week momentum dependence of the the interaction matrix elements, $V_{{\bm{q}}} \approx V_0$.
Under these assumptions substitution of Eq.~\eqref{W} in Eq.~\eqref{contrib1} yields
\begin{align}\label{omega11}
\omega_{11} = 
- & V_0 \langle A_1 | A_1 \rangle \big[ N_0 (U_{00}^{00}-U_{01}^{01} -U_{01}^{10}) 
\notag \\
& +N_1(U_{01}^{01}+U_{10}^{01} -U_{11}^{11}) \big]\, .
\end{align}
Proceeding in the similar fashion we obtain the full matrix, $\hat{\omega}$
Then by Eq.~\eqref{Mem_s1} the matrix $\hat{\Omega} \approx \hat{\omega}[\hat{\chi}_s^{(0)}]^{-1}$ reads 
\begin{align}\label{omega_2}
\hat{\Omega} = \begin{bmatrix}
\Omega_{11} & \Omega_{12} \\
\Omega_{21} & \Omega_{22} 
\end{bmatrix},
\end{align}
where
\begin{align*}
\Omega_{11} &= - V_0 N_0 (U_{00}^{00}-U_{01}^{01} -U_{01}^{10}) 
\notag \\
\Omega_{12} & = V_0 (N_0 - N_1) U_{02}^{11} 
\notag \\
\Omega_{21} & = -U_{02}^{11} N_1 
\notag \\
\Omega_{22} & = -V_0 N_0 (U^{01}_{10} - U^{02}_{20} )  -V_0 N_1 (U_{11}^{11} - U_{12}^{12} - U_{12}^{21}). 
\end{align*}
The diagonalization of the matrix, Eq.~\eqref{omega_2} gives the collective modes we are after. 
Lets first clarify the transformation properties of the matrix, $\hat{\Omega}$.
By their very definitions, Eqs.~\eqref{Mem61} and \eqref{Mem65} these matrices, $\hat{\bar{\omega}}$ and 
$\hat{\bar{\chi}}$ defined via the collective variables,
\begin{align}\label{bar_A}
\mathcal{F}_i = \sum_{j=1}^2 O_{ij} A_{j}
\end{align}
are related to the corresponding matrices in terms of the variables $A_i$ as 
$\hat{\bar{\omega}} = \hat{O} \hat{\omega} O^{tr}$, and $\hat{\bar{\chi}} = \hat{O} \hat{\chi} \hat{O}^{tr}$.
It follows that $\hat{\bar{\Omega}} = \hat{O} \hat{\Omega} \hat{O}^{-1}$.
The matrix $\hat{O}$ is fixed by KTh.
Indeed, one of the modes, Eq.~\eqref{bar_A}  must be the dipole operator, \eqref{SM}, namely $\mathcal{F}_1 = A_1 + \sqrt{2} A_2$.
The second out-of-phase mode, $\mathcal{F}_2$ should be orthogonal to $\mathcal{F}_1$ with respect to the scalar product, Eq.~\eqref{Mem55}.
This orthogonality, according to Eq.~\eqref{Mem61} means that perturbation coupled to $\mathcal{F}_2$ leaves $\langle \mathcal{F}_1 \rangle =0$. 
It finally fixes the matrix,
\begin{align}\label{O}
\hat{O} = \begin{pmatrix}
1 & \sqrt{2} \\
- \frac{ \sqrt{2}}{N_0 - N_1} & \frac{1 }{N_1}
\end{pmatrix}\, .
\end{align}
The matrix $\hat{O}$ in Eq.~\eqref{O} is independent of interactions.
This means that the different matrix elements defining the matrix, $\Omega$, Eq.~\eqref{omega_2} are not independent.
They have to satisfy general relations that rely only on the confining potential being harmonic.
Here we bring some of these identities,
\begin{align}\label{ident}
U_1 & = U_{00}^{00}-U_{01}^{01} -U_{01}^{10} =0\, , \notag \\
U_2 & = U_{01}^{01}+U_{01}^{10} -U_{11}^{11} - \sqrt{2} U_{11}^{02}=0\, , \notag \\
U_3 & = U^{11}_{02} - \sqrt{2} (U_{01}^{01} + U_{01}^{10} - U_{02}^{02} - U_{20}^{02} )=0\, , \notag \\
U_4 & = U^{11}_{02} + \sqrt{2} (U_{11}^{11} - U^{12}_{12} - U^{12}_{21})=0\, ,\notag \\
U_5 & = \sqrt{2} (U^{01}_{10} - U^{02}_{20} ) -U^{11}_{02}= 0\, .
\end{align}
The identities, Eq.~\eqref{ident} belong to an infinite set of relations between different matrix elements valid for any translational invariant interactions.
We derive and discuss these and other identities of the same kind in Sec.~\ref{sec:LT}.

With the help of Eq.~\eqref{ident} the matrix, $\hat{\Omega}$, Eq.~\eqref{omega_2} simplifies,
\begin{align}\label{omega_3}
\hat{\Omega} = - V_0 U_{02}^{11}  \begin{bmatrix}
\sqrt{2} N_1 & 
N_1 - N_0 \\
-N_1 &  -(N_1 - N_0)/\sqrt{2} 
\end{bmatrix} \, .
\end{align}
As a result,
\begin{align}\label{omega_bar}
\hat{\Omega} = - V_0 \frac{U_{02}^{11}}{\sqrt{2}}  \begin{bmatrix}
0 & 
0 \\
0 &  N_0 + N_1 
\end{bmatrix} \, .
\end{align}
The renormalization of the frequencies of the in- (DOs) and out-of-phase collective modes are given as diagonal entries of the matrix, Eq.~\eqref{omega_bar}.
The frequencies of the two modes become therefore,
\begin{align}\label{fr}
\bar{\omega}_1 = \omega_{\perp}\, ,\,\,\, 
\bar{\omega}_2 = \omega_{\perp} -  V_0 \frac{U_{02}^{11}}{\sqrt{2}}(N_0 + N_1)\, .
\end{align} 
Consider, for illustration the point-like interaction,
\be\label{def37}
f[(z_j - z_{j'})] = U_0 \ell_{\perp}\sqrt{2\pi} \delta(z_j - z_{j'}) \, .
\ee
In this case, the frequency renormalization of the out-of-phase mode reads
\begin{align}\label{split}
\bar{\omega}_2 - \omega_{\perp} = V_0 U_0(N_0+N_1)/8\, .
\end{align}
The results, Eq.~\eqref{def37} and  Eq.~\eqref{split} agree with the frequency renormalization obtained by a direct diagrammatic technique as is demonstrated in Sec.~\ref{sec:split_diag}.
\subsubsection{In- and out-of-phase modes splitting: diagrammatic analysis}
\label{sec:split_diag}
In the previous section we have evaluated the interaction induced frequency splitting between the in- (DOs)  and out-of-phase collective modes.
The goal of this section is to find both frequencies using the standard perturbation theory.
Here as in the Sec.~\ref{sec:split_mem} we do not include the direct interaction processes and set $V_q = V_0$. 
Evaluation of the frequency renormalization to the first order in interaction amounts to the summation of the diagrams of all orders that are most singular at $\omega=\omega_{\perp}$, \cite{Iqbal2014}.
This is a basis of the Random Phase Approximation (RPA)-like approaches to finding the spectrum of collective excitations at weak interactions.
For long range Coulomb interactions RPA holds at long wave lengths \cite{Li1989,Li1991}.  
Such a scheme is consistent with KTh, \cite{Li1989,Li1991,Wendler1995,Nikonov1997,Li2003}.
It yields the RPA-like expression for the correlation matrix,
$[\hat{K}]_{ij} = K_{A_i,A_j}$ defined by Eq.~\eqref{K},
\be\label{K_RPA}
\hat{K}(\omega) = [\hat{\Pi}^{-1}(\omega) + \hat{V}]^{-1}\, ,
\ee 
where the inverse polarization operator 
\be\label{Pi}
\hat{\Pi}^{-1}(\omega) =
\begin{bmatrix}
\frac{\omega - \omega_{\perp} -  \Sigma_0 }{ N_0 - N_1} & 0 \\
0 &  \frac{\omega - \omega_{\perp} - \Sigma_1 }{ N_1 } 
\end{bmatrix}
\ee 
includes the contributions of the self-energy renormalizations of the bands,
\begin{align}\label{Sigma}
\Sigma_0 &= N_0 V_0 (U_{00}^{00} - U_{01}^{10} ) + N_1 V_0( U_{01}^{10}- U_{11}^{11}  )\, , 
\notag \\
\Sigma_1 & = N_0 V_0 ( U_{01}^{10}- U_{02}^{20}   ) + N_1 V_0 (U_{11}^{11} - U_{12}^{21} )
\end{align}
and the interaction matrix reads, 
\be\label{V}
\hat{V} = V_0 \begin{bmatrix}
U_{01}^{01} & U_{01}^{12} \\
U_{01}^{12} & U_{12}^{12}
\end{bmatrix}\, .
\ee 
The expressions Eq.~\eqref{Pi}, \eqref{Sigma} and \eqref{V} are derived and discussed in details in Ref.~\cite{Iqbal2014}.

The frequencies of the collective modes are determined by the condition,
\be\label{det}
\det[\hat{\Pi}^{-1}(\omega) + \hat{V}]  = 0 \, .
\ee 
It is convenient to introduce the diagonal matrix,
\be\label{D}
\hat{D} = \begin{bmatrix}
\sqrt{N_0 - N_1} & 0 \\
0 & \sqrt{N_1} 
\end{bmatrix}
\ee
because 
\begin{align}\label{D1}
&\hat{D} (\hat{\Pi}^{-1} + \hat{V} )  \hat{D}
= \omega-\omega_{\perp} 
\notag \\
 +  & 
\begin{bmatrix}
-\Sigma_0   + V_0 U_{01}^{01}(N_0 - N_1) & V_0 U_{01}^{12} \sqrt{N_1(N_0 - N_1)} \\
V_0 U_{01}^{12} \sqrt{N_1(N_0 - N_1)} &   -\Sigma_1 + V_0 U_{12}^{12} N_1
\end{bmatrix},
\end{align}
and  therefore the condition \eqref{det} amounts to a regular eigenvalue problem.
The relations, Eq.~\eqref{ident} allow us to write Eq.~\eqref{D1} in the form,
\begin{align}\label{D3}
\hat{D}& (\hat{\Pi}^{-1} + \hat{V} )  \hat{D} = 
\omega - \omega_{\perp}
\notag \\
+V_0 U_{01}^{12} & 
\begin{bmatrix}
-\sqrt{2} N_1 &  \sqrt{N_1(N_0 - N_1)} \\
 \sqrt{N_1(N_0 - N_1)} &   -(N_0-N_1)/\sqrt{2} 
\end{bmatrix}\, .
\end{align}
The frequencies that nullify the determinant of the matrix, Eq.~\eqref{D3} coincides with Eq.~\eqref{fr} demonstrating the equivalence of the memory function and diagrammatic approaches.
The eigenvectors of the symmetric matrix, Eq.~\eqref{D3},
\be\label{v12}
 \tilde{v}_1  = \left[\frac{\sqrt{N_0-N_1}}{\sqrt{2 N_1}},1\right]^{tr}\, ,
 \tilde{v}_2   =  \left[-\frac{\sqrt{2 N_1} }{\sqrt{N_0-N_1}},1\right]^{tr}
\ee
are related to but do not coincide with the relative waits of $A_{1,2}$ in the collective modes.
To find the collective excitations we following Ref.~\cite{Forster1975} consider the time evolution of expectation values, 
$\langle A_{1,2}\rangle (t)$ given their values at  $\langle A_{1,2}\rangle (t=0)$ at initial time, $t=0$.
In terms of Fourier transformations the following relationship holds,
\be\label{R1}
\langle A_{i}\rangle (\omega) = \sum_{j=1}^2{R}_{ij}(\omega) \langle A_{j}\rangle(t=0)\, .
\ee
The response matrix, \cite{Forster1975}
\be\label{R}
\hat{R}(\omega) = \frac{ 1 }{ \omega} [ \hat{K}(\omega) \hat{K}^{-1}(\omega = 0) - 1]\, .
\ee
The collective modes are the eigenvectors of the matrix $\hat{R}$, given by $v_{1,2} = D\tilde{v}_{1,2}$.
So that the in- and out-of-phase modes read,
\begin{align}\label{v12_A}
v_{1} =\left[N_0-N_1,\sqrt{2} N_1\right]^{tr}\, ,\, \,
v_2   =  \left[-\sqrt{2} ,1\right]^{tr}\, .
\end{align}
The Eq.~\eqref{v12_A} has a transparent meaning.
Note that each of the vectors $v_{1,2}$ has two components equal to expectation values of operators $A_{1,2}$.
The expectation values $N_0-N_1$ and $\sqrt{2} N_1$ are induced by the operator $\mathcal{F}_1 = [\hat{\chi}_s^{(0)}]^{-1} v_1 \propto A_1 + \sqrt{2} A_2$ which is exactly the projection of the dipole operator, Eq.~\eqref{SM} on the first two bands.
Therefore, the in-phase collective mode, $v_1$ is the expected DOs.
 
We now turn to the significance of the second mode, $v_2$.
This vector results from the operator, $\mathcal{F}_2 =[\hat{\chi}_s^{(0)}]^{-1} v_2\propto - \sqrt{2} A_1/(N_0 - N_1) + A_2/N_1$. 
It follows that the expectation value, $\mathcal{F}_{1(2)}$ vanishes for the ground state modified by the operators $\mathcal{F}_{2(1)}$.
Equivalently, $\mathcal{F}_{1(2)}$ does not induces finite $\langle \mathcal{F}_{2(1)} \rangle$.
In terms of the memory function formalism this can be summarised as  the orthogonality statement, $\langle \mathcal{F}_1 | \mathcal{F}_2 \rangle =0$.

\subsubsection{DOs dissipation rate, $1/\tau_{\perp}$: memory function formalism}
\label{sec:LT}
\begin{figure}
\begin{center}
\includegraphics[width=0.8\columnwidth]{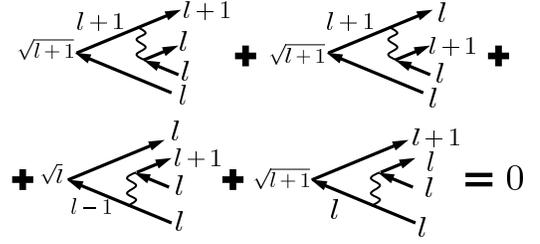}
\caption{The total amplitude for all relaxation channels of the DOs in parabolic confinement  vanishes according to KTh.
The figure illustrates this statement expressed by Eqs.~\eqref{Uc_g} and \eqref{PT46} for the decays into a particle-hole pair in the band $l$, one particle at the band $l+1$ and another hole in the band $l$. 
The fermion propagators are denoted by arrowed lines with band index specified.
The prefactors of each of the contributions are matrix elements of the dipole operator, Eq.~\eqref{SM}.}
\label{fig:KTh}
\end{center}
\end{figure}

To compute the life time of a Kohn mode, $\mathcal{F}_1 = A_1 + \sqrt{2} A_2$ in the memory function formalism we use the basic relation, \eqref{tau_lB} in the same way as in a singly occupied band.
Inspection of \eqref{A1}  and \eqref{A2} shows that the Kohn mode, $\mathcal{F}_1$ has four kinematically allowed relaxation channels labeled by index $c=1,2,3,4$.
As different channels do not interfere their contributions to the relaxation rate add,
\begin{align}\label{LT13}
\frac{1}{ \tau_{\perp}} = \frac{ 1 }{ \omega_{\perp} \langle \mathcal{F}_1 | \mathcal{F}_1 \rangle  } \sum_{c = 1}^4  U_c^2  
\Im K_{M_c,M_c}(\omega_{\perp})\, .
\end{align}
In Eq.~\eqref{LT13}, $\langle \mathcal{F}_1 | \mathcal{F}_1 \rangle = \langle A_1 | A_1 \rangle + 2\langle A_2 | A_2 \rangle = \omega_{\perp}^{-1} [(N_0 - N_1) + 2 N_1] =\omega_{\perp}^{-1} (N_0 + N_1)$ is the normalization of the DOs, 
\begin{align}\label{M_cA}
M_1 & = 
\sum_{{\bm{k}},{\bm{k}}',{\bm{q}}}  V_{\bm{q}}  \psi_{1,{\bm{k}}-{\bm{q}}}^{\dag} \psi_{0,{\bm{k}}'+{\bm{q}}}^{\dag} \psi_{0,{\bm{k}}'}  \psi_{0,{\bm{k}}} \notag \\
M_2 & = 
\sum_{{\bm{k}},{\bm{k}}',{\bm{q}}}  V_{\bm{q}}  \psi_{1,{\bm{k}}-{\bm{q}}}^{\dag} \psi_{1,{\bm{k}}'+{\bm{q}}}^{\dag} \psi_{0,{\bm{k}}'}  \psi_{1,{\bm{k}}} \notag \\
M_3 & = \sum_{{\bm{k}},{\bm{k}}',{\bm{q}}} V_{\bm{q}} \psi_{2,{\bm{k}}-{\bm{q}}}^{\dag}  \psi_{0,{\bm{k}}'+{\bm{q}}}^{\dag}  \psi_{1,{\bm{k}}'}\psi_{0,{\bm{k}}} \notag \\
M_4 & =  \sum_{{\bm{k}},{\bm{k}}',{\bm{q}}} V_{\bm{q}} \psi_{2,{\bm{k}}-{\bm{q}}}^{\dag}  \psi_{1,{\bm{k}}'+{\bm{q}}}^{\dag}  \psi_{1,{\bm{k}}'} \psi_{1,{\bm{k}}}\, ,
\end{align} 
and the matrix elements for the transitions are listed in Eq.~\eqref{ident}.
We conclude that vanishing of the combinations of the matrix elements, Eq.~\eqref{ident} follows from KT.
The identities,  Eq.~\eqref{ident} represent just few of the infinite set of relationships imposed by the KT.
For instance, we have for all $l>0$,
\begin{align}\label{Uc_g}
&U^{l+1,l-1}_{l,l} = \sqrt{\frac{l}{l+1}} ( U^{l,l-1}_{l,l-1} + U^{l-1,l}_{l,l-1}- U_{ll}^{ll} ) \, , \notag \\
&U^{l+1,l-1}_{l,l} \! = \sqrt{\frac{l+1}{l}} (  U^{l,l-1}_{l,l-1}\! +\! U^{l,l-1}_{l,l-1}\! -\!U^{l+1,l-1}_{l+1,l-1}\! -\! U^{l+1,l-1}_{l-1,l+1} ) \, , \notag \\
&U^{l+1,l-1}_{l,l} = \sqrt{\frac{l+1}{l}} ( U^{l,l+1}_{l,l+1} + U^{l+1,l}_{l,l+1} - U_{ll}^{ll} ) \, .
\end{align}
Explicitly, Eq.~\eqref{Uc_g} are the integral relations that are satisfied by the eigenstates of the harmonic oscillator, $\phi_n^{(0)}$ Eq.~\eqref{phi_n}.
For instance, the last of identities, \eqref{Uc_g} means that for any even function $f(x-y)$,
\begin{align}\label{PT46}
 \int &\!d x\!\! \int\! d y 
 \bigg\{ \sqrt{l} \phi_{l-1}^{(0)}(x) \phi_{l}^{(0)}(x) \phi_{l}^{(0)}(y) \phi_{l+1}^{(0)}(y) 
 \notag \\
 -&  \sqrt{l+1}\bigg[ \left[\phi_{l}^{(0)}(x)  \phi_{l+1}^{(0)}(y)\right]^2  - \left[\phi_{l}^{(0)}(x)  \phi_{l}^{(0)}(y) \right]^2\notag \\ 
   +&   \phi_{l}^{(0)}(x) \phi_{l}^{(0)}(y)  \phi_{l+1}^{(0)}(x)  \phi_{l+1}^{(0)}(y) \bigg]
\bigg\} f(x - y) = 0\, .
\end{align}
The relation \eqref{PT46} guarantees that the DOs do not decay into three excitations in the band $l$ and one excitation in the band $l+1$, as illustrated in Fig.~\ref{fig:KTh}.
The $\sqrt{l}$ and $\sqrt{l+1}$ coefficients naturally appear in \eqref{PT46} as the matrix elements of the center of mass operator. 

To each of the decay channels corresponds the identity similar to Eq.~\eqref{PT46}.
As an example consider the decay processes into a final state containing the two particles in the bands $l+1$ and $m$, and two holes in the bands $l$ and $m$, such that $m \neq l, l+1, l-1$.
Vanishing of the amplitude of such a processes gives rise to the following identity,
\begin{align}\label{ident_A}
\sqrt{l+1}&(U_{m,l+1}^{m,l+1} + U_{m,m}^{l+1,l+1} - U_{l,l}^{m,m} - U_{l,m}^{l,m})
\notag \\
& +\sqrt{m+1}( U_{l+1,l}^{m+1,m} + U_{m,l+1}^{m+1,l} )
\notag \\
& - \sqrt{m}( U_{l,m-1}^{l+1,m} + U_{m-1,m}^{l,l+1} )=0\, ,
\end{align}
which could be illustrated by the figure similar to Fig.~\ref{fig:KTh}.
In its explicit form, Eq.~\eqref{ident_A} reads,
\begin{align}\label{PT48}
\int \!\! d x\!\! \int \!\!d y &\bigg\{ \sqrt{l+1} \Big[ \big[\phi^{(0)}_{m}(x)\phi^{(0)}_{l+1}(y)\big]^2  
\notag \\
+ & \phi^{(0)}_m(x) \phi^{(0)}_{l+1}(x) \phi^{(0)}_{m}(y) 
 \phi^{(0)}_{l+1}(y)  
 \notag \\
 - \phi^{(0)}_l(x) &\phi^{(0)}_m(x) \phi^{(0)}_l(y) \phi^{(0)}_m(y) - \big[\phi^{(0)}_{l}(x) \phi^{(0)}_m(y)\big]^2\Big]
\notag \\
 +  \sqrt{m+1} &\Big[ \phi^{(0)}_{m+1}(x) \phi^{(0)}_{l+1}(x) \phi^{(0)}_{m}(y)\phi^{(0)}_{l}(y) 
\notag \\
& +  \phi^{(0)}_{m+1}(x) \phi^{(0)}_m(x)  \phi^{(0)}_l(y) \phi^{(0)}_{l+1}(y)  \Big]
\notag \\
- \sqrt{m} &\Big[  \phi^{(0)}_{m-1}(x) \phi^{(0)}_{m}(x) \phi^{(0)}_{l}(y)\phi^{(0)}_{l+1}(y) 
\notag \\
+  \phi^{(0)}_{m-1}&(x) \phi^{(0)}_{l}(x) \phi^{(0)}_{m}(y)\phi^{(0)}_{l+1}(y) \Big]\bigg\} U(x-y)
=0
\end{align}
for any even function $U(x-y)$.

We are not aware of an analytical proof of Eqs.~\eqref{PT46} and \eqref{PT48}.
Instead, we have verified them numerically. 
The identities, Eqs.~\eqref{Uc_g}, \eqref{ident_A} do not exhaust the list of relations imposed by KTh.
The remaining identities have the same underlying origin, and we do not bring them here for brevity. 

In the presence of finite anharmonicity, the above identities do not hold, because the eigenfunctions of the transversal motion are no longer given by Eq.~\eqref{phi_n}.
As a result the rate $1/\tau_{\perp} \neq 0$ in both quasi-1D and quasi-2D for the two occupied bands.
The actual value of the relaxation rate in this case is up to a numerical coefficient coincides with Eq.~\eqref{rate_eval}.

\section{Discussion}
\label{sec:Discussion}

In summary, we have studied the relaxation of the DOs in the quasi-1D and quasi-2D systems.
In both geometries the ultimate relaxation was produced by emission of soft particle-hole excitations along the direction orthogonal to the direction of confinement. 
We looked at this problem from three different perspectives.
Assuming the interactions are weak we have started with applying perturbation theory to compute the leading correction to the dipole-dipole correlation function.
The latter possess an infinitely sharp peak at frequency $\omega_{\perp}$ without interactions and broadens  because of the inelastic processes.
In the quasi-2D case the relaxation rate at $T=0$ is finite and scales as $1/\tau_{\perp} \propto (\delta \omega_{\perp} / \omega_{\perp})^2 (1 / \tau_{in} )$. Here $\delta \omega_{\perp}$ is the typical variation of the classical oscillation frequency due to anharmonicity, and $1/\tau_{in}$ is the usual Fermi-liquid-like inelastic relaxation rate controlled by the phase space available for scattering.
The proportionality, $1/\tau_{\perp} \propto (\delta \omega_{\perp} / \omega_{\perp})^2$ reflects the KTh satisfied by virtue of an infinite set of identities obeyed by the matrix elements of interaction computed using the wave-functions of harmonic oscillator, see {\it e.g.} Fig.~\ref{fig:KTh}.
At $\delta \omega_{\perp} \neq 0$ the transversal matrix elements are non-zero, and produce the above scaling.

The direct perturbation theory does not apply when at finding the shift of the pole of the dipole correlation function away from $\omega = \omega_{\perp}$ neither along the real, (frequency shift), nor imaginary (life time) axis in the complex $\omega$ plane.
Instead the resummation of infinite set of most singular contributions is required.
In the present case such resummation is achieved by solving the integral Bethe-Salpeter equation satisfied by the properly defined vertex function.
The solution of the Bethe-Salpeter equation revealed that the result obtained in perturbation theory is valid only if the two conditions are met.
First, as usual the interaction in units of inverse density of states has to be small.
Second requirement, $1/\tau_{\perp} \ll 1/ \tau_{in}$ amounts to the weak anharmonicity condition, $\delta \omega_{\perp} \ll \omega_{\perp}$.

The latter condition also justifies the memory function approach to the problem.
The DOs relaxation can be paralleled to the relaxation of velocity of the Brownian particle \cite{Forster1975}.
In this case, the decay time is related to the force correlation function for the Brownian particle that is infinitely heavy.
The whole procedure holds when the Brownian particle is much heavier than the particles in a surrounding liquid.  
And as a result the velocity relaxation time is much larger than the collisional time.
In our problem the role of the mass ratio is played by the anharmonicity parameter.

The kinematical constrains are substantially more restrictive in 1D than in 2D \cite{Imambekov2012}.
Only if the Fermi-energy, $E_F$ of the lowest transversal band lies below the critical value, $E_F < 3 \omega_{\perp}/4$ the same distinction holds between quasi-1D and quasi-2D systems.
Above it, $E_F > 3 \omega_{\perp}/4$ the relaxation in quasi-1D is qualitatively similar to the relaxation in quasi-2D, unless the system is close to integrability.
This happens because at $E_F > 3 \omega_{\perp}/4$ the inter-band relaxation processes become kinematically allowed.
Moreover, for $E_F < 3 \omega_{\perp}/4$ the DOs relaxation is kinematically forbidden to all orders in interaction at $T=0$ even if the trapping potential is anharmonic.
At finite temperature, $T \ll E_F$, $1/\tau_{\perp} \propto T^3$.

Let us consider the quasi-1D system in the regime, $E_F > 3 \omega_{\perp}/4$ and assume that the underlying system is close to integrability.
To be specific, consider bosons interacting via the local interaction $\propto c \delta(\bm{r} - \bm{r}')$.
If only a single band is included, the 3D local interaction translates into the 1D Lieb-Liniger model \cite{Lieb1963,Lieb1963a} with the interaction $V_B(x-x') \propto c \delta(x - x')$ that is likewise local.
To compute the boson correlation functions the Lieb-Liniger model can be mapped to the model of fermions interacting via the singular interactions, $V_F(x-x') \propto c^{-1} \delta''(x - x')$, \cite{Cheon1999}.
With such interactions the inelastic life time indeed vanishes, \cite{Lunde2007} as expected in the integrable case. We speculate that for large number of modes occupied the variation of the matrix elements with the band index may be neglected, and as a result starting with the Lieb-Liniger model we would recover $1/\tau_{\perp} \approx 0$.
 
\begin{figure}
\begin{center}
\includegraphics[width=0.8\columnwidth]{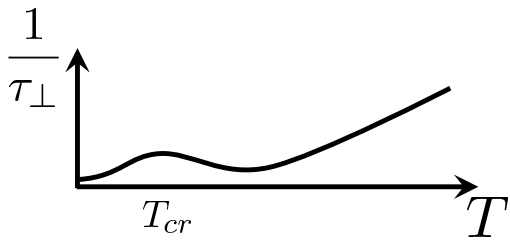}
\caption{Schematic temperature dependence of the relaxation rate, $1/\tau_{\perp}$ of the DOs mode in quasi-2D geometry.
Initially, at low $T$ the rate $1/\tau_{\perp}$ grows from finite value as the phase space for scattering opens up.
Then it reaches a maximum at the crossover to the hydrodynamic regime, $T \approx T_{cr}$.
The subsequent upturn is due to effectively more anharmonic potential felt by the expanding atomic cloud.}
\label{fig:crossover}
\end{center}
\end{figure}

Finally, in this work we have studied the regime of rare collisions, $1/\tau_{in} \ll \omega_{\perp}$ which is complementary to the hydrodynamic regime, $1/\tau_{in} \gg\omega_{\perp}$.
Since the inelastic rate $1/\tau_{in}$ grows upon heating, the crossover between the two regimes occurs at temperature $T_{cr}$ such that $1/\tau_{in}(T_{cr}) \approx \omega_{\perp}$.
As a result, the rate $1/\tau_{\perp}$ is expected to exhibit a non-monotonic temperature dependence reaching maximum at the crossover $T \approx T_{cr}$, see Fig.~\ref{fig:crossover}.
Such behaviour is typical to other collective excitations such as quadrupole mode in the quasi-2D dipolar Fermi gas, \cite{Babadi2012}. 
In addition, specifically for the dipole oscillations, the effect of anharmonicity becomes more pronounced as the atomic cloud expands with heating.
This causes the second crossover in the form of upturn in the scattering rate at higher temperatures, see Fig.~\ref{fig:crossover}.
In fact, the described temperature dependence agrees qualitatively with that reported in Ref.~\cite{Pantel2012} based on the data of Ref.~\cite{Riedl2008}.  

\begin{acknowledgements}
We are thankful to M. Raikh for discussions that stimulated us in working on this paper.
We are also grateful to D. Orgad and E. Bettelheim for discussions of separate parts of the paper. 
This work has been supported by the Binational Science Foundation (BSF Grant No. 2016317).
M.K. acknowledges the support by the Israel Science Foundation, Grant No. 1287/15. A.L. acknowledges the support from NSF CAREER Grant No. DMR-1653661, and the Wisconsin Alumni Research Foundation.

\end{acknowledgements}

\begin{appendix}
\begin{widetext}
\section{Calculation of $1/\tau_{\perp}$ for a one occupied band in quasi-2D for the model specified in Sec.~\ref{sec:PT}.}
\label{app:rates}
Our goal is to compute the relaxation rate $1/\tau_{\perp}$ as given by Eq.~\eqref{rate} for the specific choice of anharmonicity and interaction.
Namely, we consider the quartic anharmonicity with $u(x) = x^4$ in Eq.~\eqref{def12} and $f(x) = \delta(x)$ in Eq.~\eqref{def14}. 
The calculation naturally splits into two parts.
First we compute the polarization operators, Eq.~\eqref{Pi} and their convolution appearing in the second line of Eq.~\eqref{rate} 
Next we compute the prefactor in the same equation for a given anharmonicity and the interaction strength.

\subsection{Polarization operators entering the relaxation rate, Eq.~\eqref{rate}}
We start with the calculation of $\Pi_{10}(\bm{q},\omega)$.
The Fermi energy, $E_F$ lies in between the partially occupied $n=0$ and empty $n=1$ bands.
Therefore, $f_{1,\bm{k}} = 0$, and $f_{0,\bm{k}} = \theta( k_F - k)$, where Fermi momentum is $k_F = \sqrt{ 2 m E_F}$.
In result, the definition, Eq.~\eqref{Pi} yields,
\begin{align}\label{Pi10}
\Im\Pi_{10}(q,\omega) = - \frac{1 }{ 4 \pi} \int_0^{k_F} d k k \int_0^{2 \pi} d \phi \delta( \omega- \omega_{\perp} -k q \cos \phi / m -q^2/2 m) \, ,
\end{align}
where $\theta$ is the angle between the vectors $\bm{k}$ and $\bm{q}$.
Note that in Eq.~\eqref{Pi10} there is no summation over spin variables as we are considering spinless fermions throughout the paper.
It is convenient to measure the momentum and energy in units of $k_F$ and $E_F$ respectively.
We hence introduce the dimensionless variables, $\bar{\bm{k}}= \bm{k} /k_F$, $\bar{\omega} = \omega/E_F$, $\bar{\bm{q}} = q/k_F$ and $\bar{\omega}_{\perp} = \omega_{\perp}/E_F$.
In terms of new variables, Eq.~\eqref{Pi10} takes the form,
\begin{align}\label{Pi11}
\Im\Pi_{10}(q,\omega) = - \frac{m }{ 2 \pi} \int_0^{1} d \bar{k} \bar{k} \int_0^{2 \pi} d \phi 
\delta( \bar{\omega}- \bar{\omega}_{\perp} - 2 \bar{k} \bar{q} \cos \phi  - \bar{q}^2 ) \, ,
\end{align}
To evaluate Eq.~\eqref{Pi11} note first that it is non-zero only for $\bar{\omega}$ satisfying,
$\bar{\omega}_{-} < \bar{\omega} < \bar{\omega}_{+}$, where $\bar{\omega}_{\pm} = \bar{\omega} + \bar{q}^2 \pm 2 \bar{q}$. 
For these values of $\bar{\omega}$ the finite angular integration results only for $1>\bar{k}> \bar{k}_d$, where 
\begin{align}\label{kd}
\bar{k}_d =  | \bar{\omega}- \bar{\omega}_{\perp}   - \bar{q}^2  | /( 2\bar{q} ).
\end{align}
Evaluating the integral over $\theta$, we obtain,
\begin{align}\label{Pi12}
\Im\Pi_{10}(q,\omega) = - \frac{m }{ 2 \pi \bar{q}} \int_{\bar{k}_d}^{1} d \bar{k} \bar{k} 
\frac{1}{  \sqrt{  \bar{k}^2 -\bar{k}_d^2 }} 
\end{align}
The remaining integration is straightforward and yields,
\begin{align}\label{Pi13}
\Im\Pi_{10}(q,\omega) = - \frac{m }{ 2 \pi \bar{q}} \sqrt{1 - \bar{k}_d^2}
\end{align}
Returning to the original variables, we obtain,
\begin{align}\label{Pi14}
\Im\Pi_{10}(q,\omega) = - \frac{ m^2  }{ 2\pi q^2} \sqrt{( \omega_{\perp}+ v_F q + q^2/2 m -  \omega     )(\omega- \omega_{\perp}  + v_F q - q^2/2m  )  }
\end{align}
for 
For $\omega$ satisfying $\omega_{\perp} + q^2/ 2 m - v_F q < \omega < \omega_{\perp} + q^2/ 2 m + v_F q$. 
For $\omega$ outside of this interval, $\Im\Pi_{10}(q,\omega) =0$.

We now turn to the evaluation of $\Im \Pi_{00}(q,\omega)$, which according to Eq.~\eqref{Pi} is given in terms of the dimensionless variables introduced above, by
\begin{align}\label{Pi01}
\Im\Pi_{00}(q,\omega)  =  - \frac{m}{2 \pi} \int_0^{1} \bar{k} d\bar{k} \int_0^{2 \pi} d \phi
\theta(| \bar{\bm{k}} + \bar{\bm{q}}| - 1) \delta(\bar{\omega} - 2 \bar{k} \bar{q} \cos \phi - \bar{q}^2 )\, ,
\end{align}
where $| \bar{\bm{k}} + \bar{\bm{q}}| = \sqrt{\bar{k}^2 + 2 \bar{k} \bar{q} \cos \phi + \bar{q}^2}$.
It is convenient to change from the variables, $\bar{k}$ and $\phi$ to the variables,
$\bar{k}$ and $\bar{p} = | \bar{\bm{k}} + \bar{\bm{q}}|$.
In terms of new variables Eq.~\eqref{Pi01} takes the form,
\begin{align}\label{Pi01a}
\Im\Pi_{00}(q,\omega)  =  - \frac{m}{ \pi} \int_0^{1} \bar{k} d\bar{k} \int_{|\bar{k} - \bar{q}|}^{|\bar{k} + \bar{q}|} d \bar{p}
\theta( \bar{p} - 1) \frac{2 p\delta(\bar{\omega}  + \bar{k}^2 - \bar{p}^2)}{\sqrt{[\bar{p}^2 -(\bar{k}-\bar{q})^2][(\bar{k}+\bar{q})^2-\bar{p}^2]}} \, .
\end{align}
Let's denote $\bar{p}_0 = \sqrt{\bar{\omega}  + \bar{k}^2}$.
The limitation $\bar{p} > 1$ may become relevant only for $\bar{\omega}<1$.
We therefore consider the two cases separately.

First assume $\bar{\omega} > 1$.
In this case Eq.~\eqref{Pi01a} is 
\begin{align}\label{Pi01b}
\Im\Pi_{00}(q,\omega)  =  - \frac{m}{ \pi} \int_0^{1} \bar{k} d\bar{k} \int_{|\bar{k} - \bar{q}|}^{|\bar{k} + \bar{q}|} d \bar{p}
\frac{2 p\delta(\bar{\omega}  + \bar{k}^2 - \bar{p}^2)}{\sqrt{[\bar{p}^2 -(\bar{k}-\bar{q})^2][(\bar{k}+\bar{q})^2-\bar{p}^2]}}\, .
\end{align}
The limitations imposed on the $\bar{k}$ integration, are $|\bar{k}-\bar{q}|< \bar{p}_0 < |\bar{k}+\bar{q}|$ amounts to
$k  >k_1 =  |\bar{\omega} -  \bar{q}^2|/2 \bar{q}$.
It follows that the non-zero range of integration exists provided, $k_1<1$, or $|\omega -  \bar{q}^2|<2 \bar{q}$, or equivalently,
$\bar{q}^2 - 2 \bar{q} <  \bar{\omega} < \bar{q}^2 + 2 \bar{q}$.
For values of $\bar{\omega}$ in this range, Eq.~\eqref{Pi01b} gives,
\begin{align}\label{Pi01c}
\Im\Pi_{00}(q,\omega)  =  - \frac{m}{ \pi} \int_{k_1}^{1} d\bar{k} 
\frac{\bar{k} }{\sqrt{[\bar{\omega}  + 2 \bar{k} \bar{q} - \bar{q}^2][ 2 \bar{k} \bar{q} + \bar{q}^2-\bar{\omega}  ]}} \, .
\end{align}
Introducing the new variable, $\xi = \bar{k}^2$ and recalling the definition of $k_1$ we obtain,
\begin{align}\label{Pi01d}
\Im\Pi_{00}(q,\omega)  =  - \frac{m}{ 2\pi (2 \bar{q})} \int_{k_1^2}^{1} d\xi 
\frac{1 }{\sqrt{  \xi - k_1^2}  } = - \frac{m }{ 2 \pi \bar{q}} \sqrt{1 - k_1^2}\, .
\end{align}

If $\bar{\omega} < 1$ the additional constrain $\bar{p}_0>1$ in Eq.~\eqref{Pi01a} translates into the limitation on the rage of integration as
$k>k_2=\sqrt{ 1 - \bar{\omega}}$.
Since $k_2<1$ this constrain does not further limit the interval of $\bar{\omega}$.
It limits the range of $\bar{k}$ integration in addition to $k_1<1$ only if $k_2>k_1$.
The latter condition amounts to $\bar{\omega} < - \bar{q}^2 + 2 \bar{q}$.
Because $ -\bar{q}^2 + 2 \bar{q}< 1$, the condition $\bar{\omega} < 1$ is satisfied automatically.
The condition $-\bar{q}^2 + 2 \bar{q}< 1$ also implies that $|\bar{\omega} -  \bar{q}^2|<2 \bar{q}$.
It follows that for $\bar{\omega} < - \bar{q}^2 + 2 \bar{q}$, instead of Eq.~\eqref{Pi01d} we obtain,
\begin{align}\label{Pi01e}
\Im\Pi_{00}(q,\omega)  =  - \frac{m}{ 2\pi (2 \bar{q})} \int_{k_2^2}^{1} d\xi 
\frac{1 }{\sqrt{  \xi - k_1^2}  } = - \frac{m }{ 2 \pi \bar{q}}\left[ \sqrt{1 - k_1^2} -  \sqrt{k_2^2 - k_1^2}\right] \, .
\end{align}
In summary, 
\begin{align}\label{Pi01f}
\Im\Pi_{00}(q,\omega)  = - \frac{m }{ 2 \pi \bar{q}} \times
\begin{cases}
\sqrt{1 - k_1^2} &  |\bar{q}^2 - 2 \bar{q}|< \bar{\omega} < \bar{q}^2 + 2 \bar{q}  \\
\sqrt{1 - k_1^2} -  \sqrt{k_2^2 - k_1^2} & \bar{\omega} < -\bar{q}^2 + 2 \bar{q} 
\end{cases}
\end{align}
Returning to the original variables, we obtain
\begin{align}
\Im\Pi_{00}(q,\omega)  = - \frac{m^2 }{ 2 \pi q^2} \times
\begin{cases}
\sqrt{(  v_F q + q^2/2 m -  \omega     )(\omega + v_F q - q^2/2m  )  } \, ,&  |q^2/2m - v_F q|< \omega < q^2/ 2 m + v_F q  \\
\sqrt{(  v_F q + q^2/2 m -  \omega     )(\omega + v_F q - q^2/2m  )  } 
\\
-  \sqrt{(  v_F q - q^2/2 m -  \omega     )(\omega + v_F q + q^2/2m  ) } \, , & \omega < -q^2/2m +  v_F q \, .
\end{cases}
\label{Pi01g}
\end{align}

Now we compute the frequency and momentum integrals appearing in Eq.~\eqref{rate} using the explicit expressions, Eqs.~\eqref{Pi14} and \eqref{Pi01g}.
Let's first clarify the implication of the kinematical restrictions on the region of integration.
Note first that,  $\Im \Pi_{10}( \bm{q},\omega_{\perp} - \Omega) \neq 0$ only for $ - q^2/ 2 m - v_F q <  \Omega < -q^2/ 2 m + v_F q$.
As $\Omega>0$ this effectively defines the integration region as $0<  \Omega < -q^2/ 2 m + v_F q$.

\begin{align}\label{conv13}
\int_0^{\omega_{\perp}} &\frac{d \Omega}{\pi}\sum_{\bm{q}}  \Im \Pi_{10}( \bm{q},\omega_{\perp} - \Omega)  \Im \Pi_{00}( \bm{q},\Omega) =
\int_0^{2 k_F} \frac{ q d q}{2 \pi} (\frac{ m^2  }{ 2\pi q^2})^2
\int_0^{v_F q-q^2/ 2 m } \frac{d \Omega}{\pi} 
\notag \\
& \times \sqrt{( v_F q + q^2/2 m + \Omega       )(- \Omega  + v_F q - q^2/2m  )  }
\notag \\
& \times
\left[
\sqrt{(  v_F q + q^2/2 m -  \Omega     )(\Omega + v_F q - q^2/2m  )  } 
-  \sqrt{(  v_F q - q^2/2 m -  \Omega     )(\Omega + v_F q + q^2/2m  ) } 
\right]
 \, ,  
\end{align}
We change in Eq.~\eqref{conv13} to new variables, $y=\Omega/E_F$, $x = q/k_F$ and write
\begin{align}\label{conv15}
\int_0^{\omega_{\perp}} &\frac{d \Omega}{\pi}\sum_{\bm{q}}  \Im \Pi_{10}( \bm{q},\omega_{\perp} - \Omega)  \Im \Pi_{00}( \bm{q},\Omega) =
\frac{2 E_F^3 m^4}{(2 \pi)^4 k_F^2 } \int_0^{2 }   \frac{ d x}{x^3}
\int_0^{2x-x^2 } d y 
\sqrt{( 2 x + x^2 + y       )(- y  + 2 x - x^2  )  }
\notag \\
& \times
\left[
\sqrt{(  2x + x^2 -  y     )(y + 2x - x^2  )  } 
-  \sqrt{( 2x - x^2 -  y     )(y + 2x + x^2  ) } 
\right]
 \, .
\end{align}
The remaining integration in Eq.~\eqref{conv15} is trivially computed numerically giving, 
\begin{align}\label{conv17}
\int_0^{\omega_{\perp}} &\frac{d \Omega}{\pi}\sum_{\bm{q}}  \Im \Pi_{10}( \bm{q},\omega_{\perp} - \Omega)  \Im \Pi_{00}( \bm{q},\Omega) =
\frac{ k_F^4 m}{4 (2 \pi)^4}C\, ,
\end{align}
where $C \approx 0.94$.

\subsection{Calculation of the combination of the matrix elements, $U_{01}^{01} + U_{01}^{10} - U_{00}^{00}$ in the relaxation rate, Eq.~\eqref{rate}}

To compute the matrix elements of interaction in the presence of anharmonicity we consider the anharmonicity, $H_{\epsilon,\perp}$, Eq.~\eqref{def12} as a perturbation to the harmonic potential, $H_{0,\perp}$, Eq.~\eqref{H-harm}.
The unperturbed wave-functions are given by Eq.~\eqref{phi_n}.
The standard expression for the corrections to the first order in $\epsilon$ corrections wave-functions, $\phi_n^{(1)}\propto \epsilon$ reads,
\begin{align}\label{pert13}
\phi_n^{(1)}(z) = \sum_{m=0}^{\infty}{\vphantom{\sum}}'  \frac{ \langle \phi_m^{(0)} | H_{\epsilon,\perp}| \phi_n^{(0)}\rangle }{ (n - m) \omega_{\perp}} \phi_n^{(0)}(z)\, ,
\end{align} 
where the prime specifies that the summation excludes $m =n$, and we used the expressions for the energies of the harmonic oscillator, $H_{0,\perp}\phi_n^{(0)} = (n + 1/2) \phi_n^{(0)}$.
Taking for definiteness the quartic anharmonicity, $H_{\epsilon,\perp} \propto z^4$ as specified at the beginning of the section we obtain by evaluating the matrix elements in Eq.~\eqref{pert13},
\begin{align}\label{Melem}
\phi_0^{(1)} &= -\epsilon \frac{3}{2 \sqrt{2}} \phi_{2}^{(0)} -\epsilon \frac{\sqrt{3}}{4 \sqrt{2}} \phi_{4}^{(0)} \, ,
\notag \\
\phi_1^{(1)} &= -\epsilon \frac{5 \sqrt{3}}{2 \sqrt{2}} \phi_{3}^{(0)} -\epsilon \frac{\sqrt{15}}{4 \sqrt{2}} \phi_{5}^{(0)} \, .
\end{align}
Substituting Eq.~\eqref{Melem} into the definition of the interaction matrix elements, Eq.~\eqref{tr_mat} with the point-like interaction as stated at the beginning of the section gives to the first order in the anharmonicity parameter, $\epsilon$
\begin{align}\label{Melem1}
U_{01}^{01} + U_{01}^{10} - U_{00}^{00} \approx \frac{ 3 \epsilon}{2 \sqrt{ 2 \pi}}\, .
\end{align}

\subsection{Relaxation rate $1/ \tau_{\perp}$}

The occupation of a single occupied band is $N_0 = k_F^2/ (2 \pi^2)$ and clearly, $N_1 =0$.
Therefore the substitution of Eqs.~\eqref{conv17} and \eqref{Melem1} in Eq.~\eqref{rate} yields Eq.~\eqref{rate_eval}.

\section{Boundaries of continua on frequncy-wave-number, $(q,\omega)$ plane}
\label{app:edges}

In this section we find the lower edge of the the two inter-band particle-hole continuum, $\omega_{\{2\}}^{0\rightarrow1}(q)$ and four particle-hole continuum, $\omega_{\{4\}}^{0\rightarrow1}$and show that the critical filling $E^*_F = 3 \omega_{\perp}/4$.

The boundary is determined by minimization of the excitation energy, $E_{2,k_1+q_1} -  E_{1,k_1} + E_{2,k_2+q_2} -  E_{1,q_2}$ under the condition, $q_1+q_2 = q$, and $|k_{1,2}| < k_F$.
Let us first show that the minimum is achieved for $|k_{1,2}|=k_F$.
Introducing the Lagrange multiplier $\lambda$ we minimize the function  $E_{2,k_1+q_1} -  E_{1,k_1} + E_{2,k_2+q_2} -  E_{1,q_2} - \lambda (q_1 + q_2 - q)$ with the constraint on $q_{1,2}$ lifted.
Differentiating this function with respect to $k_{1,2}$ we obtain the condition $q_1 = q_2 = 0$ which can be realized only for a special case of $q=0$.
Even for $q=0$ the energy of $q_1=q_2=0$, $2\omega_{\perp}$ is lager then the energy $2 ( \omega_{\perp} - E_F)$ resulting for $q_1 = -q_2 = k_F$.
We conclude that the minimum must be found at the boundaries of the domain, i. e. at $|k_{1,2}| = k_F$. 
Which is in fact obvious as at finite $q_{1,2}$ the energy difference $E_{2,k_{1,2}+q_{1,2}} -  E_{1,k_{1,2}}$ has finite derivative with respect to $k_{1,2}$.
There are only three possibilities, $k_1 = - k_2 = k_F$, and $k_1 = k_2 = \pm k_F$.
Consider the  first option and find the $q_1$ that minimizes  $E_{2,k_F+q_1} -  E_{1,k_F} + E_{2,-k_F+(q - q_1)} -  E_{1,k_F} = 2 (\omega_{\perp} - E_F) + (k_F + q_1)^2 / 2 m - (-k_F +(q -  q_1))^2 / 2 m $.
As the function $f(x)=x^2$ is convex, we have an inequality,
$(k_F + q_1)^2 / 2 m - (-k_F +(q -  q_1))^2 / 2 m \leq 2 (q/2)^2/2m = q^2/ 4 m$.
So that the resulting condition reads $\omega^{0\rightarrow 1}_{\{2\}} \leq 2 (\omega_{\perp} - E_F) + q^2/ 4m$.
Now the possibility of  $k_1 = k_2 = \pm k_F$ results in the similar way in the condition 
$\omega^{0\rightarrow 1}_{\{2\}} \leq 2 (\omega_{\perp} - E_F) +  (q \pm 2 k_F)^2/4m$.
As a result we obtain
\begin{align}\label{boundaries2}
\omega^{0\rightarrow 1}_{\{2\}} = 2(\omega_{\perp} - E_F) + \min_{u=0,\pm 1} \left[(q + 2 u k_F)^2/4m \right]\, ,
\end{align}
which agrees with Eq.~\eqref{boundary} in the particular case $l=1$.
The generalization of Eq.~\eqref{boundaries2} to arbitrary even number of inter-band particle hole pairs proceeds using the Jensen's inequality applied to the convex function, $f(x) = x^2$ and is straightforward.

\section{Derivation of expressions for the self energies and the inelastic relaxation rates used in the text}
Consider the auxiliary sum over Matsubara frequencies of the product of the fermion Green function and the boson propagator 
\begin{align}\label{sig_13}
S_{\Sigma_{ph}} (k,\epsilon_n) = T \sum_{\epsilon_{n'}}\sum_{k'} \mathcal{G}_{m,k'}(\epsilon_{n'})\chi_{kk'}(\epsilon_n - \epsilon_{n'}) 
\end{align}
which describes the dressing of fermions by the fluctuations in the particle-hole channel with the  correlation function $\chi_{kk'}(\Omega_n)$.
Using the method of analytic continuation one arrives at the expression,
\begin{align}\label{sig_15}
\Im S^R_{\Sigma_{ph}}(k,\epsilon) = \int \frac{d \epsilon'}{ 2 \pi } \sum_{k'} \left[ \tanh\frac{\epsilon'}{2 T} - \coth\frac{\epsilon' - \epsilon}{ 2 T } \right] \Im G^R_{m,k'} ( \epsilon') \Im \chi^R_{kk'}(\epsilon - \epsilon').
\end{align}
It is also useful to write the expression, Eq.~\eqref{sig_15} in a slightly different form as,
\begin{align}\label{sig_16}
\Im S^R_{\Sigma_{ph}}(k,\epsilon+\omega) = \int \frac{d \epsilon'}{ 2 \pi } \sum_{k'}\left[ \tanh\frac{\epsilon'+\omega}{2 T} - \coth\frac{\epsilon'  - \epsilon}{ 2 T } \right] \Im G^R_{m,k'} ( \epsilon' + \omega ) \Im \chi^R_{kk'}(\epsilon- \epsilon').
\end{align}

In addition we will need to consider the effect of the quasi-particle dressing by interactions in the Cooper channel.
To this end we study the auxiliary function,
\begin{align}\label{sigC_13}
S_{\Sigma_C}(k,\epsilon_n + \omega_n) = T \sum_{\epsilon_{n'}}\sum_{k'} \chi_{kk'}(\epsilon_n + \epsilon_{n'} + \omega_n) \mathcal{G}_{mk'}(\epsilon_{n'})
\end{align}
analogous to the function $S_{\Sigma_{ph}}$ we have introduced before in Eq.~\eqref{sig_13} to analyze the effect of interactions in the particle-hole channel.
The summation in Eq.~\eqref{sigC_13} followed by the analytic continuation gives
\begin{align}\label{sigC_15}
\Im S^R_{\Sigma_C}(k,\epsilon + \omega) = \int \frac{ d \epsilon'}{ 2 \pi} \sum_{k'} \left[ \tanh\frac{\epsilon'}{2 T} - \coth\frac{\epsilon + \epsilon'+ \omega}{ 2 T} \right] \Im \chi^R_{kk'}(\epsilon+\epsilon' + \omega) \Im G_{mk'}^R(\epsilon')\, .
\end{align}

\subsection{Derivation of Eqs.~\eqref{tau_bar} for the quasi-particle inelastic relaxation rates}
\label{app:qp_rates_A}
The scattering rates $1/\bar{\tau}_{0k}$, $1/\bar{\tau}_{1a(b)k}$ are obtained by computing the on-mass-shell self energies,
\begin{align}\label{tau_sigma}
\frac{1}{\bar{\tau}_{1a(b)k}} = - 2 U_{1(\bar{1})}^{-2} \Im  \Sigma^R_{1a(b)k}(\xi_{1k})\, ,\,\, 
\frac{1}{\bar{\tau}_{0k}} = - 2 U_0^{-2} \Im  \Sigma^R_{0k}(\xi_{0k}). 
\end{align}

Let's consider first Eq.~\eqref{tau_bar_a}. 
The self energy appearing in Fig.~\ref{fig:Diagrams1}a takes the form similar to Eq.~\eqref{sig_13},
\begin{align}
\Sigma_{1ak}(i \epsilon_n) = - U_1^2 T \sum_{\epsilon_{n'}}\sum_{k'} V_q^2 \mathcal{G}_{1,k-q}(\epsilon_{n'})\Pi_{00}(q,\epsilon_n - \epsilon_{n'}) ,
\end{align}
where the intra-band polarization operator,
\begin{align}
\Pi_{00}(q,\Omega_n) = T \sum_{\epsilon_{n'},k'} \mathcal{G}_{0k'+q}(\epsilon_{n'} + \Omega_{n'}) \mathcal{G}_{0k'}(\epsilon_{n'} ) =
\sum_{k'} \frac{f_{0k'} - f_{0k'+q} }{ i \Omega_n + \xi_{0k'} - \xi_{0k'+q} }.
\end{align}
As a result Eq.~\eqref{sig_16} gives 
\begin{align}\label{34}
U_1^{-2} \Im \Sigma^R_{1ak}(\xi_{1k})  = -\frac{ \pi }{2}  \sum_{k',q }V_q^2\left[ \tanh\frac{\xi_{1,k-q}}{2 T} - \coth\frac{\xi_{1,k-q}  - \xi_{1,k}}{ 2 T } \right]  
(f_{0,k'} - f_{0,k'+q})\delta(\xi_{1k} - \xi_{1,k-q} + \xi_{0k'} -\xi_{0k'+q} ).
\end{align}
Finally using the identity 
\begin{align}
\frac{1}{2}\left[\tanh\frac{b}{2} - \coth\frac{b-a}{2}\right][f(c) - f(d)] = \frac{ 1 - f(b)}{1 - f(a)} [1 - f(d)]f(c)  
\end{align}
with $f(x) = (e^x + 1)^{-1}$ which holds provided $a-b+c-d=0$, and Eq.~\eqref{tau_sigma}, we obtain,
\begin{align}
\frac{1}{\bar{\tau}_{1ak}}  = 2 \pi   \sum_{k',q }V_q^2
\frac{ 1 - f_{1,k-q} }{ 1 - f_{1, k} } ( 1 - f_{0,k'+q} ) f_{0,k'}  
\delta(\xi_{1k} - \xi_{1,k-q} + \xi_{0k'} -\xi_{0k'+q} ),
\end{align}
which is Eq.~\eqref{tau_bar_a}.
The derivation of Eq.~\eqref{tau_bar_b} reduces to the one above upon replacement of $\xi_{0,k'+q}$ and $\xi_{1,k-q}$ with $\xi_{1,k'+q}$ and $\xi_{0,k-q}$ respectively.

For the derivation of Eq.~\eqref{tau_bar_c} we write the self-energy appearing on Fig.~\ref{fig:Diagrams1}c in the form,
\begin{align}
\Sigma_{0k}(i \epsilon) = - U_0^2 T \sum_{\epsilon_{n'}}\sum_{k'} V_q^2 \mathcal{G}_{0,k+q}(\epsilon_{n'})\Pi_{00}(-q,\epsilon_n - \epsilon_{n'}) ,
\end{align}
where the polarization operator,
\begin{align}
\Pi_{00}(-q,\Omega_n) = T \sum_{\epsilon_{n'},k'} \mathcal{G}_{0k'}(\epsilon_{n'} + \Omega_{n'}) \mathcal{G}_{0k'+q}(\epsilon_{n'} ) =
\sum_{k'} \frac{f_{0k'+q} - f_{0k'} }{ i \Omega_n + \xi_{0k'+q} - \xi_{0k'-q} }.
\end{align} 
Repeating the steps leading to Eq.~\eqref{34} we obtain
\begin{align}
U_0^{-2} \Im \Sigma^R_{0k}(\xi_{0k})  = -\frac{ \pi }{2}  \sum_{k',q }V_q^2\left[ \tanh\frac{\xi_{0,k+q}}{2 T} - \coth\frac{\xi_{0,k+q}  - \xi_{0,k}}{ 2 T } \right]  
(f_{0,k'+q} - f_{0,k'})\delta(\xi_{0k} - \xi_{0,k+q} + \xi_{0k'+q} -\xi_{0k'} ).
\end{align}
And using the identity
\begin{align}
\frac{1}{2}\left[\tanh\frac{b}{2} - \coth\frac{b-a}{2}\right][f(c) - f(d)] = \frac{ f(b) f(d) }{ f(a)} [1 - f(c)]  
\end{align}
valid under condition $a-b+c-d=0$ and Eq.~\eqref{tau_sigma} we obtain
\begin{align}
\frac{1}{\bar{\tau}_{0k}} =  2\pi   \sum_{k',q }V_q^2
\frac{ f_{0,k+q} f_{0k'} }{ f_{0k}} (1 - f_{0k'+q})
\delta(\xi_{0k} - \xi_{0,k+q} + \xi_{0k'+q} -\xi_{0k'} )
\end{align}
which is Eq.~\eqref{tau_bar_c}.

\subsection{Derivation of the expressions, Eqs.~\eqref{rate_0}, \eqref{qp_rate1a} and \eqref{qp_rate1b} for the quasi-particle inelastic relaxation rates}
\label{app:qp_rates}

Since the Eqs.~\eqref{Sigma_0} are of the form identical to Eq.~\eqref{sig_13}
we have for the analytic continuation of the hole self-energy at the $m=0$ band two equivalent expressions,
\begin{align}\label{sig_17}
\Im \Sigma^R_{0,k}(\epsilon) &= U_0^2  \int \frac{d \epsilon'}{ 2 \pi }\sum_{k'} \left[ \tanh\frac{\epsilon'}{2 T} - \coth\frac{\epsilon - \epsilon'}{ 2 T } \right] \Im G^R_{0,k'} ( \epsilon') \Im T^R_{1(2)kk'}(\epsilon - \epsilon')\, .
\end{align}
As we compute the imaginary part of the self energy to the leading order in interactions we set in Eq.~\eqref{sig_17},  
\begin{align}\label{G0}
\Im G^R_{0,k'} ( \epsilon') = - \pi \delta(\epsilon' - \xi_{k'})
\end{align}
to obtain,
\begin{align}\label{sig_18}
\Im\Sigma^R_{0,k}(\epsilon) &= - \frac{1}{2} U_0^2 \sum_{k'} \left[ \tanh\frac{\xi_{k'}}{2 T} - \coth\frac{\epsilon - \xi_{k'}}{ 2 T } \right]  \Im T^R_{1(2)kk'}(\epsilon - \xi_{k'})\, .
\end{align}
With the definition,  Eq.~\eqref{life_time} of the life time we obtain the two alternative expressions, Eqs.~\eqref{rate_0}.

By making a comparison of Eqs.~\eqref{Sigma_1a}  and \eqref{Sigma_1b} to Eq.~\eqref{sig_16}  we obtain for the self-energies of the hole, 
\begin{align}\label{sig_21}
\Im \Sigma^{(a)R}_{1,k}(\epsilon+\omega) &= U_1^2  \int \frac{d \epsilon'}{ 2 \pi }\sum_{k'} \left[ \tanh\frac{\epsilon'+\omega}{2 T} - \coth\frac{\epsilon - \epsilon'}{ 2 T } \right] \Im G^R_{1,k'} ( \epsilon'+\omega) \Im T^R_{1kk'}(\epsilon - \epsilon')\, ,
\end{align}
and in the same fashion,
\begin{align}\label{sig_23}
\Im \Sigma^{(b)R}_{1,k}(\epsilon+\omega) &= U_{\bar{1}}^2  \int \frac{d \epsilon'}{ 2 \pi }\sum_{k'} \left[ \tanh\frac{\epsilon'+\omega}{2 T} - \coth\frac{\epsilon - \epsilon'}{ 2 T } \right] \Im G^R_{1,k'} ( \epsilon'+\omega) \Im T^R_{2kk'}(\epsilon - \epsilon')\, .
\end{align}
With the approximation, $\Im G^R_{1,k'} ( \epsilon'+\omega) = - \pi \delta(\epsilon'+\omega - \xi_{k'} - \omega_{\perp})$ the definition Eq.~\eqref{life_t13} yield one of the expressions in each of the two equation Eqs.~\eqref{qp_rate1a} and \eqref{qp_rate1b}.

Making use of the definition of the scattering rates, Eq.~\eqref{life_t13}, the correspondence between Eq.~\eqref{sigC_13} and Eqs.~\eqref{Sigma_1a} and \eqref{Sigma_1b} and the approximation Eq.~\eqref{G0} we recover the other two expressions in Eqs. \eqref{qp_rate1a} and \eqref{qp_rate1b}.

\section{Derivation of expression Eq.~\eqref{M_1}}
In order to derive the expression Eq.~\eqref{M_1} we define the auxiliary function $Q_{kk'}(i \omega_n)$ to coincide with the left hand side of Eq.~\eqref{BS13},
\begin{align}\label{Q_1}
Q_{kk'}(i \omega_n ) = 
T\sum_{- i \omega_n < \epsilon_n < 0} \mathcal{G}_{1,k}(\epsilon_n + \omega_n)  \mathcal{G}_{0,k}(\epsilon_n) \Gamma_{k,\epsilon_n}(\omega_n)
\end{align}
The summation over $\epsilon_n$ in Eq.~\eqref{Q_1} is done by the method of analytic continuation, $i \epsilon_n \rightarrow z$.
The analytic continuation of the vertex function $\Gamma_{k,\epsilon_n}(\omega_n)$ is regular in the strip $- \omega < \Im z <0$.
For any function for any function $f(i \epsilon_n, i \omega_n)$ satisfying this condition we have
\begin{align}\label{sum_e}
T \sum_{- \omega_n < \epsilon_n < 0} f(i \epsilon_n,\omega_n) = \int \frac{ d \epsilon}{ 4 \pi i } \left[ \tanh\frac{ \epsilon + \omega}{ 2 T } -  \tanh\frac{ \epsilon }{ 2 T } \right]f(\epsilon, \omega)\, ,
\end{align}
where the right hand side is analytic continuation of the sum on the left hand side. 
Making use of the relation Eq.~\eqref{sum_e} we obtain for the analytic continuation of the function $Q_{kk'}(i \omega_n )$,
\begin{align}\label{Q_2}
Q_{kk'}(\omega ) = \int \frac{ d \epsilon}{ 4 \pi i } \left[ \tanh\frac{ \epsilon + \omega }{ 2 T } - \tanh\frac{ \epsilon }{ 2 T }\right] G_{1,k}^R(\epsilon + \omega) G_{0,k}^A(\epsilon ) \Gamma_{\epsilon,\omega;k}\, ,
\end{align}
The Green functions in Eq.~\eqref{Q_2} read
\begin{align}\label{Q_3}
G^R_{1,k}(\epsilon  + \omega)  =  \frac{1}{\epsilon+ \omega - \xi_{1,k} + i / 2 \tau_{1,k} }\, ,\quad
G^A_{1,k}(\epsilon )  =\frac{1}{\epsilon - \xi_{0,k} - i / 2 \tau_{0,k} }\, ,
\end{align}
where we have omitted the effects of the spectrum renormalization, and the scattering rates are assumed to be smooth functions of energy.
As such $1/\tau_{0,k,\epsilon}$ and $1/\tau_{1,k,\epsilon+\omega}$ are evaluated at the corresponding mass shells,
$\epsilon  = \xi_{0,k}$ and $\epsilon + \omega = \xi_{1,k}$ respectively.
The energy argument of the scattering rates can therefore be suppressed.
The product of the two Green functions in Eq.~\eqref{Q_3} can be transformed as follows,
\begin{align}\label{Q_4}
G^R_{1,k}(\epsilon  + \omega) G^A_{1,k}(\epsilon )=
\frac{ 1 }{ \omega - \omega_{\perp} + i / 2 \tau_{1,k} + i / 2 \tau_{0,k} }
\left[ 
\frac{1}{\epsilon - \xi_{0,k} - i / 2 \tau_{0,k} }
-
\frac{1}{\epsilon+ \omega - \xi_{1,k} + i / 2 \tau_{1,k} }
\right]\, .
\end{align}
The product of the retarded and advanced Green functions, Eq.~\eqref{Q_4} multiplies functions of energy $\epsilon$ having the scale of variation parametrically exceeding the quasi-particle relaxation rates $ 1/  \tau_{1,k},1 /  \tau_{0,k}$.
The example of such a function is the vertex function $\Gamma_{k,\epsilon}(\omega)$.
The scattering rates control the $\epsilon$ dependence of the Green function product in Eq.~\eqref{Q_4} contained entirely in the difference of the two Green functions in the square brackets in Eq.~\eqref{Q_4}.
The real part of the difference of the retarded and advanced Green functions is convergent upon $\epsilon$ integration. 
As the function multiplying this combination are smooth functions of $\epsilon$ the real part of the expression in square brackets of expression Eq.~\eqref{Q_4} gives negligible contribution to the integrals over $\epsilon$.
To the contrary, the imaginary part of this combination produces a finite contribution.
Under the above assumptions we can approximate, $(\epsilon - \xi_{0,k} - i / 2 \tau_{0,k})^{-1} \approx i \pi \delta( \epsilon - \xi_{0,k} )$ and
$(\epsilon + \omega - \xi_{1,k} + i / 2 \tau_{1,k})^{-1} \approx - i \pi \delta( \epsilon + \omega - \xi_{1,k}  )$.
These two forms are equivalent close to the resonance, $\omega \approx \omega_{\perp}$ and we write
\begin{align}\label{Q_5}
G^R_{1,k}(\epsilon  + \omega) G^A_{1,k}(\epsilon )\approx
\frac{ 2 \pi i \delta( \epsilon - \xi_{0,k} )  }{ \omega - \omega_{\perp} + i / 2 \tau_{1,k} + i / 2 \tau_{0,k} }\, .
\end{align}
Note that the strong $\omega$ dependence of the product, Eq.~\eqref{Q_4} has been isolated in the denominator of Eq.~\eqref{Q_5}.

Equation \eqref{M_1} is readily obtained from substitution of Eq.~\eqref{Q_5} in Eq.~\eqref{Q_2}. 

\section{Derivation of the expression, Eq.~\eqref{Gamma_1_13} for the matrix $T_{1kk'}$}
\label{app:M2}
To derive the Eq.~\eqref{Gamma_1_13} we obtain the analytic continuation of the expression of the generic type,
\begin{align}\label{S_kk}
S_{kk'}(i \omega_n)
=  
T^2 \sum_{\epsilon_n} \sum_{\epsilon_n'} 
\mathcal{G}_{k}^+(\epsilon_{n} + \omega_n) \mathcal{G}_{k}^-(\epsilon_{n} )   
 \varphi_{kk'}(i \epsilon_n - i \epsilon_{n'} ) 
\mathcal{G}_{k'}^+(\epsilon_{n'} + \omega_n) \mathcal{G}_{k'}^-(\epsilon_{n'}) 
\Gamma_{k',\epsilon_n',\omega_n}\, , 
\end{align}
where the summations over $\epsilon_n$ and $\epsilon_{n'}$ are limited by the $- i \omega_n$ from below and by zero from above,
the superscript of the correlation functions denote the sign of the imaginary part of the frequency argument.
Hence, the analytic continuation of $\mathcal{G}_{k}^{\pm}(\epsilon_{n})$ to the real axis, $i \epsilon_n \rightarrow \epsilon \pm i 0$ gives retarded and advanced Green functions, $G^{R(A)}_k(\epsilon)$ respectively.

We start with the summation over  $\epsilon_{n'}$.
This summation is standard and is performed by the method of analytic continuation into a complex $z'$ plane, $i \epsilon_{n'} \rightarrow z'$, see \cite{Abrikosov1963} for details.
The resulting expression contains four terms, $S_{kk'}(i \omega_n) = S_{kk'}^A(i \omega_n) + S_{kk'}^{B_+}(i \omega_n) + S_{kk'}^{B_{-}}(i \omega_n)+ S_{kk'}^C(i \omega_n)$ obtained by the transformation of the discrete sums over Matsubara  frequencies to the integrals along the the four  contours $A$, $B_{\pm}$ and $C$ parametrized by the real integration variable $\epsilon'$ in the complex $z'$ plane, respectively. 
The contour $A$ is horizontal and runs in the negative direction just below the real axis.
It is parametrized by the real integration variable $\epsilon'$ as $z' =\epsilon' - i 0$.
Similarly, the contours $B_{\pm}$ are $z' = \epsilon' + i \epsilon_n   \pm i 0$, run in positive and negative directions respectively, and finally the contour $C$ is $z' = \epsilon'- i \omega_n + i 0$ and runs in the positive direction. 
The contributions of each one of the four contours read,
\begin{align}\label{SA}
S^A=- T\sum_{\epsilon_n} \int \frac{ d \epsilon' }{ 4 \pi i } \tanh \frac{\epsilon'}{ 2 T} 
\mathcal{G}^+_{k}(\epsilon_{n} + \omega_n) \mathcal{G}^-_{k}(\epsilon_{n} )  \varphi_{kk'}^-(i\epsilon_{n}  - \epsilon' ) \mathcal{G}^+_{k'}(\epsilon' + i \omega) G^A_{k'}(\epsilon')\Gamma_{k',\epsilon',\omega}  ,
\end{align}
\begin{align}\label{SB}
S^{B_+} + S^{B_{-}} = - 2 i T\sum_{\epsilon_n} 
\int \frac{ d \epsilon' }{ 4 \pi i } \coth \frac{\epsilon'}{ 2 T} 
\mathcal{G}_{k}^+(\epsilon_{n} + \omega_n) \mathcal{G}^-_{k}(\epsilon_{n} ) 
\mathrm{Im} \chi^R_{kk'}(-\epsilon')
\mathcal{G}_{k'}^+( \epsilon' + i \epsilon_n   + \omega_n) \mathcal{G}^-_{k'}(\epsilon' + i \epsilon_n  )
\Gamma_{k',\epsilon' + \epsilon,\omega} ,
\end{align}
\begin{align}\label{SC}
S^C =T\sum_{\epsilon_n} \int \frac{ d \epsilon' }{ 4 \pi i } \tanh \frac{\epsilon'}{ 2 T} 
\mathcal{G}_{k}^+(\epsilon_{n} + \omega_n) \mathcal{G}_{k}^-(\epsilon_{n} ) \varphi_{kk'}^+(i\epsilon_{n}  - \epsilon' + i \omega_n)G^R_{k'}(\epsilon' ) \mathcal{G}^-_{k'}(\epsilon'- \omega)
\Gamma_{k',\epsilon'-\omega,\omega} 
\end{align}
In deriving Eq.~\eqref{SB} we have used the relation $\tanh(x \pm i \pi/2 ) = \coth(x)$ and the jump of the function $\varphi_{kk'}^-(i\epsilon_{n}  - z' )$ across the contours $B_{\pm}$ being equal to $- 2 i \mathrm{Im} \varphi_{kk'}^R(-\epsilon')$.

The remaining summation over $\epsilon_n$ again can be done by the same method of analytic continuation, $i \epsilon_n \rightarrow z$ in the same fashion as the summation over $\epsilon_{n'}$ above.
In this case, however the function of $z'$ is analytic in the whole strip, $- \omega_n  < \Im z <  0$ which allows us to use the general relation 
\begin{align}\label{sum_e}
T \sum_{- \omega_n < \epsilon_n < 0} f(i \epsilon_n,\omega_n) = \int \frac{ d \epsilon}{ 4 \pi i } \left[ \tanh\frac{ \epsilon + \omega}{ 2 T } -  \tanh\frac{ \epsilon }{ 2 T } \right]f(\epsilon, \omega)
\end{align}
where the right hand side is analytic continuation of the sum on the left hand side. 
The relation, Eq.~\eqref{sum_e} holds for any function $f(i \epsilon_n)$ such that its analytic continuation, $f(z)$  is analytic in the region $- \omega_n  < \Im z <  0$.
We hence perform the summation over $\epsilon_n$ in Eqs.~\eqref{SA}, \eqref{SB} and \eqref{SC} using Eq.~\eqref{sum_e},
\begin{align}\label{SA_1}
S^A = 
-  \int \frac{ d \epsilon }{ 4 \pi i } \left[ \tanh \frac{\epsilon + \omega}{ 2 T} -  \tanh \frac{\epsilon}{ 2 T} \right] \int \frac{ d \epsilon' }{ 4 \pi i } \tanh \frac{\epsilon'}{ 2 T} 
G_{k}^R(\epsilon + \omega) G_{k}^A(\epsilon) 
\varphi_{kk'}^A(\epsilon  - \epsilon' ) 
G^R_{k'}(\epsilon' + \omega) G^A_{k'}(\epsilon') \Gamma_{k',\epsilon',\omega} \, ,
\end{align}
\begin{align}\label{SB_1}
S^{B_+} + S^{B_-} =  - & 2 i
\int \frac{ d \epsilon }{ 4 \pi i } \left[ \tanh \frac{\epsilon + \omega}{ 2 T} -  \tanh \frac{\epsilon}{ 2 T} \right]  
 \int \frac{ d \epsilon' }{ 4 \pi i } \coth \frac{\epsilon'}{ 2 T} 
G_{k'}^R( \epsilon' + \epsilon   + \omega) G_{k'}^A(\epsilon' + \epsilon  )
\notag \\
& \times 
\mathrm{Im} \chi^R(-\epsilon')
G_{k}^R(\epsilon + \omega) G_{k}(\epsilon ) \Gamma_{k',\epsilon' + \epsilon,\omega} 
\end{align}
\begin{align}\label{SC_1}
S^C = &
 \int \frac{ d \epsilon }{ 4 \pi i } \left[ \tanh \frac{\epsilon + \omega}{ 2 T} -  \tanh \frac{\epsilon}{ 2 T} \right]   \int \frac{ d \epsilon' }{ 4 \pi i } \tanh \frac{\epsilon'}{ 2 T} 
  G_{k}^R(\epsilon + \omega) G_{k}^A(\epsilon )
\varphi_{kk'}^R(\epsilon  - \epsilon' +  \omega) 
G^R_{k'}(\epsilon' ) G^A_{k'}(\epsilon'- \omega)
 \Gamma_{k',\epsilon'-\omega,\omega}
\end{align}
We keep the imaginary part of the correlation function, $\varphi_{kk'}(\epsilon - \epsilon')$ by replacing   $\varphi_{kk'}^{R(A)}(\epsilon - \epsilon')$
with $\pm i \Im  \varphi_{kk'}^{R}(\epsilon - \epsilon')$ in Eqs.~\eqref{SA_1} and \eqref{SC_1} and shift the integration variable, $\epsilon'$ by $\omega$ in the latter to write,
\begin{align}\label{SA_2}
S^A = 
i \int \frac{ d \epsilon }{ 4 \pi i } \left[ \tanh \frac{\epsilon + \omega}{ 2 T} -  \tanh \frac{\epsilon}{ 2 T} \right] \int \frac{ d \epsilon' }{ 4 \pi i } \tanh \frac{\epsilon'}{ 2 T} 
G_{k}^R(\epsilon + \omega) G_{k}^A(\epsilon) 
\Im \varphi_{kk'}^R(\epsilon  - \epsilon' ) 
G^R_{k'}(\epsilon' + \omega) G^A_{k'}(\epsilon') \Gamma_{k',\epsilon',\omega} \, ,
\end{align}
\begin{align}\label{SC_2}
S^C = &
i \int \frac{ d \epsilon }{ 4 \pi i } \left[ \tanh \frac{\epsilon + \omega}{ 2 T} -  \tanh \frac{\epsilon}{ 2 T} \right]   \int \frac{ d \epsilon' }{ 4 \pi i } \tanh \frac{\epsilon'+\omega}{ 2 T} 
  G_{k}^R(\epsilon + \omega) G_{k}^A(\epsilon )
\Im \varphi_{kk'}^R(\epsilon  - \epsilon') 
G^R_{k'}(\epsilon' +\omega) G^A_{k'}(\epsilon')
 \Gamma_{k',\epsilon',\omega}
\end{align}
We rewrite Eq.~\eqref{SB_1} by shifting the $\epsilon'$ integration variable to $\epsilon ' - \epsilon$, and using the property, 
$\Im  \varphi_{kk'}^{R}(-x) = - \Im  \varphi_{kk'}^{R}(x)$ as
\begin{align}\label{SB_2}
S^{B_+} + S^{B_-} =  - & 2 i
\int \frac{ d \epsilon }{ 4 \pi i } \left[ \tanh \frac{\epsilon + \omega}{ 2 T} -  \tanh \frac{\epsilon}{ 2 T} \right]  
 \int \frac{ d \epsilon' }{ 4 \pi i } \coth \frac{\epsilon'-\epsilon}{ 2 T} 
G_{k'}^R( \epsilon' +  \omega) G_{k'}^A(\epsilon'  )
\notag \\
& \times 
\mathrm{Im} \varphi^R(\epsilon-\epsilon')
G_{k}^R(\epsilon + \omega) G_{k}(\epsilon ) \Gamma_{k',\epsilon' ,\omega} 
\end{align}
Adding up the contributions \eqref{SA_2}, \eqref{SB_2} and \eqref{SC_2} we obtain for the analytic continuation of the sum $S_{kk'}(\omega)$ defined in Eq.~\eqref{S_kk},
\begin{align}\label{Gamma_1}
S_{kk'}(\omega)  
 =
i & 
 \int \frac{ d \epsilon }{ 4 \pi i } \left[ \tanh \frac{\epsilon + \omega}{ 2 T} -  \tanh \frac{\epsilon}{ 2 T} \right] \int \frac{ d \epsilon' }{ 4 \pi i }
 \left[
\tanh \frac{\epsilon + \omega}{ 2 T} +  \tanh \frac{\epsilon'}{ 2 T} - 2\coth \frac{\epsilon' - \epsilon}{ 2 T} 
\right]  
\notag \\
& \times
G_{k}^R(\epsilon + \omega) G_{k}^A(\epsilon)
\mathrm{Im}\varphi_{kk'}^R(\epsilon  - \epsilon' ) 
G^R_{k'}(\epsilon' + \omega) G^A_{k'}(\epsilon') \Gamma_{k',\epsilon',\omega} \, .
\end{align}
Integration of \eqref{Gamma_1} over $\epsilon$ and $\epsilon'$ using the representation, \eqref{Q_5} 
yields Eq.~\eqref{Gamma_1_13}.

\end{widetext}

\section{Correlation functions and memory function formalism}
\label{app:corr-func}
In this appendix we summarize the basic definitions and properties of the correlation functions within the linear response theory.
We then summarize the memory function formalism.
Although the presentation follows closely the Ref.~\cite{Forster1975} we generalize the latter to include properties used in the present work.
We focus on the dynamics of slow variables $A_n$, $n=1,2,\ldots$.
The system is assumed to be symmetric under the time reversal operation $\mathcal{O}$, such that $\mathcal{O}^{-1} H \mathcal{O} = H$.
Furthermore, the slow variables are assumed to have a definite signature under time reversal, $\mathcal{O}$, such that $\mathcal{O}^{-1} A_n \mathcal{O} = \epsilon_{A_n} A_n$.
The variables $A_n$ are not in general Hermitian and the general scheme requires generalization as outlined below. 
In the main text we work with variables defined by \eqref{PT1}.
These variables are non-Hermitian and time reversal even.
Indeed, they are proportional to the raising operator of the harmonic operator, $z + i p_z$ projected onto a pair of adjacent bands. 
In the text the variables $A_n$ are singled out thanks to their simple harmonic time dynamics, $A_n(t) = A_n  e^{- i \omega_{\perp} t}$ in the leading approximation.
As a result these variables would be more properly referred to as resonant rather than slow variables.
Since the crucial element of the whole analysis is the time scales separation this distinction plays no role, and we follow the standard terminology.
For the sake of generality we allow for the $\bm{r}$-dependence of the operators, $A_n(\bm{r},t)$.

The response functions are introduced via modifying the Hamiltonian, $H$ to include the perturbation,
\be\label{Mem10}
 K = H -\sum_i \int d \bm{r} A_{i}(\bm{r}) \delta a_i^{ext}(\bm{r},t)\, ,
\ee 
where $\delta a_i^{ext}(\bm{r},t)$ is a time dependent perturbation field coupled to a variable $A_{i}(\bm{r})$.
The response obtained from the equations of motion,
\be\label{Mem11}
\delta \langle A^{\dag}_{i}(\bm{r},t) \rangle = \sum_j \int_{-\infty}^t d t' \int d \bm{r}' 2 i \chi''_{ij}(\bm{r}t,\bm{r}'t') \delta a_j^{ext}(\bm{r}' t')
\ee 
where we have defined the function,
\be\label{Mem13}
\chi^{''}_{ij}(\bm{r}t,\bm{r}'t')  = \frac{ 1 }{ 2 \hbar} \langle [ A_i^{\dag}(\bm{r}t),A_{j}(\bm{r}'t') ]_- \rangle
\ee
and the time dependence, $A_n(\bm{r},t) = e^{ i H t/\hbar} A_{n}(\bm{r}) e^{- i H t/\hbar}$
is according to the Heisenberg representation.
In the stationary, and translationally invariant system, $\chi^{''}_{ij}(\bm{r}t,\bm{r}'t') = \chi^{''}_{ij}(\bm{r}-\bm{r}',t-t')$.

With the definition of the Fourier transform of a function, $F(\bm{r},t)$ as 
\be\label{Fourier}
F(\bm{k},\omega) = \int_{-\infty}^{\infty} dt \int d \bm{r} e^{i \omega t - \bm{k} \bm{r}} F(\bm{r},t)
\ee
the response function $\chi_{ij}(\bm{k},\omega)$ satisfying
\be\label{Mem15}
\delta \langle A_i^{\dag} \rangle (\bm{k},\omega)  = \sum_j \chi_{ij}(\bm{k},\omega) \delta a_j^{ext}(\bm{k},\omega)\, .
\ee
reads
\be\label{Mem22}
\lim_{\epsilon \rightarrow 0} \chi_{ij}(k,\omega + i \epsilon)  = \chi_{ij}(k,\omega) = - K^R_{A_i,A_j}(k,\omega)
\ee
where 
\begin{align}\label{Mem23}
K^R_{i,j}(k,\omega) =  - i \int d t e^{ i \omega t } \theta(t) \langle [A_i^{\dag}(k,t), A_j(k,0)]_- \rangle
\end{align}

It is often useful to define the Laplace transformation of a function $F(t)$ for $\Im z \geq 0$,
\be\label{Laplace}
F(z) = \int_0^{\infty} d t e^{ i z t } F(t)
\ee
The Laplace transformation, Eq.~\eqref{Laplace} is related to the Fourier transformation, Eq.~\eqref{Fourier} as ($\Im z >0$),
\be
F(z) = \int_{-\infty}^{\infty} \frac{d \omega}{2 \pi i} \frac{ F(\omega)}{\omega - z}  \, .
\ee
The response function in \eqref{Mem15} is related to the function of the complex variable $z$ ($\Im z \neq 0$) defined as 
\be\label{Mem17}
\chi_{ij}(\bm{k},z) = \int \frac{d \omega}{\pi} \frac{ \chi''_{ij}(\bm{k},\omega)}{\omega - z}
\ee
via the analytic continuation,
\be\label{Mem19}
\chi_{ij}(\bm{k},\omega) = \lim_{\epsilon \rightarrow 0} \chi_{ij} ( \bm{k} ,z)\vert_{z = \omega + i \epsilon}
\ee
from the upper complex half plain, $\Im z >0$, where the definition \eqref{Mem17} is equivalent to
\be\label{Mem21}
\chi_{ij}(k,z) = 2 i \int_0^{\infty} d t e^{ i zt} \chi''_{ij}(k,t)\, .
\ee

Now introduce the Kubo relaxation function,
\begin{align}\label{Mem27}
C_{ij}(r,r',t-t') = \int_0^{\beta} d \beta' 
\big[ & \langle A_i^{\dag} (r,t) A_j (r', t' + i \hbar \beta') \rangle 
\notag \\
- & \langle A_i^{\dag} \rangle \langle A_j \rangle \big]\, .
\end{align}
This function has the several properties crucial in constructing the formalism.
\subsection{Equivalence of equal time Kubo correlation function and the scalar product in the space of observables}
The function \eqref{Mem27} defines the scalar product at coinciding times,
\begin{align}\label{Mem27a}
& C_{ij}(r,r',t=t')  \equiv \langle A_i  | A_j  \rangle 
\notag \\
& = \int_0^{\beta} d \beta' 
\left[ \langle A_i^{\dag} (r) A_j (r',  i \hbar \beta') \rangle -  \langle A_i^{\dag} \rangle \langle A_j \rangle \right]
\end{align}
provided that all participating variables are normal ordered, i.e. their equilibrium or invariant part has beed subtracted out,
in such a way that $\langle A_i \rangle = 0$.
This property is important to construct the memory matrix formalism.
However, the use of the original operators, $A_{i}$ and their normal ordered counterpart, $A_{i} - \langle A_i \rangle$ leads to identical expressions.
We will comment on this below. 
For now let us show the property \eqref{Mem27a}.
The properties, $\langle A | c_1 B_1 + c_2 B_2\rangle =c_1   \langle A | B_1\rangle  + c_2\langle A| B_2\rangle $ and
$\langle c_1 A_1 + c_2 A_2 | B \rangle = c_1^*  \langle A_1 | B \rangle + c_2^* \langle A_2 | B \rangle $ are obvious.
Another property, $\langle A|B\rangle^* =  \langle B | A \rangle$ is shown as follows
\begin{align}
Z^{-1}&\int_0^{\beta}  d \tau \Tr 
[ e^{-\beta \hat{H} } A^{\dag} e^{- \tau \hat{H}}  B  e^{+ \tau \hat{H}}  ]^*
\notag \\
& =
Z^{-1}\int_0^{\beta} d \tau \Tr 
[ e^{-\beta \hat{H} } A^{\dag} e^{- \tau H}  B  e^{+ \tau \hat{H}}  ]^{\dag}
\notag \\
& =
Z^{-1}\int_0^{\beta} d \tau \Tr 
\left[ e^{+ \tau H} [e^{-\beta \hat{H} } A^{\dag} e^{- \tau \hat{H}}  B ] ^{\dag}   \right]
\notag \\
& =
Z^{-1}\int_0^{\beta} d \tau \Tr 
\left[ e^{+ \tau H} B^{\dag} [e^{-\beta \hat{H} } A^{\dag} e^{- \tau \hat{H}} ] ^{\dag}   \right]
\notag \\
& =
Z^{-1}\int_0^{\beta} d \tau \Tr 
[ e^{+ \tau H}         B^{\dag}   e^{- \tau H} A  e^{-\beta \hat{H} }       ]^{\dag}
\notag \\
& =
Z^{-1}\int_0^{\beta} d \tau \Tr 
[ e^{-\beta \hat{H} } B^{\dag} e^{- \tau \hat{H}}  A  e^{+ \tau \hat{H}}  ]\, .
\end{align}
And since $[\langle A^{\dag} \rangle \langle B \rangle ]^* = \langle B^{\dag} \rangle \langle A \rangle$ the postulated property follows,
where $Z = \mathrm{Tr} e^{ -\beta H}$ is the partition function.

It remains to show that $\langle A | A \rangle =0$ implies $A=0$.
To show this we denote $p_n = Z^{-1} e^{-\beta E_n}$.
We have,
\begin{align}
\langle A | A \rangle & = 
\int_0^{\beta} d \tau \Big[ \sum_{n,m}  p_n A^*_{mn} A_{mn} e^{ - (E_m -E_n)\tau} 
\notag \\
&- \sum_{n} p_n A_{nn} \sum_m p_m A_{mm} \Big]
\notag \\
 = &
\int_0^{\beta} d \tau  \sum_{n,m} p_n \left| A_{nm} - \delta_{nm} \sum_{k} p_k A_{kk} \right|^2 e^{ - (E_m - E_n) \tau } 
\notag \\
&\geq 0
\end{align}
as it should be.
However, $\langle A | A \rangle  = 0$ implies $A = \langle A \rangle$ rather than $A =0$.
We see, therefore that strictly speaking $\langle A | A \rangle$ is not a scalar product.
To remedy for this, one can decide to work with quantities that have zero equilibrium statistical averages.
Or in other words with their fluctuation part subtracted, \cite{Mori1965}.
Alternatively one could refer to normal ordered operators only to avoid the above complication.
Then \eqref{Mem27} is a scalar product.
The reason why this formal point is usually omitted,  \cite{Forster1975} is that computing the frequency shift and the memory matrix with observables, $A_i - \langle A_i \rangle$ and $A_i $ gives exactly the same answers.
This is physically obvious, and formally is a consequence of the structure of the memory function formalism.
Namely that all the correlation functions are expressed in terms of the commutators of the observables with Hamiltonian, and clearly subtracting the invariant part,   $\langle A_i \rangle$, as referred by  \cite{Mori1965} does not affect final expressions even though it is formally important at the intermediate stages of the derivations.
\subsection{Asymptotic behavior of the Kubo correlation function at long times}
\be\label{Mem29}
\lim_{t-t' \rightarrow \infty} C_{ij}(r,r',t-t') = 0\, .
\ee
This follows from the subtraction of the product of the expectation $\langle A_i^{\dag} \rangle \langle A_j \rangle$ values in \eqref{Mem27}.
\subsection{Time derivative of the Kubo correlation function}
Lets demonstrate the important property,
\be\label{Mem31}
i \partial_t C_{ij}(r,r',t-t') = 2 \chi''_{ij}(r,r',t-t')\, .
\ee
To see this property we perform the transformations,
\begin{align}\label{Mem33}
&i  \partial_t C_{ij}(r,r',t-t') = i \partial_t \int_0^{\beta} d \beta' 
 \langle A_i^{\dag} (r,t) A_j (r', t' + i \hbar \beta') \rangle 
\notag \\ 
& =
i  \int_0^{\beta} d \beta' 
 \langle \partial_t A_i^{\dag} (r,t) A_j (r', t' + i \hbar \beta') \rangle 
 \notag \\
& =
-  i  \int_0^{\beta} d \beta' 
 \langle A_i^{\dag} (r,t) \partial_t  A_j (r', t' + i \hbar \beta') \rangle 
\notag \\
& =
- \frac{ i }{ i \hbar} \int_0^{\beta} d \beta' 
 \langle A_i^{\dag} (r,t) \partial_{\beta'}  A_j (r', t' + i \hbar \beta') \rangle 
 \notag \\
& =\!
\frac{ 1}{  \hbar}\! \left[\! \langle \! A_i^{\dag} (r,t) A_j (r', t') \rangle \! - \! \langle A_i^{\dag} (r,t)  A_j (r', t' \!+\! i \hbar \beta)\! \rangle\! \right]
\end{align}
Now observe that
\begin{align}\label{Mem35}
&\langle A_i^{\dag} (r,t)  A_j (r', t' + i \hbar \beta) \rangle = 
\notag \\
& Z^{-1}
\mathrm{Tr} \left[ e^{ - \beta H} A_i^{\dag} (r,t)  e^{ i (i\beta \hbar)  H/\hbar} A_j (r', t')e^{ -i (i\beta \hbar)  H/\hbar}\right]
\notag \\
& = \langle A_j (r', t' )  A_i^{\dag} (r,t)  \rangle \, .
\end{align}
Therefore, substituting \eqref{Mem35} in \eqref{Mem33} gives by comparison with \eqref{Mem13},
\be\label{Mem37}
i  \partial_t C_{ij}(r,r',t-t') = \frac{ 1 }{ \hbar} \langle [ A_i^{\dag} (r,t),  A_j (r', t') ]_- \rangle = 2 \chi^{''}_{ij}(rt,r't').
\ee
\subsection{Equivalence of the equal time Kubo correlation function and static, zero frequency response function}
Equal time Kubo relaxation function, \eqref{Mem27} is equal to the zero frequency response function \eqref{Mem19},
\begin{align}\label{Mem39}
C_{ij}(r,r',& t=0)  = \langle A_i | A_j \rangle= \chi_{ij}(r,r',\omega = 0) 
\notag \\
& = \chi_{ij}^s(r,r') = -K_{A_i,A_j}^R(r,r',\omega = 0)\, ,
\end{align}
where the superscript $s$ stands for the static correlation function.
Integrate \eqref{Mem37} using that our system is time translational invariant, so all the correlation functions depend on the time difference $t-t'$, and use the property \eqref{Mem29}, 
 to write
\begin{align}\label{Mem41}
i  \int_0^{\infty} & \partial_t C_{ij}(r,r',t') =  - i  C_{ij}(r,r',t' = 0)  
 =  2 \int_{0}^{\infty} d t'\chi^{''}_{ij}(r,r't') 
 \notag \\
&= \lim_{\epsilon \rightarrow 0} 
2 \int_{0}^{\infty} d t'\chi^{''}_{ij}(r,r't') e^{ - \epsilon t}\, ,
\end{align}
where in the last infinity we used the fact that $\chi^{''}_{ij}(r,r't')$ vanishes in the $t' \rightarrow \infty$ limit because it is defined via the commutator in \eqref{Mem13}.
With \eqref{Mem21}, \eqref{Mem41} becomes 
\begin{align}\label{Mem43}
& C_{ij}(r,r',t' = 0)  = 2 i  \lim_{\epsilon \rightarrow 0}  \int_{0}^{\infty} d t'\chi^{''}_{ij}(r,r't') e^{ - \epsilon t} 
\notag \\
& =\lim_{\epsilon \rightarrow 0}  \chi_{ij}(k,z= i \epsilon ) 
= \chi_{ij}(r,r',\omega = 0) 
\notag \\
&
= \chi_{ij}^s(r,r') = -K^R(r,r',\omega = 0)\, .
\end{align}
\subsection{Laplace transformation of the Kubo relaxation function}
We introduce the Laplace transform of the Kubo relaxation function, \eqref{Mem27} as
\be\label{Mem45}
C_{ij}(r,r',z)  = \int_0^{\infty} d t e^{ i z t}  C_{ij}(r,r',t).
\ee
The crucial relationship is
\be\label{Mem47}
C_{ij}(r,r',z)  = \frac{ 1 }{ i z } \left[ K^R_{A_i,A_j}(r,r';z = i 0) - K^R_{A_i,A_j}(r,r'; z) \right], 
\ee 
where $\Im z > 0$ and $K^R(r,r';z)$ is a function that is analytic in the upper complex plain, $\Im z >0$ and coincides with the retarded Green function at the real axis, $\lim_{\epsilon \rightarrow 0} K^R(r,r';z=\omega + i \epsilon ) = K^R(r,r';\omega)$.
Clearly this function coincides with the Matsubara Green function at the discrete Matsubara frequencies, $z = i \Omega_n$ in the same region $\Im z >0$, $\Omega_n > 0$. 
In fact we will always use the relationship \eqref{Mem47}  for $z = \omega + i \epsilon$ in the limit $\lim_{\epsilon \rightarrow 0}$.

To show \eqref{Mem47} integrate \eqref{Mem45} by parts,
\begin{align}\label{Mem51}
C_{ij}&(r,r',z)   =  \int_0^{\infty} d t \frac{ 1 }{ i z }\frac{ d }{ d t} [e^{ i z t}]  C_{ij}(r,r',t) 
\notag \\
&
= 
- \frac{ 1 }{ i z} C_{ij}(r,r',t=0) - \frac{ 1 }{ i z}  \int_0^{\infty} e^{i z t } \partial_t   C_{ij}(r,r',t) 
\notag \\
& =
 \frac{ 1 }{ i z} K^R_{ij}(r,r',\omega=0) - \frac{ 1 }{ i z}  K^R_{ij}(r,r',z),
\end{align}
where we used \eqref{Mem21}, \eqref{Mem22}, \eqref{Mem37} and \eqref{Mem39}.

\section{Derivation of the expression \eqref{MM} for the correlation function $K^{(0)}_{\bar{M},\bar{M}}$}
\label{der:MM}
Equation \eqref{MM} is most simply derived by using the fluctuation dissipation relation between the imaginary part of the (retarded) response and correlation functions,
\begin{align}\label{MM_app1}
\Im K_{\bar{M},\bar{M}} = - \frac{1}{2} ( 1 - e^{- \omega/T} )i K^>_{\bar{M}\bar{M}}(\omega)\, ,
\end{align}
where the $iK^>_{\bar{M}\bar{M}}(t) = \langle \bar{M}(t) \bar{M}(0) \rangle$
Within the Keldysh diagrammatic technique  \cite{Kamenev2011}
\begin{align}\label{MM_app2}
& i K^>_{\bar{M}\bar{M}}(\omega) = \sum_{\bm{k},\bm{k}',\bm{q}} \int \frac{ d \epsilon d \epsilon' d \Omega}{ (2\pi)^3} (V_q^2 - V_q V_{q - k' + k})i G^<_{0,k}(\epsilon)
\notag \\
& \times
 i G^>_{1,k-q}(\epsilon + \omega - \Omega)
  iG^<_{0,k'+q} (\epsilon' + \Omega) i G^>_{0,k'}(\epsilon') .
\end{align}
The functions $G^{>,<}$ defined in the standard way,  \cite{Bruus2004} in the non-interacting limit take the form,
$G^>_{m,k}(\epsilon) = 2 \pi ( 1 - f_{m,k}) \delta(\epsilon - \xi_{m,k})$, and $G^<_{m,k}(\epsilon) = - 2 \pi f_{m,k} \delta(\epsilon - \xi_{m,k})$
With these definitions combination of Eqs.~\eqref{MM_app1} and \eqref{MM_app2} reproduces Eq.~\eqref{MM}.

\section{Commutation relations}
\label{app:commute}
The commutations relations used in the main text can be conveniently evaluated using the relation,
$[A, B C]_- = [A,B]_- C + B[A,C]_-$, $[A, B C]_- = [A,B]_+ C - B[A,C]_+$.
And for the fermions
$[\psi_1^{\dag}\psi_2,  \psi_3^{\dag}\psi_4]_- = [\psi_1^{\dag}\psi_2,  \psi_3^{\dag}]_- \psi_4  + \psi_3^{\dag}[\psi_1^{\dag}\psi_2,  \psi_4]_- =
\psi_1^{\dag}[\psi_2,  \psi_3^{\dag}]_+ \psi_4 -\psi_3^{\dag} [\psi_1^{\dag},  \psi_4]_ + \psi_2 = \psi_1^{\dag} \psi_4 \delta_{2,3}  - \psi_3^{\dag}\psi_2 \delta_{1,4}$

\subsection{Commutators with the observables, $A_1 = \sum_p \psi_{p,1}^{\dag} \psi_{p,0}$ and $A_2 = \sum_p \psi_{p,2}^{\dag} \psi_{p,1}$.}
\begin{subequations}
\be\label{mem45}
[A_1,  \frac{1}{2}  \psi_{0,k-q}^{\dag}  \psi_{0,k'+q}^{\dag}    \psi_{0,k'} \psi_{0,k}]_-=
 \psi_{0,k-q}^{\dag} \psi_{1,k'+q}^{\dag}    \psi_{0,k'} \psi_{0,k} 
\ee
\be\label{mem45b}
[A_2,  \frac{1}{2}  \psi_{0,k'+q}^{\dag} \psi_{0,k-q}^{\dag} \psi_{0,k}   \psi_{0,k'}]_-=
0
\ee
\end{subequations}
\begin{subequations}
\be\label{comm_13}
[A_1 ,\frac{1}{2} \psi_{1,k'+q}^{\dag}\psi_{1,k-q}^{\dag}  \psi_{1,k} \psi_{1,k'}   ]_-
= 
- \psi^{\dag}_{1,k'+q} \psi_{1,k-q}^{\dag} \psi_{1,k} \psi_{0,k'}
\ee
\be\label{comm_15}
[A_2 ,\frac{1}{2} \psi_{1,k'+q}^{\dag}\psi_{1,k-q}^{\dag}  \psi_{1,k} \psi_{1,k'}   ]_-
= 
\psi_{2,k'+q}^{\dag} \psi_{1,k-q}^{\dag} \psi_{1,k}   \psi_{1,k'}
\ee
\end{subequations}
\eqref{comm_15} is obtained from \eqref{mem45} by shifting all the band indices by one up.
\begin{subequations}
\be\label{comm_17}
[A_1 ,  \psi_{2,k'+q}^{\dag}\psi_{2,k-q}^{\dag}  \psi_{2,k} \psi_{2,k'}   ]_-
= 
0
\ee
\be\label{comm_17}
[A_2 ,  \psi_{2,k'+q}^{\dag}\psi_{2,k-q}^{\dag}  \psi_{2,k} \psi_{2,k'}   ]_-
=
- \psi^{\dag}_{2,k'+q} \psi_{2,k-q}^{\dag} \psi_{2,k} \psi_{1,k'}
\ee
\end{subequations}
\eqref{comm_17} is obtained from \eqref{comm_13} by shifting all the band indices by one up.
\begin{subequations}
\begin{align}\label{mem38}
[A_1 ,& \psi_{1,k-q}^{\dag} \psi_{0,k'+q}^{\dag}  \psi_{0,k'}   \psi_{1,k}]_-  =
- \psi_{1,k-q}^{\dag} \psi_{0,k'+q}^{\dag} \psi_{0,k'}  \psi_{0,k} 
\notag \\
&
+  \psi_{1,k-q}^{\dag} \psi_{1,k'+q}^{\dag} \psi_{0,k'} \psi_{1,k}  
\end{align}
\begin{align}\label{mem40}
[A_2 , \psi_{1,k-q}^{\dag} \psi_{0,k'+q}^{\dag}  \psi_{0,k'}   \psi_{1,k}]_-  =
 \psi_{2,k-q}^{\dag} \psi_{0,k'+q}^{\dag}  \psi_{0,k'}  \psi_{1,k} 
\end{align}
\end{subequations}
\begin{subequations}
\begin{align}\label{V2a}
[A_1 & , 
 \psi_{1,k-q}^{\dag}   \psi_{0,k'+q}^{\dag} \psi_{1,k'} \psi_{0,k}  ]_- 
= - \psi_{1,k-q}^{\dag} \psi_{0,k'+q}^{\dag}  \psi_{0,k'} \psi_{0,k}
\notag \\
&
+  \psi_{1,k-q}^{\dag} \psi_{1,k'+q}^{\dag}  \psi_{1,k'} \psi_{0,k}
\end{align}
\begin{align}
[A_2 , 
 \psi_{1,k-q}^{\dag}   \psi_{0,k'+q}^{\dag} \psi_{1,k'} \psi_{0,k}  ]_- 
=  \psi_{2,k-q}^{\dag} \psi_{0,k'+q}^{\dag}  \psi_{1,k'}  \psi_{0,k} 
\end{align}
\end{subequations}
\begin{subequations}
\begin{align}
[A_1 & , \psi_{2,k-q}^{\dag}  \psi_{0,k'+q}^{\dag}  \psi_{1,k'} \psi_{1,k} ]_- 
= -\psi_{2,k-q}^{\dag} \psi_{0,k'+q}^{\dag} \psi_{1,k'} \psi_{0,k}   
\notag \\
&
+ \psi_{2,k-q}^{\dag} \psi_{1,k'+q}^{\dag}   \psi_{1,k'}\psi_{1,k}
\end{align}
\begin{align}
[A_2 , \psi_{2,k-q}^{\dag}  \psi_{0,k'+q}^{\dag}  \psi_{1,k'} \psi_{1,k} ]_- =0
\end{align}
\end{subequations}
\begin{subequations}
\be
[A_1 ,  \psi_{1,k-q}^{\dag}   \psi_{1,k'+q}^{\dag}   \psi_{0,k'}\psi_{2,k} ]_-=0
\ee
\begin{align}
&[A_2  ,  \psi_{1,k-q}^{\dag}   \psi_{1,k'+q}^{\dag}   \psi_{0,k'}\psi_{2,k} ]_-=
-\psi_{1,k-q}^{\dag}   \psi_{1,k'+q}^{\dag}   \psi_{0,k'}\psi_{1,k}
\notag \\
&
+
\psi_{1,k-q}^{\dag}   \psi_{2,k'+q}^{\dag}   \psi_{0,k'}\psi_{2,k}
+
\psi_{2,k-q}^{\dag}   \psi_{1,k'+q}^{\dag}   \psi_{0,k'}\psi_{2,k}
\end{align}
\end{subequations}
\begin{subequations}
\be
[A_1 ,  \psi_{2,k-q}^{\dag} \psi_{0,k'+q}^{\dag} \psi_{0,k'}     \psi_{2,k}  ]_-=
\psi_{2,k-q}^{\dag} \psi_{1,k'+q}^{\dag}  \psi_{0,k'}  \psi_{2,k}
\ee
\be
[A_2,  \psi_{2,k-q}^{\dag} \psi_{0,k'+q}^{\dag} \psi_{0,k'}     \psi_{2,k}  ]_-=
-\psi_{2,k-q}^{\dag} \psi_{0,k'+q}^{\dag} \psi_{0,k'}    \psi_{1,k}
\ee
\end{subequations}
\begin{subequations}
\be
[A_1 ,  \psi_{2,k-q}^{\dag} \psi_{0,k'+q}^{\dag} \psi_{2,k'}     \psi_{0,k}  ]_-=
\psi_{2,k-q}^{\dag} \psi_{1,k'+q}^{\dag}  \psi_{2,k'}  \psi_{0,k}
\ee
\be
[A_2 ,  \psi_{2,k-q}^{\dag} \psi_{0,k'+q}^{\dag} \psi_{2,k'}     \psi_{0,k}  ]_-=
-
\psi_{2,k-q}^{\dag} \psi_{0,k'+q}^{\dag}  \psi_{1,k'}  \psi_{0,k}
\ee
\end{subequations}
\begin{subequations}
\be
[A_1 ,  \psi_{2,k'+q}^{\dag}  \psi_{1,k-q}^{\dag} \psi_{1,k}  \psi_{2,k'} ]_-=
-
\psi_{1,k-q}^{\dag} \psi_{2,k'+q}^{\dag}  \psi_{2,k'}  \psi_{0,k}
\ee
\begin{align}
[A_2 & ,   \psi_{1,k-q}^{\dag}  \psi_{2,k'+q}^{\dag} \psi_{2,k'} \psi_{1,k}   ]_-=
\psi_{2,k-q}^{\dag} \psi_{2,k'+q}^{\dag}  \psi_{2,k'}  \psi_{1,k}
\notag \\
& -
\psi_{1,k-q}^{\dag} \psi_{2,k'+q}^{\dag}  \psi_{1,k'}  \psi_{1,k}
\end{align}
\end{subequations}
\begin{subequations}
\be
[A_1 ,  \psi_{2,k-q}^{\dag}  \psi_{1,k'+q}^{\dag}  \psi_{2,k'}  \psi_{1,k} ]_-=
- \psi_{2,k-q}^{\dag} \psi_{1,k'+q}^{\dag}  \psi_{2,k'}  \psi_{0,k}
\ee
\begin{align}
[A_2 &,  \psi_{2,k-q}^{\dag}  \psi_{1,k'+q}^{\dag}  \psi_{2,k'}  \psi_{1,k} ]_-=
-\psi_{2,k-q}^{\dag} \psi_{1,k'+q}^{\dag}  \psi_{1,k'}  \psi_{1,k}
\notag \\
& + \psi_{2,k-q}^{\dag} \psi_{2,k'+q}^{\dag}  \psi_{2,k'}  \psi_{1,k}\, .
\end{align}
\end{subequations}
\end{appendix}

\end{document}